\documentclass[journal=jacsat,manuscript=article]{achemso}

\usepackage{chemformula} % Formula subscripts using \ch{}
\usepackage[T1]{fontenc} % Use modern font encodings

\usepackage[english]{babel}
\usepackage[utf8]{inputenc}
\usepackage{microtype}

\usepackage{subfig}

\usepackage{bm}

\author{Lori\'en L\'opez-Villellas}
\affiliation[Unizar]{Departamento de Inform\'atica e Ingenier{\'i}a de Sistemas / Arag\'on Institute for Engineering Research (I3A), Universidad de Zaragoza, Zaragoza, Spain}
\email{lorien.lopez@unizar.es}

\author{Carl Christian Kjelgaard Mikkelsen}
\affiliation[Umea]{Department of Computing Science and HPC2N, Ume{\aa}, Sweden}

\author{Juan Jos\'e Galano-Frutos}
\affiliation[BIFI]{Instituto de Biocomputaci\'on y F\'isica de Sistemas Complejos, Zaragoza, Spain}
\altaffiliation{Present address: Certest Biotec S.L., 50840 San Mateo de G\'allego, Zaragoza, Spain}

\author{Santiago Marco-Sola}
\affiliation[BSC]{Barcelona Supercomputing Center, Spain}
\alsoaffiliation[UPC]{Departament d'Arquitectura de Computadors, Universitat Polit\`ecnica de Catalunya, Spain}

\author{Jes\'us Alastruey-Bened\'e}
\affiliation[Unizar]{Departamento de Inform\'atica e Ingenier{\'i}a de Sistemas / Arag\'on Institute for Engineering Research (I3A), Universidad de Zaragoza, Zaragoza, Spain}

\author{Pablo Ib\'a\~{n}ez}
\affiliation[Unizar]{Departamento de Inform\'atica e Ingenier{\'i}a de Sistemas / Arag\'on Institute for Engineering Research (I3A), Universidad de Zaragoza, Zaragoza, Spain}

\author{Pablo Echenique}
\affiliation[CSIC]{Instituto de Qu\'imica F\'isica Blas Cabrera (CSIC), Madrid, Spain}

\author{Miquel Moret\'o}
\affiliation[BSC]{Barcelona Supercomputing Center, Spain}
\alsoaffiliation[UPC]{Departament d'Arquitectura de Computadors, Universitat Polit\`ecnica de Catalunya, Spain}

\author{Maria Cristina De Rosa}
\affiliation{Istituto di Scienze e Tecnologie Chimiche ``Giulio Natta'' (SCITEC) -- National Research Council (CNR), Italy}

\author{Pablo Garc\'ia-Risue\~{n}o}
\affiliation{Instituto de Biocomputaci\'on y F\'isica de Sistemas Complejos, Zaragoza, Spain}
\email{risueno@unizar.es}

\title[ILVES]{ILVES: Accurate and Efficient Bond Length and Angle Constraints in Molecular Dynamics}


\begin{document}

\begin{tocentry}

\includegraphics[width=3.27in, height=1.8in]{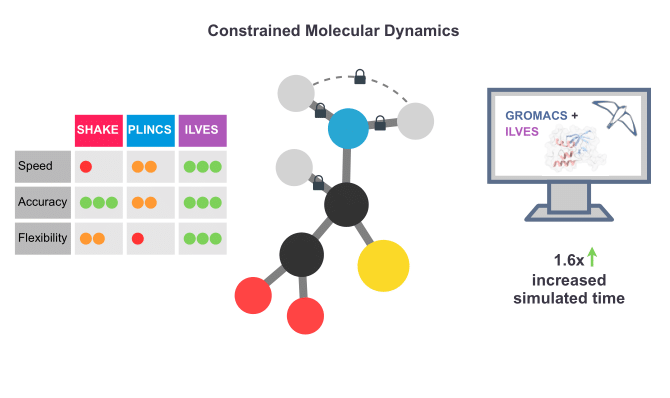}

\end{tocentry}

\begin{abstract}
All-atom, force field-based molecular dynamics simulations are essential tools in computational chemistry, enabling the prediction and analysis of biomolecular systems with atomic-level resolution. However, as system sizes and simulation timescales increase, so does the associated computational cost. To extend simulated time using the same resources, a common strategy is to constrain the fastest degrees of freedom, such as bond lengths, allowing for larger integration time steps without compromising accuracy. The {\it de facto} state-of-the-art algorithms for this purpose---SHAKE, LINCS, and P-LINCS---are integrated into most molecular dynamics packages and widely adopted across the field. Despite their impact, these methods exhibit limitations: all converge slowly when high numerical accuracy is required, and the LINCS and P-LINCS algorithms cannot handle general angular constraints, limiting further increases in time step.

In this article, we introduce ILVES, a family of parallel algorithms that converge so rapidly that it is now practical to solve bond length and associated angular constraint equations as accurately as the hardware will allow. We have integrated ILVES into {\textsc{Gromacs}} and our analysis demonstrates that it is superior to the state-of-the-art when constraining bond lengths. Due to its better convergence properties, we also show that if the time step is increased up to 3.5 fs by enforcing angular constraints, ILVES enables a 1.65$\times$ increase in simulated time using the same computational resources and wall-clock time, an outcome unattainable with current methods. This advance can significantly reduce the computational cost of most all-atom molecular dynamics simulations while improving their accuracy and extending access to larger systems and longer timescales.

\end{abstract}

\section{Introduction}

Molecular dynamics simulations (MD)~\cite{Leimkuhlerbook, karplus1977dynamics} have greatly impacted a wide range of fields in science and technology~\cite{Karplus2002, Hospital2015}. They are of special importance in chemistry and medicine, with applications including the design of drugs and catalysts~\cite{shan2024rational,coronavirus_shake,coronavirus_gromacs1,jacdrugdesign1,jacdrugdesign2}, e.g., helping to understand interaction- or mutation-driven biological processes~\cite{galano-review,Hollingsworth2018,scienceunderstanding2023}. One of the biggest advantages of MD-based approaches is that they provide information on the simulated systems at the atomic level (positions, velocities, forces), which enables the study of phenomena whose analysis in a laboratory is often not feasible or affordable~\cite{Alo2004Book, GalanoFrutos2024}. Thanks to recent advances, particularly in artificial intelligence, the impact of MD is set to grow. For example, the prediction of protein structures through AlphaFold~\cite{alphafold_main_paper} is already boosting massive-scale analysis of the behaviour of proteins and their interactions for a variety of relevant areas~\cite{li2024single,gromacscancer1,gromacsalphafold2,gromacsalphafold3,gromacsalphafold4,gromacsalphafold5,jacsalphafold}.

It is commonly accepted that, to be reliable, the discrete integration of the equations of motion in molecular dynamics must include at least five steps per vibration period of every degree of freedom (e.g., bond lengths, bond angles, or dihedral angles). This sets an upper limit for the value of the {\it time step}, i.e., for the separation between consecutive simulated times. Since the calculation at every time point requires a certain number of arithmetic operations, the size of the time step limits the total real-time that can be simulated using a given amount of computational resources. Due to this, in MD simulations it is customary to constrain some of the fastest internal degrees of freedom to fixed values. If the physical model and the resulting dynamics and thermodynamics are not distorted by doing so, the removal of the shortest vibrational periods allows an increase in the time step, thus reaching longer total times with the same computational effort.

Taking into account the well-known hierarchy that organizes vibrational periods in proteins and other biological molecules~\cite{hierarchy4}, the imposition of constraints begins with the fastest degrees of freedom and proceeds gradually to slower ones, thus allowing to increase the time step at each stage. It is very common in production MD simulations to constrain all bond lengths or those involving a hydrogen atom to increase the time step up to 2 fs. However, the situation around constraining bond angles is more heterogeneous in the literature. Although it is commonly mentioned that imposing constraints on the bond angles of hydrogen atoms is the reasonable next stage for further increasing the time step, it is difficult to find actual production simulations that do so. Instead, a variety of techniques are used, such as dummy hydrogens~\cite{hierarchy3}, redistribution of mass to make hydrogen atoms heavier and their vibrations slower~\cite{hierarchy3}, the artificial enlargement of the angular vibrational constants in the force field~\cite{hierarchy4} or the use of united atoms, i.e., the assimilation of the mass and the charge of hydrogen atoms into the heavy atoms to which they are bonded, thus effectively removing them from the model~\cite{vGu1982MM}. All these techniques are useful for increasing the time step, but they introduce alterations to the model that are not justifiable a priori from physical or chemical considerations. The actual constraining of hydrogen bond angles---which is physically justifiable if we accept that this vibrational degree of freedom can be modeled as a quantum harmonic oscillator at its ground state---is performed only in a selected set of works \cite{Gon2009JCP,Maz1998JPCB,pechlaner2021method,pechlaner2022use,thallmair2023} and not without difficulties. This has been done, for example, using internal coordinates to integrate the equations of motion instead of Cartesian coordinates, which introduces an important computational overhead~\cite{Maz1998JPCB}. Hydrogen bond angles have been constrained in production simulations using (P-)LINCS, but the bond lengths of heavy atoms had to remain unconstrained due to (P-)LINCS' convergence problems~\cite{kriegerlincs}. Finally, successful simulations with the GROMOS package have been reported using a modified version of the SHAKE algorithm to handle angles~\cite{pechlaner2022use}. However, no in-depth assessment of the computational cost is provided, and convergence difficulties appear in the study when a small set of new constraints is added~\cite{pechlaner2022use}.

In this work, we present ILVES-M and ILVES-F, two parallel algorithms that solve the same system of differential-algebraic equations as SHAKE, but the constraint equations are solved using either Newton's method or a quasi-Newton method rather than the nonlinear Gauss-Seidel method used by SHAKE.
Our algorithms and software outperform the state-of-the-art algorithms, SHAKE~\cite{shake_paper} and (P-)LINCS~\cite{lincs_paper, plincs_paper}.
A statement of all relevant algorithms, SHAKE, (P)-LINCS, ILVES-M, and ILVES-F, can be found in Sec.~2 of the Supplementary Materials. 
In particular, we show that, in most tests involving bond length constraints, the ILVES algorithms converge so rapidly that solving the constraint equations with high accuracy is not only possible but eminently practical. 
Most importantly, by leveraging the existing \textsc{Gromacs} framework, we show that ILVES-M and ILVES-F can also constrain bond angles in parallel with low computational overhead, in contrast to SHAKE and (P-)LINCS. Our analysis shows that by constraining the bond angles of hydrogen atoms and increasing the time step to 3.5 fs, ILVES enables a 1.65$\times$ increase in simulated time using the same computational resources and wall-clock time as a simulation with the default 2 fs time step. 
These results establish, for the first time as far as we are aware, that constraining hydrogen bond angles enables a substantial increase in simulation throughput.
The ILVES-M and ILVES-F code, integrated into \textsc{Gromacs}, is publicly available at \url{https://github.com/LorienLV/_PAPER_ILVES}.

\section{Limitations of the State-of-the-Art Constraint Solvers}

SHAKE and (P-)LINCS are decades-old algorithms. SHAKE~\cite{shake_paper} is nearly 50 years old, while LINCS~\cite{lincs_paper} was presented in 1997 and P-LINCS~\cite{plincs_paper} appeared in 2008. 
Though their contribution to science has been tremendous, they have specific limitations that we seek to address. 
The constraint solver in the original SHAKE algorithm converges slowly
and is not considered a good candidate for parallelization \cite{shakeparallel2}. 
Parallel versions of SHAKE~\cite{shakeparallel1,shakeparallel2} have not been widely used, and the implementation of SHAKE in {\textsc{Gromacs}} is sequential.
The {\textsc{Gromacs}} library for molecular simulation is so widely used that we have chosen it to serve as a baseline for our analysis. 
The successful application of (P-)LINCS hinges on the convergence of a specific infinite series and this condition can be violated in the context of coupled angular constraints (see the Supplementary Materials and the original LINCS paper~\cite{lincs_paper}), and the use of (P-)LINCS for this purpose is actively discouraged in the GROMACS manual~\cite{note_LINCS_no_angles} itself.
An interesting example can be found in the paper~\cite{hishaketol1} where the issues were so severe that LINCS had to be abandoned in favor of SHAKE.

In general, SHAKE, LINCS, and P-LINCS are rarely used to solve the constraint equations as accurately as the hardware will allow, as this goal can only be achieved using significant time and computational resources \cite{lincs_fail_chol,ilves_2023}. 
Superficially, this issue might appear insignificant, as there are many other sources of error in a simulation of molecular dynamics. However, there are cases where the error introduced in the constraints phase can result in severe distortions of the simulated system's physics.

\section{The Need to Solve the Constraint Equations Accurately}

In a recent study~\cite{ilves_2023}, we demonstrated that solving constraints inaccurately introduces distortions that can make the simulation unreliable. Insufficient accuracy when solving the constraints
is equivalent to applying undesired, spurious, and random external forces~\cite{ilves_2023}, which generates a non-negligible drift in the energy of the simulated system that consequently ruins the trustworthiness of simulations in the microcanonical (NVE) ensemble. 
This has led several studies to state that constraint equations must be solved {\it down to the limit of computational arithmetic/machine precision}~\cite{hammonds,machineprecisionconstraints2}. 
Simulations with a thermostat (NVT, NPT ensembles) also present such undesired energetic drifts, which contribute to making the conserved quantity (also called {\it conserved energy}) of the thermostat (e.g., Nos\'e-Hoover, V-rescale) become non-conserved. Due to this, there is no guarantee that the equations of the thermostat are satisfactorily solved, hence there is no guarantee that the simulation corresponds to the sought ensemble, which makes its reliability drop~\cite{okumuraneedconservedquantity,papervrescale}. 
Moreover, the drift introduced by the inaccurate solving of the constraints distorts the time ($\tau_T$) for reaching the sought temperature ($T$). 
This can be observed in the V-rescale thermostat~\cite{papervrescale}, which calculates a rescaling factor for the velocities that, on average, is expected to make the temperature of the system approximately equal to the desired temperature $T$ after a simulated time $\tau_T$ (being $\tau_T$ an input parameter of the simulation). However, due to the inaccurate constraint solving, an additional amount of energy is injected into, or extracted from, the system, which makes the average time for reaching $T$ deviate from $\tau_T$ in an unknown manner.
In addition, imposing constraints inaccurately systematically misestimates bond lengths and 
makes them randomly change their values in an irregular manner. 
Moreover, artifactual regimes arise as periods where the averages of the lengths of the bonds differ from the values set by constraints, which alternate with periods where the bond lengths remain nearly unchanged~\cite{ilves_2023}.

In {\textsc{Gromacs}}, the default SHAKE tolerance ($\texttt{shake-tol}$, defined as the maximum relative error allowed when solving constraints) is $10^{-4}$. There exists no such demanded tolerance for P-LINCS, which has been said to cause unphysical dynamics and temperatures of thousands of Kelvin due to fast rotation of NH$_3$ groups~\cite{shakefailure2}. Nevertheless, it is generally assumed that the average accuracy of P-LINCS
with the default {\textsc{Gromacs}} parameters is typically similar to SHAKE's. Such default settings lead to the non-negligible distorting effects on energy drifts and bond lengths mentioned above; in contrast, solving the constraints more accurately strongly dampens these undesired effects~\cite{ilves_2023}.
Other research works have also found non-negligible distorting effects due to inaccurate constraint solving: Ref.~\cite{lincs_fail_chol} stressed that 
{\textsc{Gromacs}}' default parameters lead to non-converged results and make temperatures of the simulated system unreliable,
which is fixed if constraints are accurately solved.
Other research indicates that inaccuracy in constraints can lead to wrong densities~\cite{hessongonzalves} or to collective motion artifacts, like spurious phase transitions from liquid to an icy state~\cite{hishaketol2}.
In Sec.~4 of the Supplementary Materials, we shall argue further in favor of solving the constraint equations as accurately as the hardware will allow.

The array of inconveniences due to inaccurate constraint solving can be largely mitigated if the constraint forces are calculated with the largest possible accuracy (for the chosen numerical precision) instead of the default values in \textsc{Gromacs} of $10^{-4}$---for SHAKE---or undetermined---for LINCS---.
Nevertheless, doing so has been precluded to date, most likely due to numerical complexity issues.

\section{The ILVES Algorithms}

ILVES is a family of algorithms for imposing constraints in the context of molecular dynamics. The ILVES algorithms compute discrete approximations of the solution to the same system of differential-algebraic equations as the SHAKE algorithm. However, whereas SHAKE relies on the nonlinear Gauss-Seidel method, which converges locally and linearly, the ILVES algorithms are based on Newton's method combined with direct solvers, resulting in drastically faster convergence rates. In general, applying direct solvers to linear systems requires $\mathcal{O}(n^3)$ floating-point operations, being $n$ the
number of equations (which is equal to the number of constraints in our case \cite{ilves_original}). Nonetheless, the particular structure of the linear systems that arise when applying constraints in MD is directly tied to the linear and sparse topology of molecular structures, so direct solvers can be applied in $\mathcal{O}(n)$ time for general molecules~\cite{ilves_original, GR2010JCoP}. The ILVES algorithms exploit this property to dramatically accelerate convergence relative to SHAKE.

In this paper, we present two algorithms, ILVES-M (``main'') and ILVES-F (``fast''), both of which leverage distributed-memory parallelism, shared-memory parallelism, and SIMD vectorization. ILVES-M solves the same system of differential-algebraic equations as SHAKE but employs Newton's method and a direct solver. To exploit shared-memory parallelism, ILVES-M uses a custom thread-parallel $LU$ factorization based on the Schur complement method~\cite{Gol1989BOOK}. For distributed-memory parallelism, it extends this thread-parallel $LU$ factorization with the Overlapping Partitioning Method (OPM)~\cite{lou1989}. Consequently, when executed across multiple domains, ILVES-M behaves as a quasi-Newton method. However, its convergence remains extremely fast, typically requiring very few (usually zero) additional iterations compared to single-domain execution.

ILVES-F is a variant of ILVES-M that reduces the computational cost by using a fixed symmetric approximation of the coordinate matrix~\cite{Eas2010JCTC1}. The symmetry of this matrix allows for replacing $LU$ factorization with $LDLT$ factorization, which improves efficiency. Moreover, since the $LDLT$ factorization needs only be computed once per time step, the total computational cost is nearly halved. Due to its symmetric approximation, ILVES-F behaves as a quasi-Newton method even in shared-memory executions. Nonetheless, its convergence is exceptionally fast~\cite{ppam2_paper,ppam2b_paper}, and it delivers better performance than ILVES-M in most scenarios.

In a previous article~\cite{ilves_2023}, we introduced ILVES-PC, a proof-of-concept implementation applying direct solvers and Newton's method to calculate constraint forces in biological molecules, specifically peptides and proteins. For completeness, we include ILVES-PC in our performance analysis in this paper.

We present a detailed description of the mathematical foundations of ILVES, as well as implementation details of ILVES-M and ILVES-F, in Secs.~2 and 3 of the Supplementary Materials.

\section{Results}

We conducted an extensive set of simulations to assess the efficiency and reliability of ILVES, covering five representative systems: two solvated proteins (barnase, referred to as the BARN system, and the COVID-19 main protease, referred to as the COVID system), a solvated protein–DNA complex (the DNAP system), a system of 2,000 benzene molecules (the BENZ system), and a tetrameric protein embedded in a lipid bilayer (the LIPID system). Full details of these systems, along with the procedures used for their preparation and simulation are provided in Secs.~5 and 6 of the Supplementary Materials.
Our reliability analysis---based on the calculation of observable quantities---as well as complementary performance results can also be found in the Supplementary Materials (Secs.~7 and 8). 
Below, we summarise the outcome of our performance study, comparing ILVES-M, ILVES-F, and ILVES-PC with state-of-the-art constraint solvers. The time spent in the initialization of the solvers can be high in distributed memory simulations. For this reason, the execution times in our analysis include both initialization and processing for all solvers---except ILVES-PC, which was released as a proof of concept without distributed memory support and optimized initialization. Additionally, to ensure a fair comparison, we developed a modified version of P-LINCS that guarantees constraints are satisfied within a given tolerance, as SHAKE does; we refer to this variant as MP-LINCS. This was accomplished by repeating P-LINCS' correction phase, controlled by the \texttt{lincs-iter} parameter in the original implementation until the desired maximum relative error in solving the constraints is met. Although this modification introduces additional synchronization points in parallel executions, potentially affecting performance, it is important to note that it is straightforward to modify ILVES to execute a fixed number of iterations without checking the tolerance, thereby eliminating the same synchronization points introduced in MP-LINCS. While \textsc{Gromacs}' implementation of SHAKE itself is not parallelized, it can still be used in parallel simulations without domain decomposition. In such cases, SHAKE runs on a single thread, while the rest of the simulation proceeds in parallel. This approach was used to obtain the results reported for SHAKE in parallel simulations.

The speed of the constraint solving is closely related to the minimum accuracy demanded.
We thus considered three values of the referred tolerance (maximum allowed relative error for every constraint):
$\textrm{Tol} = 10^{-4}$ (which is the default in {\textsc{Gromacs}}), $\textrm{Tol} = 10^{-8}$ and $\textrm{Tol} = 10^{-12}$. Simulations for $\textrm{Tol} = 10^{-4}$ are performed using \textsc{Gromacs} compiled in single-precision mode (FP32), whereas simulations with stricter tolerances are conducted with \textsc{Gromacs} compiled in double-precision mode (FP64).

We have measured the performance of the algorithms in cases where constraints are imposed on either: i) hydrogen bonds, ii) all bonds, or iii) all bonds together with certain angles of hydrogen atoms (specifically, H–X–H and X–O–H angles, where X represents a generic atomic species). The choice of constraint settings is generally determined—or at least recommended—by the force field used. CHARMM36~\cite{charmm36} and CHARMM36m~\cite{charmm36m} support hydrogen bond constraints, while other force fields such as AMBER~\cite{amberff} and OPLS/AA~\cite{OPLS/AA} support constraints on hydrogen bonds and all bonds. Constraining angles remain uncommon, likely because none of the widely used constraint algorithms, SHAKE or (P-)LINCS, can satisfactorily handle coupled angle constraints.

Although solving constraints is sometimes assumed to require a relatively low fraction of the total execution time of simulations, some authors inform that it can be as high as 50\% or 60\%~\cite{hammonds1991convergence,hammonds}. 
In our simulations constraining all bonds, SHAKE accounts for up to 92\% of the total execution time, MP-LINCS up to 42\%, and ILVES up to 16\%. These percentages depend on the number of cores and are lower in H-bonds simulations, where SHAKE accounts for up to 60\%, MP-LINCS up to 6\%, and ILVES up to 5\%. 
A detailed figure of the solvers' relative execution times is provided in Sec.~8 of the Supplementary Materials.

Fig.~\ref{fig:speedup_over_shake_nt_double} presents the speedup over SHAKE for different numbers of threads and tasks when constraining all bonds and when constraining H-bonds. This metric is defined as the ratio of the execution time of SHAKE with a single thread to the execution time of the given solver using $N$ threads. For the barnase (BARN), Covid-19 main protease (COVID), benzene (BENZ), and DNA-protein complex (DNAP), our simulations employed up to 56 threads and 1 task on a single Intel Xeon Platinum 8480+ chip. The larger (390K atoms, 149K ex water) lipid bilayer with proteins (LIPID) simulation was executed using up to 8 tasks and 56 threads per task, i.e., up to 4 nodes and 8 chips. The production stage of each of the simulations consisted of 50k steps of size 2 fs. In all-bonds simulations (Fig.~\ref{fig:speedup_over_shake_nt_double}-a), ILVES-M and ILVES-F achieve speedups over MP-LINCS across all simulations and tolerances, with a maximum of $158\times$ over SHAKE and $14\times$ over MP-LINCS. In H-bonds simulations (Fig.~\ref{fig:speedup_over_shake_nt_double}-b), MP-LINCS delivers better parallel performance than in all-bonds simulations, thus narrowing the performance gap between the solvers. In these simulations, MP-LINCS only surpasses ILVES-M and ILVES-F in the LIPID simulation at $\textrm{Tol}=10^{-4}$. On the other hand, ILVES-F delivers better performance than MP-LINCS in the rest of the simulations for a maximum speedup of $134\times$ over SHAKE and speedups over MP-LINCS up to $1.8\times$.

\begin{figure}[htbp]
    \centering
    \begin{minipage}[b]{0.98\textwidth}
        \centering
        \includegraphics[width=\textwidth]{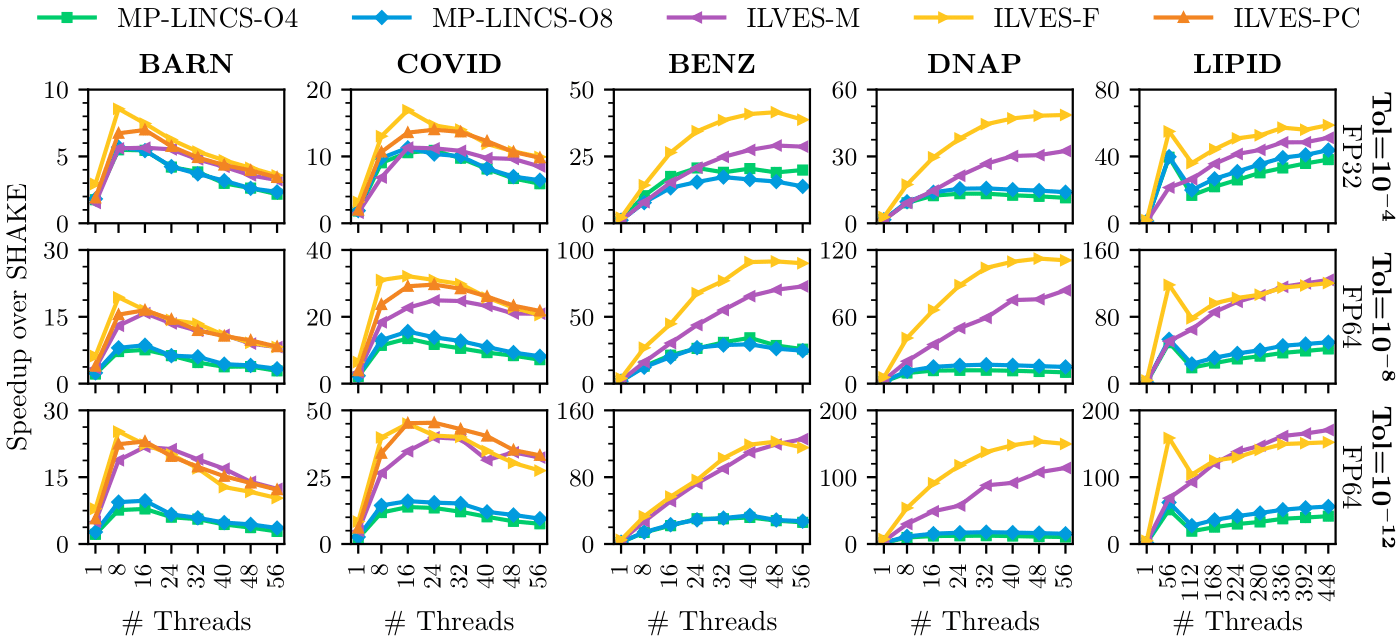}
        \renewcommand{\thefigure}{a}
        \\ (a) All bonds
        \label{fig:speedup_over_shake_nt_double_all_bonds}
    \end{minipage}

    \medskip

    \begin{minipage}[b]{0.98\textwidth}
        \centering
        \includegraphics[width=\textwidth]{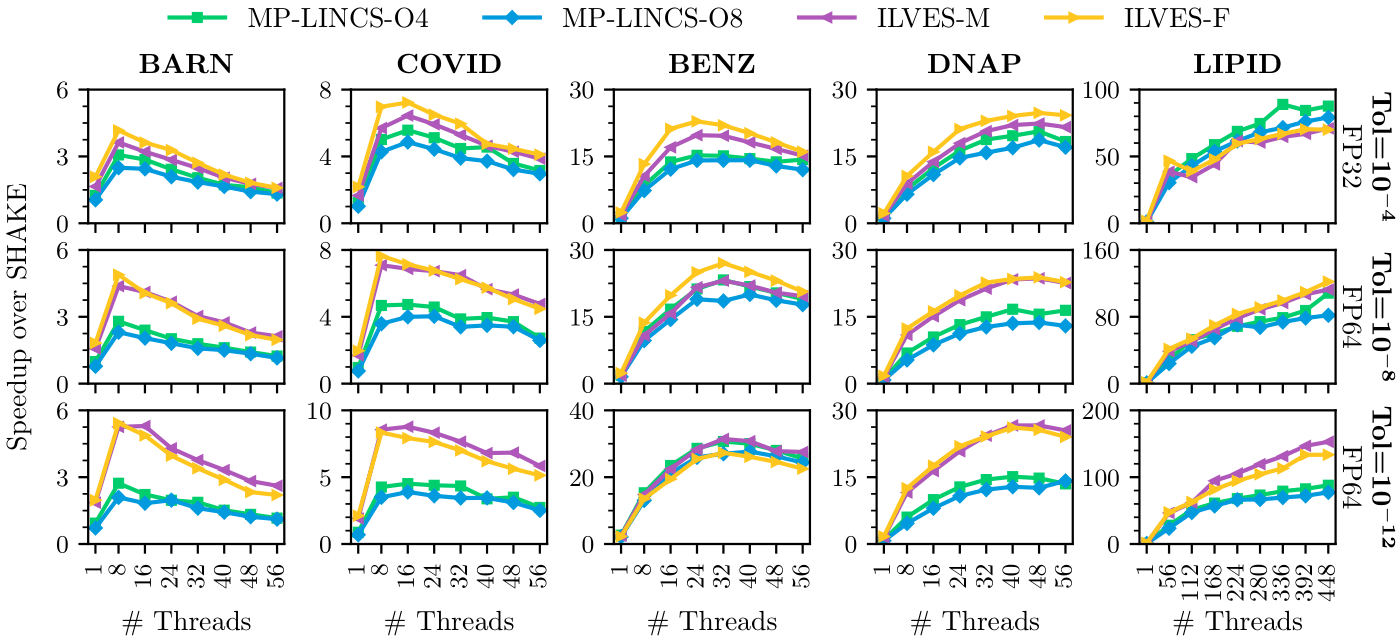}
        \renewcommand{\thefigure}{b}
        \\ (b) H-bonds
        \label{fig:speedup_over_shake_nt_double_h_bonds}
    \end{minipage}
    \caption{Multi-thread speedup over SHAKE of MP-LINCS, ILVES-M, ILVES-F and ILVES-PC~\cite{ilves_2023}. The BARN, COVID, BENZ, and DNAP simulations are executed using a single task in a single chip of a computing node. The LIPID simulation is executed using up to 8 tasks (up to 4 nodes and 8 chips). MP-LINCS tested for \texttt{lincs-order=4} and \texttt{lincs-order=8}. (a) Constraints imposed on all bonds; (b) Constraints imposed on H-bonds.}
    \label{fig:speedup_over_shake_nt_double}
\end{figure}

\begin{figure}[htbp]
    \centering
    \includegraphics[width=0.98\textwidth]{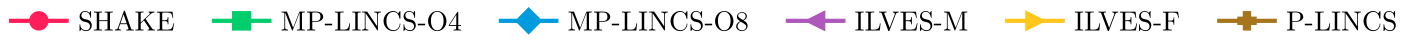} 
    
    \begin{minipage}[b]{0.48\textwidth}
        \centering
        \includegraphics[width=\textwidth]{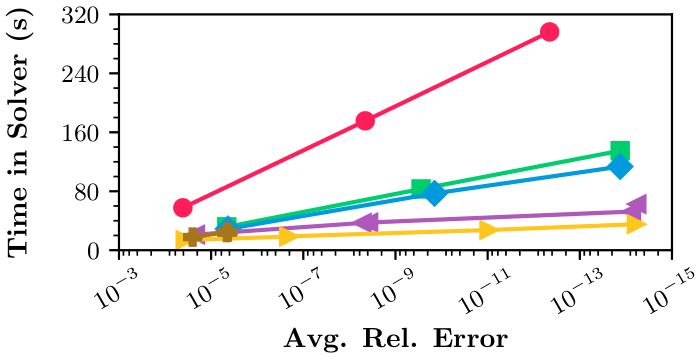}
        \renewcommand{\thefigure}{a}
        \\ (a) All bonds
        \label{fig:acc_vs_time_double_all_bonds}
    \end{minipage}
    \hfill
    \begin{minipage}[b]{0.48\textwidth}
        \centering
        \includegraphics[width=\textwidth]{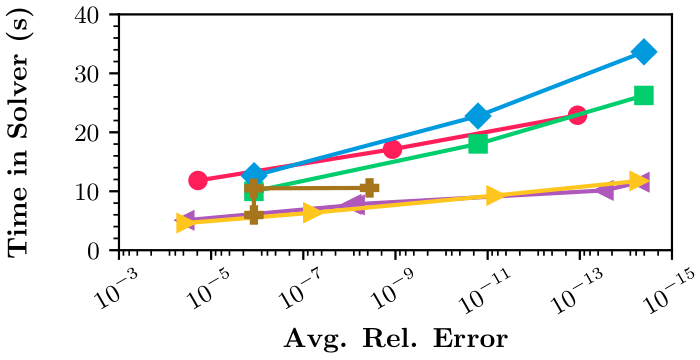}
        \renewcommand{\thefigure}{b}
        \\ (b) H-bonds
        \label{fig:acc_vs_time_double_h_bonds}
    \end{minipage}

    \caption{Execution time of the block for solving constraints as a function of the average relative error in satisfying them for different constraint solvers. (a) Constraints imposed on all bonds; (b) Constraints imposed on H-bonds. Note that the y-axis is logarithmic and that the tolerance decreases from left to right.}
    \label{fig:acc_vs_time_double}
\end{figure}

Even though the tolerance defines the maximum acceptable error, the rapid convergence of the ILVES algorithm often yields errors significantly below this threshold. 
This results in accuracy gains, providing a compelling reason to choose ILVES over MP-LINCS in simulations where their performance is similar. 
In Fig.~\ref{fig:acc_vs_time_double} we display the execution time required for imposing the constraints as a function of the average relative error, which is defined as the average for $N_s$ steps and $n$ constraints that follows:
$\left(\frac{1}{2 \, n\cdot N_s} \right) \cdot \left(  
\sum_{k=1}^{N_s} \, \sum_{i=1}^n \, 
|d_i^2 - (\boldsymbol{q}_{a_i}(t_k) - \boldsymbol{q}_{b_i}(t_k))^2|
/d_i^2
\right)$, 
where $\boldsymbol{q}_{a_i}(t_k)$, $\boldsymbol{q}_{b_i}(t_k)$ are the positions of both atoms joined by the $i$th constraint after applying the constraint forces corresponding to the $k$th step, and $d_i$ are the bond length constants. Every point of Fig.~\ref{fig:acc_vs_time_double} corresponds to a simulation performed with different parameters (values of the constraint tolerance for SHAKE; values of the constraint tolerance or number of iterations for ILVES-M and ILVES-F; values of the number of iterations ---{\texttt{lincs-iter}}--- and truncation of the Neumann series ---{\texttt{lincs-order}}--- for P-LINCS;
values of the constraint tolerance and {\texttt{lincs-order}} for MP-LINCS).
The results displayed in Fig.~\ref{fig:acc_vs_time_double} correspond to the Covid main protease (4697 constraints) simulated for 50K steps in a single core. 
If we compare algorithms that ensure that a minimum accuracy is satisfied, like MP-LINCS-O4 and ILVES-F,
we observe that, for approximately the same execution time, the latter is far more accurate than the former. For example, for constraints on all bonds, 
Fig.~\ref{fig:acc_vs_time_double} displays a point for MP-LINCS-O4 whose execution time is 32 s and whose average relative error is $5\cdot 10^{-6} $; it also displays a point for ILVES-F whose execution time is 27 s and whose average relative error is $9\cdot 10^{-12}$.
%MPLINCS-O4: 32 4.6E-6
%ILVES-F     27 9.34E-12
%ILVES-F     35 5.9E-15
This feature also holds for constraints on hydrogen bonds: examples of points displayed in
Fig.~\ref{fig:acc_vs_time_double} are
(10 s, $10^{-6}$) for MP-LINCS-O4,
(9 s, $7\cdot 10^{-12}$) for ILVES-F and
(10 s, $3\cdot 10^{-14}$) for ILVES-M.
%
%MPLINCS-O4: 10 1.2E-6
%ILVES-F      9 6.6E-12
%ILVES       10 3E-14
This example indicates that, for similar execution times, the ILVES algorithms are between 500,000 and 30,000,000 times more accurate than P-LINCS algorithms. 
Fig.~\ref{fig:acc_vs_time_double} indicates that the fast convergence of the ILVES methods makes much more accurate solutions possible requiring very low execution times, making it affordable to solve constraints near the limit of machine precision. 
We stress that increasing the accuracy of constraint solving is also desirable in simulations made with numerical single precision. 
In such cases, our tests indicate that the maximum enforceable tolerance is about $\textrm{Tol}=10^{-6}$ instead of the value $\textrm{Tol}=10^{-12}$ which corresponds to double precision.

The discussion presented in this section so far corresponds to the case of imposing constraints on just bond lengths, which limits the maximum time step to 2 fs. But, in addition to the possibility of achieving a higher degree of accuracy and computational savings that the ILVES family of algorithms provides for this very common set of constraints, its better convergence properties also allow us to cross a line unprecedented in the literature as far as we are aware. In what follows, we demonstrate that ILVES can be utilized to efficiently impose constraints on specific bond angles of hydrogen atoms, thereby enabling an increase in both the time step and the simulation throughput.

First of all, let us note that at present \textsc{Gromacs} offers time steps beyond 2 fs by applying several techniques such as {\it mass repartitioning} or {\it virtual sites}~\cite{gromacs_manual,thallmair2023}. Mass repartitioning involves assigning hydrogen atom masses greater than 1 atomic mass unit, which is compensated by withdrawing part of the mass of heavy atoms. Virtual sites consist of determining the position of hydrogen atoms as a function of the position of three nearby heavy atoms, i.e. not using forces on the hydrogen atoms. Both approaches can be viewed as unphysical, and increase the number of assumptions behind the simulation. In contrast, imposing constraints on H-angles is chemically and physically justifiable because quantum harmonic oscillators are more similar to constraints than to classical harmonic oscillators. Literature indicates that the time step can be safely increased up to 4 fs by constraining all covalent bonds and the angles involving hydrogen atoms~\cite{hierarchy4,Maz1998JPCB,hierarchy3,pechlaner2022use}. However, the option {\texttt{constraints=h-angles}} in {\textsc{Gromacs}} only imposes constraints on a subset of all the bond angles related to hydrogen atoms, namely those defined between two hydrogen atoms connected to the same heavy atom X in a H-X-H scheme and the angle between a hydrogen atom connected to an oxygen atom and the heavy atom X connected to the oxygen in a X-O-H scheme. This freezes some of the vibrations associated with the angular degrees of freedom of hydrogen atoms, but not all of them, and this is the reason why {\texttt{constraints=h-angles}} in {\textsc{Gromacs}} allows an increase of the time step to 3.5 fs but not to 4 fs. The implementation of full bond angle constraints will be a matter of future research.

Despite its availability, the {\texttt{constraints=h-angles}} option of {\textsc{Gromacs}} has not been successfully used in the literature to increase the time step, most likely due to the limitations of the state-of-the-art constraint solvers: (P-)LINCS is usually unable to impose constraints on coupled angles~\cite{hishaketol1}, and SHAKE converges extremely slowly~\cite{usageBAconstraints3}. This is shown in Fig.~\ref{fig:performance}, in which we increase the time step of the LIPID simulation to 3.5 fs (this is the identified upper bound for stable simulations when using the {\texttt{constraints=h-angles}} setup). The referred figure reports the simulation performance in nanoseconds simulated per day and the percentage of execution time spent on the constraint solver using SHAKE, P-LINCS, ILVES-M, and ILVES-F, under two configurations: the default {\textsc{Gromacs}} settings ({\texttt{constraints=h-bonds}}, {\texttt{ts=2 fs}}, {\texttt{Tol=$10^{-4}$}}) and the new settings ({\texttt{constraints=h-angles}}, {\texttt{ts=3.5 fs}}, {\texttt{Tol=$10^{-4}$}}). The simulation was performed using all 56 cores of an Intel Xeon Platinum 8480+ processor and ran for 1.5 million steps. The results show that increasing the time step from 2 fs to 3.5 fs by introducing angle constraints causes SHAKE to dominate the simulation time, accounting for 93\% of the total runtime and severely limiting overall performance. In addition, P-LINCS does not work with {\texttt{constraints=h-angles} (marked as N/A in the figure). In contrast, ILVES enables the simulation to run significantly faster, increasing performance from 23 ns/day to 38 ns/day, which translates to a 1.65$\times$ improvement. Furthermore, ILVES accounts for a small fraction of the total runtime: 14\% for ILVES-M and 7\% for ILVES-F.

The previous results demonstrate that ILVES paves the way for a new approach to increasing the time step without relying on potentially unphysical approximations, such as mass repartitioning. Nevertheless, further research is required. Currently, increasing the time step beyond 3.5 fs is not possible within {\textsc{Gromacs}}, as it lacks support for constraining a larger set of angles involving hydrogen atoms. Additionally, existing force fields have not been parameterized for use with angle constraints. Further investigation is, therefore, necessary to establish how to correctly impose angle constraints---and potentially dihedral constraints---so that ILVES can enable time steps well above 4 fs and to evaluate how introducing such constraints would affect the physical accuracy and stability of molecular dynamics simulations.

\begin{figure}[htbp]
    \centering

    \begin{minipage}[b]{0.8\textwidth}
        \centering
        \includegraphics[width=\textwidth]{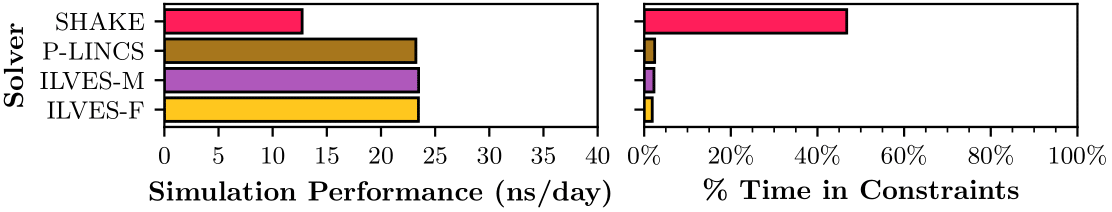}
        \renewcommand{\thefigure}{a}
        \\ (a) H-bonds, 2 fs
        \label{fig:performance-low-ts}
    \end{minipage}

    \medskip

    \begin{minipage}[b]{0.8\textwidth}
        \centering
        \includegraphics[width=\textwidth]{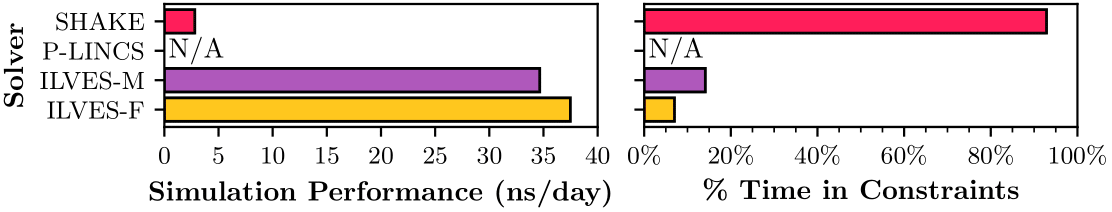}
        \renewcommand{\thefigure}{a}
        \\ (b) H-angles, 3.5 fs
        \label{fig:performance-high-ts}
    \end{minipage}
    \caption{LIPID simulation performance (left) and percentage of the execution time spent on the constraint solver (right) using a 2 fs time step with {\texttt{constraints=h-bonds}} (a) and a 3.5 fs time step with {\texttt{constraints=h-angles}} (b), across four constraint solvers: SHAKE, P-LINCS, ILVES-M, and ILVES-F. The performance is reported in nanoseconds simulated per day, using all 56 cores of an Intel Xeon Platinum 8480+ chip. P-LINCS is not compatible with {\texttt{constraints=h-angles}} and is marked as N/A in the figure.}
    \label{fig:performance}
\end{figure}

\section*{Conclusions and Future Work}\label{conclusions}

In this work, we introduce novel parallel algorithms based on Newton's method and direct linear solvers, designed to impose constraints on molecular systems. These algorithms demonstrate significant improvements in both accuracy and efficiency compared to the current state-of-the-art methods when constraining bond lengths. Moreover, they enable the efficient constraining of additional degrees of freedom in parallel, establishing the foundation for increasing the time step of simulations. We show that when the time step is increased from 2 fs to 3.5 fs by constraining some hydrogen angles (leveraging \textsc{Gromacs'} existing framework for this task), P-LINCS does not work, and SHAKE dominates the total execution time of \textsc{Gromacs}, significantly degrading performance. In contrast, our solvers enable a 1.65$\times$ increase in simulated time using the same computational resources and wall-clock time.

In this article, we have shown that increasing the time step by applying angle constraints in combination with ILVES can yield substantial performance gains. Looking toward future research, we plan to investigate how to further and reliably extend the time step by constraining all hydrogen angles~\cite{kriegerlincs}, dihedral angles~\cite{pechlaner2021method}, and other internal degrees of freedom.
ILVES is expected to solve the equations associated with these additional constraints efficiently, thus enabling higher time steps in simulations with and without virtual sites~\cite{thallmair2023}, which we also intend to explore.
Further work will also focus on integrating these constraints into molecular dynamics packages and assessing their effects on simulation accuracy and stability.
The current implementation of P-LINCS found in \textsc{Gromacs} supports GPU execution when synchronization between threads or nodes is not required, such as when constraining hydrogen bonds without domain decomposition. In a similar vein, we are working on developing GPU-accelerated versions of the ILVES-M and ILVES-F. Further planned research will consist of optimizing ILVES for water molecules~\cite{ppam1_paper}, which ---from preliminary tests--- is expected to improve performance over the widely used SETTLE algorithm~\cite{settle}.

\begin{acknowledgement}

The authors thank Jos\'e Alejandre and Edgar N\'u\~nez (UAM-Iztapalapa, Mexico) for providing input files of organic solvents, and Prof. Berk Hess for useful discussion.

This work has been partially supported by the Spanish Ministry of Science and Innovation MCIN/AEI/10.13039/501100011033 (contracts PID2022-136454NB-C22, PID2023-146193OB-I00, and PID2023-146511NB-I00), by the Generalitat de Catalunya (contract 2021-SGR-00763 and 2021-SGR-00574), by the Gobierno de Aragón (E45\_20R T58\_23R research groups), and by Lenovo-BSC Contract-Framework Contract (2020). Carl Christian Kjelgaard Mikkelsen is supported by eSSENCE, a collaborative e-Science programme funded by the Swedish Research Council within the framework of the strategic research areas designated by the Swedish Government. The funders had no role in study design, data collection and analysis, decision to publish, or preparation of the manuscript.

\end{acknowledgement}

\begin{suppinfo}

\begin{itemize}
\item ilves-sup.pdf: Overview of constraints solvers for molecular dynamics, mathematical background, description of the ILVES implementation, extended study of the importance of accuracy when solving constraints, experimental setup, reliability study, and extended performance analysis.
\end{itemize}

\end{suppinfo}

\bibliography{ilves}

\providecommand{\latin}[1]{#1}
\makeatletter
\providecommand{\doi}
  {\begingroup\let\do\@makeother\dospecials
  \catcode`\{=1 \catcode`\}=2 \doi@aux}
\providecommand{\doi@aux}[1]{\endgroup\texttt{#1}}
\makeatother
\providecommand*\mcitethebibliography{\thebibliography}
\csname @ifundefined\endcsname{endmcitethebibliography}  {\let\endmcitethebibliography\endthebibliography}{}
\begin{mcitethebibliography}{65}
\providecommand*\natexlab[1]{#1}
\providecommand*\mciteSetBstSublistMode[1]{}
\providecommand*\mciteSetBstMaxWidthForm[2]{}
\providecommand*\mciteBstWouldAddEndPuncttrue
  {\def\EndOfBibitem{\unskip.}}
\providecommand*\mciteBstWouldAddEndPunctfalse
  {\let\EndOfBibitem\relax}
\providecommand*\mciteSetBstMidEndSepPunct[3]{}
\providecommand*\mciteSetBstSublistLabelBeginEnd[3]{}
\providecommand*\EndOfBibitem{}
\mciteSetBstSublistMode{f}
\mciteSetBstMaxWidthForm{subitem}{(\alph{mcitesubitemcount})}
\mciteSetBstSublistLabelBeginEnd
  {\mcitemaxwidthsubitemform\space}
  {\relax}
  {\relax}

\bibitem[Leimkuhler and Reich(2004)Leimkuhler, and Reich]{Leimkuhlerbook}
Leimkuhler,~B.; Reich,~S. \emph{Simulating Hamiltonian dynamics}, 1st ed.; Cambridge University Press - Cambridge Monographs on Applied and Computational Mathematics, 2004\relax
\mciteBstWouldAddEndPuncttrue
\mciteSetBstMidEndSepPunct{\mcitedefaultmidpunct}
{\mcitedefaultendpunct}{\mcitedefaultseppunct}\relax
\EndOfBibitem
\bibitem[Karplus(1978)]{karplus1977dynamics}
Karplus,~M. Dynamics of folded proteins. \emph{Abstracts of papers of the {A}merican {C}hemical {S}ociety} \textbf{1978}, \emph{175}, 70\relax
\mciteBstWouldAddEndPuncttrue
\mciteSetBstMidEndSepPunct{\mcitedefaultmidpunct}
{\mcitedefaultendpunct}{\mcitedefaultseppunct}\relax
\EndOfBibitem
\bibitem[Karplus and McCammon(2002)Karplus, and McCammon]{Karplus2002}
Karplus,~M.; McCammon,~J.~A. Molecular dynamics simulations of biomolecules. \emph{Nature Structural Biology} \textbf{2002}, \emph{9}, 646--652\relax
\mciteBstWouldAddEndPuncttrue
\mciteSetBstMidEndSepPunct{\mcitedefaultmidpunct}
{\mcitedefaultendpunct}{\mcitedefaultseppunct}\relax
\EndOfBibitem
\bibitem[Hospital \latin{et~al.}(2015)Hospital, Goñi, Orozco, and Gelpí]{Hospital2015}
Hospital,~A.; Goñi,~J.~R.; Orozco,~M.; Gelpí,~J.~L. Molecular dynamics simulations: advances and applications. \emph{Advances and Applications in Bioinformatics and Chemistry} \textbf{2015}, \emph{8}, 37--47\relax
\mciteBstWouldAddEndPuncttrue
\mciteSetBstMidEndSepPunct{\mcitedefaultmidpunct}
{\mcitedefaultendpunct}{\mcitedefaultseppunct}\relax
\EndOfBibitem
\bibitem[Shan \latin{et~al.}(2024)Shan, Cai, Zhu, Zhou, Sun, Xu, Yin, Wang, and Li]{shan2024rational}
Shan,~X.; Cai,~Y.; Zhu,~B.; Zhou,~L.; Sun,~X.; Xu,~X.; Yin,~Q.; Wang,~D.; Li,~Y. Rational strategies for improving the efficiency of design and discovery of nanomedicines. \emph{Nature Communications} \textbf{2024}, \emph{15}, 9990\relax
\mciteBstWouldAddEndPuncttrue
\mciteSetBstMidEndSepPunct{\mcitedefaultmidpunct}
{\mcitedefaultendpunct}{\mcitedefaultseppunct}\relax
\EndOfBibitem
\bibitem[Andreano \latin{et~al.}(2021)Andreano, Piccini, Licastro, Casalino, Johnson, Paciello, Dal~Monego, Pantano, Manganaro, Manenti, Manna, Casa, Hyseni, Benincasa, Montomoli, Amaro, McLellan, and Rappuoli]{coronavirus_shake}
Andreano,~E. \latin{et~al.}  {S}{A}{R}{S}-{C}o{V}-2 escape a highly neutralizing {C}{O}{V}{I}{D}-19 convalescent plasma. \emph{Proceedings of the National Academy of Sciences} \textbf{2021}, \emph{118}\relax
\mciteBstWouldAddEndPuncttrue
\mciteSetBstMidEndSepPunct{\mcitedefaultmidpunct}
{\mcitedefaultendpunct}{\mcitedefaultseppunct}\relax
\EndOfBibitem
\bibitem[Bhardwaj \latin{et~al.}(2021)Bhardwaj, Singh, Sharma, Rajendran, Purohit, and Kumar]{coronavirus_gromacs1}
Bhardwaj,~V.~K.; Singh,~R.; Sharma,~J.; Rajendran,~V.; Purohit,~R.; Kumar,~S. Identification of bioactive molecules from tea plant as {SARS}-{C}o{V}-2 main protease inhibitors. \emph{Journal of Biomolecular Structure and Dynamics} \textbf{2021}, \emph{39}, 3449 -- 3458\relax
\mciteBstWouldAddEndPuncttrue
\mciteSetBstMidEndSepPunct{\mcitedefaultmidpunct}
{\mcitedefaultendpunct}{\mcitedefaultseppunct}\relax
\EndOfBibitem
\bibitem[Araki \latin{et~al.}(2024)Araki, Ekimoto, Takemura, Matsumoto, Tamura, Kokubo, Bekker, Yamane, Isaka, Sagae, \latin{et~al.} others]{jacdrugdesign1}
Araki,~M.; Ekimoto,~T.; Takemura,~K.; Matsumoto,~S.; Tamura,~Y.; Kokubo,~H.; Bekker,~G.-J.; Yamane,~T.; Isaka,~Y.; Sagae,~Y.; others Molecular dynamics unveils multiple-site binding of inhibitors with reduced activity on the surface of dihydrofolate reductase. \emph{Journal of the American Chemical Society} \textbf{2024}, \emph{146}, 28685--28695\relax
\mciteBstWouldAddEndPuncttrue
\mciteSetBstMidEndSepPunct{\mcitedefaultmidpunct}
{\mcitedefaultendpunct}{\mcitedefaultseppunct}\relax
\EndOfBibitem
\bibitem[Qiu \latin{et~al.}(2024)Qiu, Chao, Song, Wang, Chen, Rouse, Yen, and Robinson]{jacdrugdesign2}
Qiu,~X.; Chao,~K.; Song,~S.; Wang,~Y.-Q.; Chen,~Y.-A.; Rouse,~S.~L.; Yen,~H.-Y.; Robinson,~C.~V. Coupling and Activation of the $\beta$1 Adrenergic Receptor-The Role of the Third Intracellular Loop. \emph{Journal of the American Chemical Society} \textbf{2024}, \emph{146}, 28527--28537\relax
\mciteBstWouldAddEndPuncttrue
\mciteSetBstMidEndSepPunct{\mcitedefaultmidpunct}
{\mcitedefaultendpunct}{\mcitedefaultseppunct}\relax
\EndOfBibitem
\bibitem[Galano-Frutos \latin{et~al.}(2019)Galano-Frutos, García-Cebollada, and Sancho]{galano-review}
Galano-Frutos,~J.~J.; García-Cebollada,~H.; Sancho,~J. {Molecular dynamics simulations for genetic interpretation in protein coding regions: where we are, where to go and when}. \emph{Briefings in Bioinformatics} \textbf{2019}, \emph{22}, 3--19\relax
\mciteBstWouldAddEndPuncttrue
\mciteSetBstMidEndSepPunct{\mcitedefaultmidpunct}
{\mcitedefaultendpunct}{\mcitedefaultseppunct}\relax
\EndOfBibitem
\bibitem[Hollingsworth and Dror(2018)Hollingsworth, and Dror]{Hollingsworth2018}
Hollingsworth,~S.~A.; Dror,~R.~O. Molecular Dynamics Simulation for All. \emph{Neuron} \textbf{2018}, \emph{99}, 1129--1143\relax
\mciteBstWouldAddEndPuncttrue
\mciteSetBstMidEndSepPunct{\mcitedefaultmidpunct}
{\mcitedefaultendpunct}{\mcitedefaultseppunct}\relax
\EndOfBibitem
\bibitem[Kim \latin{et~al.}(2023)Kim, Shin, Wang, and Nowakowski]{scienceunderstanding2023}
Kim,~C.~N.; Shin,~D.; Wang,~A.; Nowakowski,~T.~J. Spatiotemporal molecular dynamics of the developing human thalamus. \emph{Science} \textbf{2023}, \emph{382}, eadf9941\relax
\mciteBstWouldAddEndPuncttrue
\mciteSetBstMidEndSepPunct{\mcitedefaultmidpunct}
{\mcitedefaultendpunct}{\mcitedefaultseppunct}\relax
\EndOfBibitem
\bibitem[Alonso \latin{et~al.}(2004)Alonso, Chass, Csizmadia, Echenique, and Taranc{\'o}n]{Alo2004Book}
Alonso,~J.~L.; Chass,~G.~A.; Csizmadia,~I.~G.; Echenique,~P.; Taranc{\'o}n,~A. In \emph{Meeting on Fundamental Physics `Alberto Galindo'}; Alvarez-Estrada,~R.~F., others, Eds.; Aula Documental: Madrid, 2004\relax
\mciteBstWouldAddEndPuncttrue
\mciteSetBstMidEndSepPunct{\mcitedefaultmidpunct}
{\mcitedefaultendpunct}{\mcitedefaultseppunct}\relax
\EndOfBibitem
\bibitem[Galano-Frutos and Sancho(2024)Galano-Frutos, and Sancho]{GalanoFrutos2024}
Galano-Frutos,~J.~J.; Sancho,~J. Energy, water, and protein folding: A molecular dynamics-based quantitative inventory of molecular interactions and forces that make proteins stable. \emph{Protein Science} \textbf{2024}, \emph{33}, e4905\relax
\mciteBstWouldAddEndPuncttrue
\mciteSetBstMidEndSepPunct{\mcitedefaultmidpunct}
{\mcitedefaultendpunct}{\mcitedefaultseppunct}\relax
\EndOfBibitem
\bibitem[Jumper \latin{et~al.}(2021)Jumper, Evans, Pritzel, Green, \latin{et~al.} others]{alphafold_main_paper}
Jumper,~J.; Evans,~R.; Pritzel,~A.; Green,~T.; others Highly accurate protein structure prediction with {A}lpha{F}old. \emph{Nature} \textbf{2021}, \emph{596}, 583--589\relax
\mciteBstWouldAddEndPuncttrue
\mciteSetBstMidEndSepPunct{\mcitedefaultmidpunct}
{\mcitedefaultendpunct}{\mcitedefaultseppunct}\relax
\EndOfBibitem
\bibitem[Li \latin{et~al.}(2024)Li, Arghittu, Dietz, Hella, Ha{\ss}e, Ferraris, Freund, Barth, Iamele, de~Jonge, \latin{et~al.} others]{li2024single}
Li,~Y.; Arghittu,~S.~M.; Dietz,~M.~S.; Hella,~G.~J.; Ha{\ss}e,~D.; Ferraris,~D.~M.; Freund,~P.; Barth,~H.-D.; Iamele,~L.; de~Jonge,~H.; others Single-molecule imaging and molecular dynamics simulations reveal early activation of the MET receptor in cells. \emph{Nature Communications} \textbf{2024}, \emph{15}, 9486\relax
\mciteBstWouldAddEndPuncttrue
\mciteSetBstMidEndSepPunct{\mcitedefaultmidpunct}
{\mcitedefaultendpunct}{\mcitedefaultseppunct}\relax
\EndOfBibitem
\bibitem[Lehrer and Rheinstein(2023)Lehrer, and Rheinstein]{gromacscancer1}
Lehrer,~S.; Rheinstein,~P.~H. Predicted Configuration and Stability of the ATAD2/SOX10 Complex Using Molecular Dynamics Simulations. \emph{Cancer Diagnosis \& Prognosis} \textbf{2023}, \emph{3}, 398\relax
\mciteBstWouldAddEndPuncttrue
\mciteSetBstMidEndSepPunct{\mcitedefaultmidpunct}
{\mcitedefaultendpunct}{\mcitedefaultseppunct}\relax
\EndOfBibitem
\bibitem[Zheng \latin{et~al.}(2022)Zheng, Long, Zhang, Zan, Osire, Zhou, and Xia]{gromacsalphafold2}
Zheng,~N.; Long,~M.; Zhang,~Z.; Zan,~Q.; Osire,~T.; Zhou,~H.; Xia,~X. Protein-Glutaminase engineering based on isothermal compressibility perturbation for enhanced modification of soy protein isolate. \emph{Journal of Agricultural and Food Chemistry} \textbf{2022}, \emph{70}, 13969--13978\relax
\mciteBstWouldAddEndPuncttrue
\mciteSetBstMidEndSepPunct{\mcitedefaultmidpunct}
{\mcitedefaultendpunct}{\mcitedefaultseppunct}\relax
\EndOfBibitem
\bibitem[Osifalujo \latin{et~al.}(2023)Osifalujo, Rutkowski, Satterwhite, Betts, Nkosi, and Froese]{gromacsalphafold3}
Osifalujo,~E.~A.; Rutkowski,~B.~N.; Satterwhite,~L.~R.; Betts,~P.~C.; Nkosi,~A.~K.; Froese,~J.~T. Production of novel Rieske dioxygenase metabolites enabled by enzyme engineering. \emph{Catalysis Science \& Technology} \textbf{2023}, \emph{13}, 3784--3790\relax
\mciteBstWouldAddEndPuncttrue
\mciteSetBstMidEndSepPunct{\mcitedefaultmidpunct}
{\mcitedefaultendpunct}{\mcitedefaultseppunct}\relax
\EndOfBibitem
\bibitem[Meng \latin{et~al.}(2022)Meng, Abdullahi, Ferreira, Goonawardane, Saito, Kimura, Yamasoba, Gerber, Fatihi, Rathore, \latin{et~al.} others]{gromacsalphafold4}
Meng,~B.; Abdullahi,~A.; Ferreira,~I.~A.; Goonawardane,~N.; Saito,~A.; Kimura,~I.; Yamasoba,~D.; Gerber,~P.~P.; Fatihi,~S.; Rathore,~S.; others Altered TMPRSS2 usage by SARS-CoV-2 Omicron impacts infectivity and fusogenicity. \emph{Nature} \textbf{2022}, \emph{603}, 706--714\relax
\mciteBstWouldAddEndPuncttrue
\mciteSetBstMidEndSepPunct{\mcitedefaultmidpunct}
{\mcitedefaultendpunct}{\mcitedefaultseppunct}\relax
\EndOfBibitem
\bibitem[Mosalaganti \latin{et~al.}(2022)Mosalaganti, Obarska-Kosinska, Siggel, Taniguchi, Turo{\v{n}}ov{\'a}, Zimmerli, Buczak, Schmidt, Margiotta, Mackmull, \latin{et~al.} others]{gromacsalphafold5}
Mosalaganti,~S.; Obarska-Kosinska,~A.; Siggel,~M.; Taniguchi,~R.; Turo{\v{n}}ov{\'a},~B.; Zimmerli,~C.~E.; Buczak,~K.; Schmidt,~F.~H.; Margiotta,~E.; Mackmull,~M.-T.; others AI-based structure prediction empowers integrative structural analysis of human nuclear pores. \emph{Science} \textbf{2022}, \emph{376}, eabm9506\relax
\mciteBstWouldAddEndPuncttrue
\mciteSetBstMidEndSepPunct{\mcitedefaultmidpunct}
{\mcitedefaultendpunct}{\mcitedefaultseppunct}\relax
\EndOfBibitem
\bibitem[Wu \latin{et~al.}(2024)Wu, Gu, Modica, Chen, Mrksich, and Voth]{jacsalphafold}
Wu,~J.; Gu,~Z.; Modica,~J.~A.; Chen,~S.; Mrksich,~M.; Voth,~G.~A. Megamolecule Self-Assembly Networks: A Combined Computational and Experimental Design Strategy. \emph{Journal of the American Chemical Society} \textbf{2024}, \emph{146}, 30553--30564\relax
\mciteBstWouldAddEndPuncttrue
\mciteSetBstMidEndSepPunct{\mcitedefaultmidpunct}
{\mcitedefaultendpunct}{\mcitedefaultseppunct}\relax
\EndOfBibitem
\bibitem[Stocker \latin{et~al.}(2003)Stocker, Juchli, and van Gunsteren]{hierarchy4}
Stocker,~U.; Juchli,~D.; van Gunsteren,~W.~F. Increasing the time step and efficiency of molecular dynamics simulations: optimal solutions for equilibrium simulations or structure refinement of large biomolecules. \emph{Molecular Simulation} \textbf{2003}, \emph{29}, 123--138\relax
\mciteBstWouldAddEndPuncttrue
\mciteSetBstMidEndSepPunct{\mcitedefaultmidpunct}
{\mcitedefaultendpunct}{\mcitedefaultseppunct}\relax
\EndOfBibitem
\bibitem[Feenstra \latin{et~al.}(1999)Feenstra, Hess, and Berendsen]{hierarchy3}
Feenstra,~K.~A.; Hess,~B.; Berendsen,~H. J.~C. Improving efficiency of large time-scale molecular dynamics simulations of hydrogen-rich Systems. \emph{J. Comput. Chem.} \textbf{1999}, \emph{20}, 786--798\relax
\mciteBstWouldAddEndPuncttrue
\mciteSetBstMidEndSepPunct{\mcitedefaultmidpunct}
{\mcitedefaultendpunct}{\mcitedefaultseppunct}\relax
\EndOfBibitem
\bibitem[Van~Gunsteren and Karplus(1982)Van~Gunsteren, and Karplus]{vGu1982MM}
Van~Gunsteren,~W.~F.; Karplus,~M. Effects of constraints on the dynamics of macromolecules. \emph{Macromolecules} \textbf{1982}, \emph{15}, 1528--1544\relax
\mciteBstWouldAddEndPuncttrue
\mciteSetBstMidEndSepPunct{\mcitedefaultmidpunct}
{\mcitedefaultendpunct}{\mcitedefaultseppunct}\relax
\EndOfBibitem
\bibitem[Gonnet \latin{et~al.}(2009)Gonnet, Walther, and Koumoutsakos]{Gon2009JCP}
Gonnet,~P.; Walther,~J.~H.; Koumoutsakos,~P. Theta {S}{H}{A}{K}{E}: {A}n extension to {S}{H}{A}{K}{E} for the explicit treatment of angular constraints. \emph{Comput. Phys. Comm.} \textbf{2009}, \emph{180}, 360--364\relax
\mciteBstWouldAddEndPuncttrue
\mciteSetBstMidEndSepPunct{\mcitedefaultmidpunct}
{\mcitedefaultendpunct}{\mcitedefaultseppunct}\relax
\EndOfBibitem
\bibitem[Mazur(1998)]{Maz1998JPCB}
Mazur,~A. Hierarchy of Fast Motions in Protein Dynamics. \emph{J. Phys. Chem. B} \textbf{1998}, \emph{102}, 473--479\relax
\mciteBstWouldAddEndPuncttrue
\mciteSetBstMidEndSepPunct{\mcitedefaultmidpunct}
{\mcitedefaultendpunct}{\mcitedefaultseppunct}\relax
\EndOfBibitem
\bibitem[Pechlaner \latin{et~al.}(2021)Pechlaner, Dorta, Lin, Rusu, and van Gunsteren]{pechlaner2021method}
Pechlaner,~M.; Dorta,~A.~P.; Lin,~Z.; Rusu,~V.~H.; van Gunsteren,~W.~F. A method to apply bond-angle constraints in molecular dynamics simulations. \emph{J. Comput. Chem.} \textbf{2021}, \emph{42}, 418--434\relax
\mciteBstWouldAddEndPuncttrue
\mciteSetBstMidEndSepPunct{\mcitedefaultmidpunct}
{\mcitedefaultendpunct}{\mcitedefaultseppunct}\relax
\EndOfBibitem
\bibitem[Pechlaner and van Gunsteren(2022)Pechlaner, and van Gunsteren]{pechlaner2022use}
Pechlaner,~M.; van Gunsteren,~W.~F. On the use of intra-molecular distance and angle constraints to lengthen the time step in molecular and stochastic dynamics simulations of proteins. \emph{Proteins: Structure, Function, and Bioinformatics} \textbf{2022}, \emph{90}, 543--559\relax
\mciteBstWouldAddEndPuncttrue
\mciteSetBstMidEndSepPunct{\mcitedefaultmidpunct}
{\mcitedefaultendpunct}{\mcitedefaultseppunct}\relax
\EndOfBibitem
\bibitem[F{\'a}bi{\'a}n \latin{et~al.}(2023)F{\'a}bi{\'a}n, Thallmair, and Hummer]{thallmair2023}
F{\'a}bi{\'a}n,~B.; Thallmair,~S.; Hummer,~G. Optimal bond constraint topology for molecular dynamics simulations of cholesterol. \emph{J. Chem. Theory and Comput.} \textbf{2023}, \emph{19}, 1592--1601\relax
\mciteBstWouldAddEndPuncttrue
\mciteSetBstMidEndSepPunct{\mcitedefaultmidpunct}
{\mcitedefaultendpunct}{\mcitedefaultseppunct}\relax
\EndOfBibitem
\bibitem[Krieger and Vriend(2015)Krieger, and Vriend]{kriegerlincs}
Krieger,~E.; Vriend,~G. New ways to boost molecular dynamics simulations. \emph{{J}ournal of {C}omputational {C}hemistry} \textbf{2015}, \emph{36}, 996--1007\relax
\mciteBstWouldAddEndPuncttrue
\mciteSetBstMidEndSepPunct{\mcitedefaultmidpunct}
{\mcitedefaultendpunct}{\mcitedefaultseppunct}\relax
\EndOfBibitem
\bibitem[Ryckaert \latin{et~al.}(1977)Ryckaert, Ciccotti, and Berendsen]{shake_paper}
Ryckaert,~J.~P.; Ciccotti,~G.; Berendsen,~H. J.~C. Numerical integration of the {C}artesian equations of motion of a system with constraints: {M}olecular dynamics of n-alkanes. \emph{J. Comput. Phys.} \textbf{1977}, \emph{23}, 327--341\relax
\mciteBstWouldAddEndPuncttrue
\mciteSetBstMidEndSepPunct{\mcitedefaultmidpunct}
{\mcitedefaultendpunct}{\mcitedefaultseppunct}\relax
\EndOfBibitem
\bibitem[Hess \latin{et~al.}(1997)Hess, Bekker, Berendsen, and Fraaije]{lincs_paper}
Hess,~B.; Bekker,~H.; Berendsen,~H. J.~C.; Fraaije,~J. G. E.~M. LINCS: A Linear constraint solver for molecular simulations. \emph{J. Comput. Chem.} \textbf{1997}, \emph{18}, 1463--1472\relax
\mciteBstWouldAddEndPuncttrue
\mciteSetBstMidEndSepPunct{\mcitedefaultmidpunct}
{\mcitedefaultendpunct}{\mcitedefaultseppunct}\relax
\EndOfBibitem
\bibitem[Hess(2008)]{plincs_paper}
Hess,~B. {P-LINCS}: {A} parallel linear constraint solver for molecular simulation. \emph{J. Chem. Theory Comput.} \textbf{2008}, \emph{4}, 116--122\relax
\mciteBstWouldAddEndPuncttrue
\mciteSetBstMidEndSepPunct{\mcitedefaultmidpunct}
{\mcitedefaultendpunct}{\mcitedefaultseppunct}\relax
\EndOfBibitem
\bibitem[Elber \latin{et~al.}(2011)Elber, Ruymgaart, and Hess]{shakeparallel2}
Elber,~R.; Ruymgaart,~A.; Hess,~B. SHAKE parallelization. \emph{Eur. Phys. J. Spec. Top.} \textbf{2011}, \emph{200}, 211--223\relax
\mciteBstWouldAddEndPuncttrue
\mciteSetBstMidEndSepPunct{\mcitedefaultmidpunct}
{\mcitedefaultendpunct}{\mcitedefaultseppunct}\relax
\EndOfBibitem
\bibitem[Weinbach and Elber(2005)Weinbach, and Elber]{shakeparallel1}
Weinbach,~Y.; Elber,~R. Revisiting and parallelizing {SHAKE}. \emph{J. Comput. Phys.} \textbf{2005}, \emph{209}, 193--206\relax
\mciteBstWouldAddEndPuncttrue
\mciteSetBstMidEndSepPunct{\mcitedefaultmidpunct}
{\mcitedefaultendpunct}{\mcitedefaultseppunct}\relax
\EndOfBibitem
\bibitem[not()]{note_LINCS_no_angles}
The {\sc Gromacs} manual \cite{gromacs_manual_new} states that ``LINCS should not be used with coupled angle constraints''; further information about this limitation can be found in \cite{thallmair2023}.\relax
\mciteBstWouldAddEndPunctfalse
\mciteSetBstMidEndSepPunct{\mcitedefaultmidpunct}
{}{\mcitedefaultseppunct}\relax
\EndOfBibitem
\bibitem[Broadbent \latin{et~al.}(2014)Broadbent, Spencer, Mostofi, and Sutton]{hishaketol1}
Broadbent,~R.~J.; Spencer,~J.~S.; Mostofi,~A.~A.; Sutton,~A.~P. Accelerated simulations of aromatic polymers: application to polyether ether ketone ({PEEK}). \emph{Molecular Physics} \textbf{2014}, \emph{112}, 2672--2680\relax
\mciteBstWouldAddEndPuncttrue
\mciteSetBstMidEndSepPunct{\mcitedefaultmidpunct}
{\mcitedefaultendpunct}{\mcitedefaultseppunct}\relax
\EndOfBibitem
\bibitem[Thallmair \latin{et~al.}(2021)Thallmair, Javanainen, Fábián, Mart\'inez-Seara, and Marrink]{lincs_fail_chol}
Thallmair,~S.; Javanainen,~M.; Fábián,~B.; Mart\'inez-Seara,~H.; Marrink,~S.~J. Nonconverged constraints cause artificial temperature gradients in lipid bilayer simulations. \emph{J. Phys. Chem. B} \textbf{2021}, \emph{125}, 9537--9546\relax
\mciteBstWouldAddEndPuncttrue
\mciteSetBstMidEndSepPunct{\mcitedefaultmidpunct}
{\mcitedefaultendpunct}{\mcitedefaultseppunct}\relax
\EndOfBibitem
\bibitem[L\'opez-Villellas \latin{et~al.}(2023)L\'opez-Villellas, Mikkelsen, Galano, Marco-Sola, Alastruey, Ib\'a\~nez Mar\'in, Moret\'o, Sancho, and Garc\'ia-Risue\~no]{ilves_2023}
L\'opez-Villellas,~L.; Mikkelsen,~C.; Galano,~J.; Marco-Sola,~S.; Alastruey,~J.; Ib\'a\~nez Mar\'in,~P.; Moret\'o,~M.; Sancho,~J.; Garc\'ia-Risue\~no,~P. Accurate and efficient constrained molecular dynamics of polymers using Newton's method and special purpose code. \emph{Computer Physics Communications} \textbf{2023}, \emph{288}, 108742\relax
\mciteBstWouldAddEndPuncttrue
\mciteSetBstMidEndSepPunct{\mcitedefaultmidpunct}
{\mcitedefaultendpunct}{\mcitedefaultseppunct}\relax
\EndOfBibitem
\bibitem[Hammonds and Heyes(2020)Hammonds, and Heyes]{hammonds}
Hammonds,~K.; Heyes,~D. Shadow Hamiltonian in classical NVE molecular dynamics simulations: A path to long time stability. \emph{J. Chem. Phys.} \textbf{2020}, \emph{152}\relax
\mciteBstWouldAddEndPuncttrue
\mciteSetBstMidEndSepPunct{\mcitedefaultmidpunct}
{\mcitedefaultendpunct}{\mcitedefaultseppunct}\relax
\EndOfBibitem
\bibitem[Toxvaerd \latin{et~al.}(2009)Toxvaerd, Heilmann, Ingebrigtsen, Schr{\o}der, and Dyre]{machineprecisionconstraints2}
Toxvaerd,~S.; Heilmann,~O.~J.; Ingebrigtsen,~T.; Schr{\o}der,~T.~B.; Dyre,~J.~C. Time-reversible molecular dynamics algorithms with bond constraints. \emph{The Journal of {C}hemical {P}hysics} \textbf{2009}, \emph{131}\relax
\mciteBstWouldAddEndPuncttrue
\mciteSetBstMidEndSepPunct{\mcitedefaultmidpunct}
{\mcitedefaultendpunct}{\mcitedefaultseppunct}\relax
\EndOfBibitem
\bibitem[Okumura \latin{et~al.}(2014)Okumura, Itoh, Ito, Nakamura, and Fukushima]{okumuraneedconservedquantity}
Okumura,~H.; Itoh,~S.~G.; Ito,~A.~M.; Nakamura,~H.; Fukushima,~T. Manifold correction method for the {N}os{\'e}--{H}oover and {N}os{\'e}--{P}oincare {M}olecular {D}ynamics simulations. \emph{Journal of the Physical Society of Japan} \textbf{2014}, \emph{83}, 024003\relax
\mciteBstWouldAddEndPuncttrue
\mciteSetBstMidEndSepPunct{\mcitedefaultmidpunct}
{\mcitedefaultendpunct}{\mcitedefaultseppunct}\relax
\EndOfBibitem
\bibitem[Bussi \latin{et~al.}(2009)Bussi, Zykova-Timan, and Parrinello]{papervrescale}
Bussi,~G.; Zykova-Timan,~T.; Parrinello,~M. Isothermal-isobaric {M}olecular {D}ynamics using stochastic velocity rescaling. \emph{J. Chem. Phys.} \textbf{2009}, \emph{130}, 074101\relax
\mciteBstWouldAddEndPuncttrue
\mciteSetBstMidEndSepPunct{\mcitedefaultmidpunct}
{\mcitedefaultendpunct}{\mcitedefaultseppunct}\relax
\EndOfBibitem
\bibitem[van Gunsteren \latin{et~al.}(2021)van Gunsteren, Daura, Fuchs, Hansen, Horta, H{\"u}nenberger, Mark, Pechlaner, Riniker, and Oostenbrink]{shakefailure2}
van Gunsteren,~W.~F.; Daura,~X.; Fuchs,~P.~F.; Hansen,~N.; Horta,~B.~A.; H{\"u}nenberger,~P.~H.; Mark,~A.~E.; Pechlaner,~M.; Riniker,~S.; Oostenbrink,~C. On the effect of the various assumptions and approximations used in molecular simulations on the properties of bio-molecular systems: overview and perspective on issues. \emph{ChemPhysChem} \textbf{2021}, \emph{22}, 264--282\relax
\mciteBstWouldAddEndPuncttrue
\mciteSetBstMidEndSepPunct{\mcitedefaultmidpunct}
{\mcitedefaultendpunct}{\mcitedefaultseppunct}\relax
\EndOfBibitem
\bibitem[Hess \latin{et~al.}(2019)Hess, van~der Spoel, Abraham, and Lindahl]{hessongonzalves}
Hess,~B.; van~der Spoel,~D.; Abraham,~M.; Lindahl,~E. \emph{On The Importance of Accurate Algorithms for Reliable {M}olecular {D}ynamics Simulations}; ChemRxiv. Cambridge: Cambridge Open Engage, 2019\relax
\mciteBstWouldAddEndPuncttrue
\mciteSetBstMidEndSepPunct{\mcitedefaultmidpunct}
{\mcitedefaultendpunct}{\mcitedefaultseppunct}\relax
\EndOfBibitem
\bibitem[Chiu \latin{et~al.}(2000)Chiu, Clark, Subramaniam, and Jakobsson]{hishaketol2}
Chiu,~S.-W.; Clark,~M.; Subramaniam,~S.; Jakobsson,~E. Collective motion artifacts arising in long-duration {M}olecular {D}ynamics simulations. \emph{J. Comput. Chem.} \textbf{2000}, \emph{21}, 121--131\relax
\mciteBstWouldAddEndPuncttrue
\mciteSetBstMidEndSepPunct{\mcitedefaultmidpunct}
{\mcitedefaultendpunct}{\mcitedefaultseppunct}\relax
\EndOfBibitem
\bibitem[Garc\'ia-Risue\~no \latin{et~al.}(2011)Garc\'ia-Risue\~no, Echenique, and Alonso]{ilves_original}
Garc\'ia-Risue\~no,~P.; Echenique,~P.; Alonso,~J.~L. Exact and efficient calculation of {L}agrange multipliers in constrained biological polymers: Proteins and nucleic acids as example cases. \emph{J. Comput. Chem.} \textbf{2011}, \emph{32}, 3039--3046\relax
\mciteBstWouldAddEndPuncttrue
\mciteSetBstMidEndSepPunct{\mcitedefaultmidpunct}
{\mcitedefaultendpunct}{\mcitedefaultseppunct}\relax
\EndOfBibitem
\bibitem[Garc\'ia-Risue\~no and Echenique(2012)Garc\'ia-Risue\~no, and Echenique]{GR2010JCoP}
Garc\'ia-Risue\~no,~P.; Echenique,~P. Linearly scaling direct method for accurately inverting sparse banded matrices. \emph{J. Phys. A: Math. and Theor.} \textbf{2012}, \emph{45}, 065204\relax
\mciteBstWouldAddEndPuncttrue
\mciteSetBstMidEndSepPunct{\mcitedefaultmidpunct}
{\mcitedefaultendpunct}{\mcitedefaultseppunct}\relax
\EndOfBibitem
\bibitem[Golub and Van~Loan(1993)Golub, and Van~Loan]{Gol1989BOOK}
Golub,~G.~H., Van~Loan,~C.~F., Eds. \emph{Matrix Computations}, 2nd ed.; The Johns Hopkins University Press: Baltimore and London, 1993\relax
\mciteBstWouldAddEndPuncttrue
\mciteSetBstMidEndSepPunct{\mcitedefaultmidpunct}
{\mcitedefaultendpunct}{\mcitedefaultseppunct}\relax
\EndOfBibitem
\bibitem[Lou(1989)]{lou1989}
Lou,~G. Parallel methods for solving linear systems via overlapping decomposition. M.Sc.\ thesis, University of Illinois at Urbana-Champaign, 1989\relax
\mciteBstWouldAddEndPuncttrue
\mciteSetBstMidEndSepPunct{\mcitedefaultmidpunct}
{\mcitedefaultendpunct}{\mcitedefaultseppunct}\relax
\EndOfBibitem
\bibitem[Eastman and Pande(2010)Eastman, and Pande]{Eas2010JCTC1}
Eastman,~P.; Pande,~V.~S. Constant Constraint Matrix Approximation: A robust, parallelizable constraint method for molecular simulations. \emph{J. Chem. Theory Comput.} \textbf{2010}, \emph{6}, 434--437\relax
\mciteBstWouldAddEndPuncttrue
\mciteSetBstMidEndSepPunct{\mcitedefaultmidpunct}
{\mcitedefaultendpunct}{\mcitedefaultseppunct}\relax
\EndOfBibitem
\bibitem[Kjelgaard~Mikkelsen \latin{et~al.}(2022)Kjelgaard~Mikkelsen, L{\'o}pez-Villellas, and Garc{\'\i}a-Risue{\~n}o]{ppam2_paper}
Kjelgaard~Mikkelsen,~C.~C.; L{\'o}pez-Villellas,~L.; Garc{\'\i}a-Risue{\~n}o,~P. How accurate does Newton have to be? \emph{Proceedings of the 14$^{\textrm{th}}$ International Conference on Parallel Processing and Applied Mathematics (PPAM)} \textbf{2022}, \emph{I}, 3--15\relax
\mciteBstWouldAddEndPuncttrue
\mciteSetBstMidEndSepPunct{\mcitedefaultmidpunct}
{\mcitedefaultendpunct}{\mcitedefaultseppunct}\relax
\EndOfBibitem
\bibitem[Kjelgaard~Mikkelsen \latin{et~al.}(2024)Kjelgaard~Mikkelsen, L{\'o}pez-Villellas, and Garc{\'\i}a-Risue{\~n}o]{ppam2b_paper}
Kjelgaard~Mikkelsen,~C.~C.; L{\'o}pez-Villellas,~L.; Garc{\'\i}a-Risue{\~n}o,~P. Newton's method revisited: How accurate do we have to be? \emph{Concurrency and Computation: Practice and Experience} \textbf{2024}, \emph{36}, e7853\relax
\mciteBstWouldAddEndPuncttrue
\mciteSetBstMidEndSepPunct{\mcitedefaultmidpunct}
{\mcitedefaultendpunct}{\mcitedefaultseppunct}\relax
\EndOfBibitem
\bibitem[Huang and MacKerell~Jr(2013)Huang, and MacKerell~Jr]{charmm36}
Huang,~J.; MacKerell~Jr,~A.~D. CHARMM36 all-atom additive protein force field: Validation based on comparison to NMR data. \emph{Journal of Computational Chemistry} \textbf{2013}, \emph{34}, 2135--2145\relax
\mciteBstWouldAddEndPuncttrue
\mciteSetBstMidEndSepPunct{\mcitedefaultmidpunct}
{\mcitedefaultendpunct}{\mcitedefaultseppunct}\relax
\EndOfBibitem
\bibitem[Huang \latin{et~al.}(2017)Huang, Rauscher, Nawrocki, Ran, Feig, de~Groot, Grubmüller, and MacKerell]{charmm36m}
Huang,~J.; Rauscher,~S.; Nawrocki,~G.; Ran,~T.; Feig,~M.; de~Groot,~B.~L.; Grubmüller,~H.; MacKerell,~A. D.~J. {CHARMM36m: an improved force field for folded and intrinsically disordered proteins}. \emph{Nature Methods} \textbf{2017}, \emph{14}, 71--73\relax
\mciteBstWouldAddEndPuncttrue
\mciteSetBstMidEndSepPunct{\mcitedefaultmidpunct}
{\mcitedefaultendpunct}{\mcitedefaultseppunct}\relax
\EndOfBibitem
\bibitem[Hornak \latin{et~al.}(2006)Hornak, Abel, Okur, Strockbine, Roitberg, and Simmerling]{amberff}
Hornak,~V.; Abel,~R.; Okur,~A.; Strockbine,~B.; Roitberg,~A.; Simmerling,~C. Comparison of multiple Amber force fields and development of improved protein backbone parameters. \emph{Proteins: Structure, Function, and Bioinformatics} \textbf{2006}, \emph{65}, 712--725\relax
\mciteBstWouldAddEndPuncttrue
\mciteSetBstMidEndSepPunct{\mcitedefaultmidpunct}
{\mcitedefaultendpunct}{\mcitedefaultseppunct}\relax
\EndOfBibitem
\bibitem[Jorgensen \latin{et~al.}(1996)Jorgensen, Maxwell, and Tirado-Rives]{OPLS/AA}
Jorgensen,~W.~L.; Maxwell,~D.~S.; Tirado-Rives,~J. Development and Testing of the OPLS All-Atom Force Field on Conformational Energetics and Properties of Organic Liquids. \emph{Journal of the American Chemical Society} \textbf{1996}, \emph{118}, 11225--11236\relax
\mciteBstWouldAddEndPuncttrue
\mciteSetBstMidEndSepPunct{\mcitedefaultmidpunct}
{\mcitedefaultendpunct}{\mcitedefaultseppunct}\relax
\EndOfBibitem
\bibitem[Hammonds and Ryckaert(1991)Hammonds, and Ryckaert]{hammonds1991convergence}
Hammonds,~K.~D.; Ryckaert,~J.-P. On the convergence of the {S}{H}{A}{K}{E} algorithm. \emph{Comput. Phys. Comm.} \textbf{1991}, \emph{62}, 336--351\relax
\mciteBstWouldAddEndPuncttrue
\mciteSetBstMidEndSepPunct{\mcitedefaultmidpunct}
{\mcitedefaultendpunct}{\mcitedefaultseppunct}\relax
\EndOfBibitem
\bibitem[Apol \latin{et~al.}(2024)Apol, Apostolov, Berendsen, van Buuren, Bjelkmar, van Drunen, Feenstra, Groenhor, Kasson, Larsson, Meulenhoff, Murtola, Pall, Pronk, Schulz, Shirts, Sijbers, Tieleman, Hess, van~der Spoel, and Lindahl]{gromacs_manual}
Apol,~E. \latin{et~al.}  GROMACS Documentation Release 2024.1. \textbf{2024}, \relax
\mciteBstWouldAddEndPunctfalse
\mciteSetBstMidEndSepPunct{\mcitedefaultmidpunct}
{}{\mcitedefaultseppunct}\relax
\EndOfBibitem
\bibitem[Wacha \latin{et~al.}(2023)Wacha, Varga, and Beke-Somfai]{usageBAconstraints3}
Wacha,~A.; Varga,~Z.; Beke-Somfai,~T. Comparative Study of Molecular Mechanics Force Fields for $\beta$-Peptidic Foldamers: Folding and Self-Association. \emph{Journal of Chemical Information and Modeling} \textbf{2023}, \emph{63}, 3799--3813\relax
\mciteBstWouldAddEndPuncttrue
\mciteSetBstMidEndSepPunct{\mcitedefaultmidpunct}
{\mcitedefaultendpunct}{\mcitedefaultseppunct}\relax
\EndOfBibitem
\bibitem[Mikkelsen \latin{et~al.}(2016)Mikkelsen, Alastruey-Bened\'e, Ib\'a\~nez Mar\'in, and Garc\'ia-Risue\~no]{ppam1_paper}
Mikkelsen,~C. C.~K.; Alastruey-Bened\'e,~J.; Ib\'a\~nez Mar\'in,~P.; Garc\'ia-Risue\~no,~P. Accelerating sparse arithmetic in the context of {N}ewton's method for small molecules with bond constraints. \emph{Proceedings of the 11$^{\textrm{th}}$ International Conference on Parallel Processing and Applied Mathematics (PPAM)} \textbf{2016}, \emph{I}, 160--171\relax
\mciteBstWouldAddEndPuncttrue
\mciteSetBstMidEndSepPunct{\mcitedefaultmidpunct}
{\mcitedefaultendpunct}{\mcitedefaultseppunct}\relax
\EndOfBibitem
\bibitem[Miyamoto and Kollman(1992)Miyamoto, and Kollman]{settle}
Miyamoto,~S.; Kollman,~P.~A. Settle: An analytical version of the {SHAKE} and {RATTLE} algorithm for rigid water models. \emph{J. Comput. Chem.} \textbf{1992}, \emph{13}, 952--962\relax
\mciteBstWouldAddEndPuncttrue
\mciteSetBstMidEndSepPunct{\mcitedefaultmidpunct}
{\mcitedefaultendpunct}{\mcitedefaultseppunct}\relax
\EndOfBibitem
\bibitem[Abraham \latin{et~al.}(2014)Abraham, Berk~Hess, van~der Spoel, and Lindahl]{gromacs_manual_new}
Abraham,~M.; Berk~Hess,~B.; van~der Spoel,~D.; Lindahl,~E. \emph{GROMACS user manual version 5.0.4}; http://www.gromacs.org/Documentation/Manual, 2014\relax
\mciteBstWouldAddEndPuncttrue
\mciteSetBstMidEndSepPunct{\mcitedefaultmidpunct}
{\mcitedefaultendpunct}{\mcitedefaultseppunct}\relax
\EndOfBibitem
\end{mcitethebibliography}


\providecommand{\latin}[1]{#1}
\makeatletter
\providecommand{\doi}
  {\begingroup\let\do\@makeother\dospecials
  \catcode`\{=1 \catcode`\}=2 \doi@aux}
\providecommand{\doi@aux}[1]{\endgroup\texttt{#1}}
\makeatother
\providecommand*\mcitethebibliography{\thebibliography}
\csname @ifundefined\endcsname{endmcitethebibliography}  {\let\endmcitethebibliography\endthebibliography}{}
\begin{mcitethebibliography}{84}
\providecommand*\natexlab[1]{#1}
\providecommand*\mciteSetBstSublistMode[1]{}
\providecommand*\mciteSetBstMaxWidthForm[2]{}
\providecommand*\mciteBstWouldAddEndPuncttrue
  {\def\EndOfBibitem{\unskip.}}
\providecommand*\mciteBstWouldAddEndPunctfalse
  {\let\EndOfBibitem\relax}
\providecommand*\mciteSetBstMidEndSepPunct[3]{}
\providecommand*\mciteSetBstSublistLabelBeginEnd[3]{}
\providecommand*\EndOfBibitem{}
\mciteSetBstSublistMode{f}
\mciteSetBstMaxWidthForm{subitem}{(\alph{mcitesubitemcount})}
\mciteSetBstSublistLabelBeginEnd
  {\mcitemaxwidthsubitemform\space}
  {\relax}
  {\relax}

\bibitem[Andersen(1983)]{And1983JCOP}
Andersen,~H.~C. {Rattle}: {A} {``velocity''} version of the {SHAKE} algorithm for {M}olecular {D}ynamics calculations. \emph{J. Comput. Phys.} \textbf{1983}, \emph{52}, 24--34\relax
\mciteBstWouldAddEndPuncttrue
\mciteSetBstMidEndSepPunct{\mcitedefaultmidpunct}
{\mcitedefaultendpunct}{\mcitedefaultseppunct}\relax
\EndOfBibitem
\bibitem[Miyamoto and Kollman(1992)Miyamoto, and Kollman]{settle}
Miyamoto,~S.; Kollman,~P.~A. Settle: An analytical version of the {SHAKE} and {RATTLE} algorithm for rigid water models. \emph{J. Comput. Chem.} \textbf{1992}, \emph{13}, 952--962\relax
\mciteBstWouldAddEndPuncttrue
\mciteSetBstMidEndSepPunct{\mcitedefaultmidpunct}
{\mcitedefaultendpunct}{\mcitedefaultseppunct}\relax
\EndOfBibitem
\bibitem[Lambrakos \latin{et~al.}(1989)Lambrakos, Boris, Oran, Chandrasekhar, and Nagumo]{lambrakosshake}
Lambrakos,~S.~G.; Boris,~J.; Oran,~E.; Chandrasekhar,~I.; Nagumo,~M. A modified {S}{H}{A}{K}{E} algorithm for maintaining rigid bonds in molecular dynamics simulations of large molecules. \emph{J. Comput. Phys.} \textbf{1989}, \emph{85}, 473--486\relax
\mciteBstWouldAddEndPuncttrue
\mciteSetBstMidEndSepPunct{\mcitedefaultmidpunct}
{\mcitedefaultendpunct}{\mcitedefaultseppunct}\relax
\EndOfBibitem
\bibitem[Lee \latin{et~al.}(2005)Lee, Palmo, and Krimm]{wiggle}
Lee,~S.-H.; Palmo,~K.; Krimm,~S. {W}{I}{G}{G}{L}{E}: A new constrained molecular dynamics algorithm in Cartesian coordinates. \emph{J. Comput. Phys.} \textbf{2005}, \emph{210}, 171--182\relax
\mciteBstWouldAddEndPuncttrue
\mciteSetBstMidEndSepPunct{\mcitedefaultmidpunct}
{\mcitedefaultendpunct}{\mcitedefaultseppunct}\relax
\EndOfBibitem
\bibitem[Gonnet(2006)]{Gon2006JCP}
Gonnet,~P. P-{S}{H}{A}{K}{E}: a quadratically convergent SHAKE in {O}(n-squared). \emph{J. Chem. Phys.} \textbf{2006}, \emph{220}, 740--750\relax
\mciteBstWouldAddEndPuncttrue
\mciteSetBstMidEndSepPunct{\mcitedefaultmidpunct}
{\mcitedefaultendpunct}{\mcitedefaultseppunct}\relax
\EndOfBibitem
\bibitem[Ciccotti and Ryckaert(1986)Ciccotti, and Ryckaert]{ciccotti1986molecular}
Ciccotti,~G.; Ryckaert,~J.-P. Molecular dynamics simulation of rigid molecules. \emph{Computer Physics Reports} \textbf{1986}, \emph{4}, 346--392\relax
\mciteBstWouldAddEndPuncttrue
\mciteSetBstMidEndSepPunct{\mcitedefaultmidpunct}
{\mcitedefaultendpunct}{\mcitedefaultseppunct}\relax
\EndOfBibitem
\bibitem[Ryckaert \latin{et~al.}(2001)Ryckaert, Arialdi, and Melchionna]{ryckaert2001molecular}
Ryckaert,~J.-P.; Arialdi,~G.; Melchionna,~S. Molecular dynamics of polymers with explicit but frozen hydrogens. \emph{Molecular Physics} \textbf{2001}, \emph{99}, 155--165\relax
\mciteBstWouldAddEndPuncttrue
\mciteSetBstMidEndSepPunct{\mcitedefaultmidpunct}
{\mcitedefaultendpunct}{\mcitedefaultseppunct}\relax
\EndOfBibitem
\bibitem[Eastman and Pande(2010)Eastman, and Pande]{Eas2010JCTC1}
Eastman,~P.; Pande,~V.~S. Constant Constraint Matrix Approximation: A robust, parallelizable constraint method for molecular simulations. \emph{J. Chem. Theory Comput.} \textbf{2010}, \emph{6}, 434--437\relax
\mciteBstWouldAddEndPuncttrue
\mciteSetBstMidEndSepPunct{\mcitedefaultmidpunct}
{\mcitedefaultendpunct}{\mcitedefaultseppunct}\relax
\EndOfBibitem
\bibitem[Krautler \latin{et~al.}(2001)Krautler, Van~Gunsteren, and Hunenberger]{Kra2000JCC}
Krautler,~V.; Van~Gunsteren,~W.~F.; Hunenberger,~P.~H. A fast {S}{H}{A}{K}{E} Algorithm to Solve Distance Constraint Equations for Small Molecules in {M}olecular {D}ynamics Simulations. \emph{J. Comput. Chem.} \textbf{2001}, \emph{22}, 501--508\relax
\mciteBstWouldAddEndPuncttrue
\mciteSetBstMidEndSepPunct{\mcitedefaultmidpunct}
{\mcitedefaultendpunct}{\mcitedefaultseppunct}\relax
\EndOfBibitem
\bibitem[Christen and van Gunsteren(2005)Christen, and van Gunsteren]{flexshake}
Christen,~M.; van Gunsteren,~W.~F. An approximate but fast method to impose flexible distance constraints in molecular dynamics simulations. \emph{J. Chem. Phys.} \textbf{2005}, \emph{122}\relax
\mciteBstWouldAddEndPuncttrue
\mciteSetBstMidEndSepPunct{\mcitedefaultmidpunct}
{\mcitedefaultendpunct}{\mcitedefaultseppunct}\relax
\EndOfBibitem
\bibitem[Bailey \latin{et~al.}(2008)Bailey, Lowe, and Sutton]{Bai2008JCoP}
Bailey,~A.~G.; Lowe,~C.~P.; Sutton,~A.~P. Efficient constraint dynamics using {MILC SHAKE}. \emph{J. Comput. Phys.} \textbf{2008}, \emph{227}, 8949--8959\relax
\mciteBstWouldAddEndPuncttrue
\mciteSetBstMidEndSepPunct{\mcitedefaultmidpunct}
{\mcitedefaultendpunct}{\mcitedefaultseppunct}\relax
\EndOfBibitem
\bibitem[Bailey and Lowe(2009)Bailey, and Lowe]{Bai2009JCC}
Bailey,~A.~G.; Lowe,~C.~P. {MILCH SHAKE}: An efficient method for constraint dynamics applied to alkanes. \emph{J. Comput. Chem.} \textbf{2009}, \emph{30}, 2485--2493\relax
\mciteBstWouldAddEndPuncttrue
\mciteSetBstMidEndSepPunct{\mcitedefaultmidpunct}
{\mcitedefaultendpunct}{\mcitedefaultseppunct}\relax
\EndOfBibitem
\bibitem[Mazars(2007)]{Maz2007JPAMT}
Mazars,~M. Holonomic constraints, an analytical result. \emph{J. Phys. A: Math. Theor.} \textbf{2007}, \emph{49}, 1747--1755\relax
\mciteBstWouldAddEndPuncttrue
\mciteSetBstMidEndSepPunct{\mcitedefaultmidpunct}
{\mcitedefaultendpunct}{\mcitedefaultseppunct}\relax
\EndOfBibitem
\bibitem[Weinbach and Elber(2005)Weinbach, and Elber]{shakeparallel1}
Weinbach,~Y.; Elber,~R. Revisiting and parallelizing {SHAKE}. \emph{J. Comput. Phys.} \textbf{2005}, \emph{209}, 193--206\relax
\mciteBstWouldAddEndPuncttrue
\mciteSetBstMidEndSepPunct{\mcitedefaultmidpunct}
{\mcitedefaultendpunct}{\mcitedefaultseppunct}\relax
\EndOfBibitem
\bibitem[Elber \latin{et~al.}(2011)Elber, Ruymgaart, and Hess]{shakeparallel2}
Elber,~R.; Ruymgaart,~A.; Hess,~B. SHAKE parallelization. \emph{Eur. Phys. J. Spec. Top.} \textbf{2011}, \emph{200}, 211--223\relax
\mciteBstWouldAddEndPuncttrue
\mciteSetBstMidEndSepPunct{\mcitedefaultmidpunct}
{\mcitedefaultendpunct}{\mcitedefaultseppunct}\relax
\EndOfBibitem
\bibitem[Barth \latin{et~al.}(1995)Barth, Kuczera, Leimkuhler, and Skeel]{Bar1995JCoP}
Barth,~E.; Kuczera,~K.; Leimkuhler,~B.; Skeel,~R.~D. Algorithms for constrained {M}olecular {D}ynamics. \emph{J. Comput. Phys.} \textbf{1995}, \emph{16 (10)}, 1192--1209\relax
\mciteBstWouldAddEndPuncttrue
\mciteSetBstMidEndSepPunct{\mcitedefaultmidpunct}
{\mcitedefaultendpunct}{\mcitedefaultseppunct}\relax
\EndOfBibitem
\bibitem[Gonnet \latin{et~al.}(2009)Gonnet, Walther, and Koumoutsakos]{Gon2009JCP}
Gonnet,~P.; Walther,~J.~H.; Koumoutsakos,~P. Theta {S}{H}{A}{K}{E}: {A}n extension to {S}{H}{A}{K}{E} for the explicit treatment of angular constraints. \emph{Comput. Phys. Comm.} \textbf{2009}, \emph{180}, 360--364\relax
\mciteBstWouldAddEndPuncttrue
\mciteSetBstMidEndSepPunct{\mcitedefaultmidpunct}
{\mcitedefaultendpunct}{\mcitedefaultseppunct}\relax
\EndOfBibitem
\bibitem[Pechlaner \latin{et~al.}(2021)Pechlaner, Dorta, Lin, Rusu, and van Gunsteren]{pechlaner2021method}
Pechlaner,~M.; Dorta,~A.~P.; Lin,~Z.; Rusu,~V.~H.; van Gunsteren,~W.~F. A method to apply bond-angle constraints in molecular dynamics simulations. \emph{J. Comput. Chem.} \textbf{2021}, \emph{42}, 418--434\relax
\mciteBstWouldAddEndPuncttrue
\mciteSetBstMidEndSepPunct{\mcitedefaultmidpunct}
{\mcitedefaultendpunct}{\mcitedefaultseppunct}\relax
\EndOfBibitem
\bibitem[Pechlaner and van Gunsteren(2020)Pechlaner, and van Gunsteren]{pechlaner2020algorithms}
Pechlaner,~M.; van Gunsteren,~W.~F. Algorithms to apply dihedral-angle constraints in molecular or stochastic dynamics simulations. \emph{J. Chem. Phys} \textbf{2020}, \emph{152}\relax
\mciteBstWouldAddEndPuncttrue
\mciteSetBstMidEndSepPunct{\mcitedefaultmidpunct}
{\mcitedefaultendpunct}{\mcitedefaultseppunct}\relax
\EndOfBibitem
\bibitem[Dubbeldam \latin{et~al.}(2010)Dubbeldam, Oxford, Krishna, Broadbelt, and Snurr]{dihedral_solver_2}
Dubbeldam,~D.; Oxford,~G.~A.; Krishna,~R.; Broadbelt,~L.~J.; Snurr,~R.~Q. Distance and angular holonomic constraints in molecular simulations. \emph{The Journal of chemical physics} \textbf{2010}, \emph{133}\relax
\mciteBstWouldAddEndPuncttrue
\mciteSetBstMidEndSepPunct{\mcitedefaultmidpunct}
{\mcitedefaultendpunct}{\mcitedefaultseppunct}\relax
\EndOfBibitem
\bibitem[Hayashi \latin{et~al.}(2022)Hayashi, Shiomi, Morikawa, and Yoshida]{usageBAconstraints4}
Hayashi,~Y.; Shiomi,~J.; Morikawa,~J.; Yoshida,~R. {R}adon{P}y: automated physical property calculation using all-atom classical molecular dynamics simulations for polymer informatics. \emph{npj Computational Materials} \textbf{2022}, \emph{8}, 222\relax
\mciteBstWouldAddEndPuncttrue
\mciteSetBstMidEndSepPunct{\mcitedefaultmidpunct}
{\mcitedefaultendpunct}{\mcitedefaultseppunct}\relax
\EndOfBibitem
\bibitem[Huda and Rai(2020)Huda, and Rai]{usageBAconstraints1}
Huda,~M.~M.; Rai,~N. Probing early-stage aggregation of low molecular weight gelator in an organic solvent. \emph{J. Phys. Chem. B} \textbf{2020}, \emph{124}, 2277--2288\relax
\mciteBstWouldAddEndPuncttrue
\mciteSetBstMidEndSepPunct{\mcitedefaultmidpunct}
{\mcitedefaultendpunct}{\mcitedefaultseppunct}\relax
\EndOfBibitem
\bibitem[Ciulla \latin{et~al.}(2023)Ciulla, Fontana, Lorenzi, Marchini, Campone, Sadeghi, Paleari, Sattin, and Gelain]{usageBAconstraints2}
Ciulla,~M.~G.; Fontana,~F.; Lorenzi,~R.; Marchini,~A.; Campone,~L.; Sadeghi,~E.; Paleari,~A.; Sattin,~S.; Gelain,~F. Novel self-assembling cyclic peptides with reversible supramolecular nanostructures. \emph{Materials Chemistry Frontiers} \textbf{2023}, \emph{7}, 3680--3692\relax
\mciteBstWouldAddEndPuncttrue
\mciteSetBstMidEndSepPunct{\mcitedefaultmidpunct}
{\mcitedefaultendpunct}{\mcitedefaultseppunct}\relax
\EndOfBibitem
\bibitem[Wacha \latin{et~al.}(2023)Wacha, Varga, and Beke-Somfai]{usageBAconstraints3}
Wacha,~A.; Varga,~Z.; Beke-Somfai,~T. Comparative Study of Molecular Mechanics Force Fields for $\beta$-Peptidic Foldamers: Folding and Self-Association. \emph{Journal of Chemical Information and Modeling} \textbf{2023}, \emph{63}, 3799--3813\relax
\mciteBstWouldAddEndPuncttrue
\mciteSetBstMidEndSepPunct{\mcitedefaultmidpunct}
{\mcitedefaultendpunct}{\mcitedefaultseppunct}\relax
\EndOfBibitem
\bibitem[Marrink \latin{et~al.}(2007)Marrink, Risselada, Yefimov, Tieleman, and De~Vries]{martinicoarsegrain}
Marrink,~S.~J.; Risselada,~H.~J.; Yefimov,~S.; Tieleman,~D.~P.; De~Vries,~A.~H. The MARTINI force field: coarse grained model for biomolecular simulations. \emph{J. Phys. Chem. B} \textbf{2007}, \emph{111}, 7812--7824\relax
\mciteBstWouldAddEndPuncttrue
\mciteSetBstMidEndSepPunct{\mcitedefaultmidpunct}
{\mcitedefaultendpunct}{\mcitedefaultseppunct}\relax
\EndOfBibitem
\bibitem[Hammonds and Heyes(2020)Hammonds, and Heyes]{hammonds}
Hammonds,~K.; Heyes,~D. Shadow Hamiltonian in classical NVE molecular dynamics simulations: A path to long time stability. \emph{J. Chem. Phys.} \textbf{2020}, \emph{152}\relax
\mciteBstWouldAddEndPuncttrue
\mciteSetBstMidEndSepPunct{\mcitedefaultmidpunct}
{\mcitedefaultendpunct}{\mcitedefaultseppunct}\relax
\EndOfBibitem
\bibitem[not()]{note2024a}
Another reason may be the influence of Ref. \cite{vGu1982MM}; however, note that such reference did not analyse explicit hydrogen atoms, hence its conclusions do not apply to H-angles.\relax
\mciteBstWouldAddEndPunctfalse
\mciteSetBstMidEndSepPunct{\mcitedefaultmidpunct}
{}{\mcitedefaultseppunct}\relax
\EndOfBibitem
\bibitem[Ryckaert \latin{et~al.}(1977)Ryckaert, Ciccotti, and Berendsen]{shake_paper}
Ryckaert,~J.~P.; Ciccotti,~G.; Berendsen,~H. J.~C. Numerical integration of the {C}artesian equations of motion of a system with constraints: {M}olecular dynamics of n-alkanes. \emph{J. Comput. Phys.} \textbf{1977}, \emph{23}, 327--341\relax
\mciteBstWouldAddEndPuncttrue
\mciteSetBstMidEndSepPunct{\mcitedefaultmidpunct}
{\mcitedefaultendpunct}{\mcitedefaultseppunct}\relax
\EndOfBibitem
\bibitem[Hess \latin{et~al.}(1997)Hess, Bekker, Berendsen, and Fraaije]{lincs_paper}
Hess,~B.; Bekker,~H.; Berendsen,~H. J.~C.; Fraaije,~J. G. E.~M. LINCS: A Linear constraint solver for molecular simulations. \emph{J. Comput. Chem.} \textbf{1997}, \emph{18}, 1463--1472\relax
\mciteBstWouldAddEndPuncttrue
\mciteSetBstMidEndSepPunct{\mcitedefaultmidpunct}
{\mcitedefaultendpunct}{\mcitedefaultseppunct}\relax
\EndOfBibitem
\bibitem[Hess(2008)]{plincs_paper}
Hess,~B. {P-LINCS}: {A} parallel linear constraint solver for molecular simulation. \emph{J. Chem. Theory Comput.} \textbf{2008}, \emph{4}, 116--122\relax
\mciteBstWouldAddEndPuncttrue
\mciteSetBstMidEndSepPunct{\mcitedefaultmidpunct}
{\mcitedefaultendpunct}{\mcitedefaultseppunct}\relax
\EndOfBibitem
\bibitem[Verlet(1967)]{Ver1967PR}
Verlet,~L. Computer {''experiments''} on classical fluids. {I}. {T}hermodynamical properties of {L}ennard-{J}ones molecules. \emph{Phys. Rev.} \textbf{1967}, \emph{159}, 98--103\relax
\mciteBstWouldAddEndPuncttrue
\mciteSetBstMidEndSepPunct{\mcitedefaultmidpunct}
{\mcitedefaultendpunct}{\mcitedefaultseppunct}\relax
\EndOfBibitem
\bibitem[Hairer \latin{et~al.}(2006)Hairer, Wanner, and Lubich]{hairer2006geometric}
Hairer,~E.; Wanner,~G.; Lubich,~C. \emph{Geometric Numerical Integration}; Springer, 2006\relax
\mciteBstWouldAddEndPuncttrue
\mciteSetBstMidEndSepPunct{\mcitedefaultmidpunct}
{\mcitedefaultendpunct}{\mcitedefaultseppunct}\relax
\EndOfBibitem
\bibitem[Mor{\'e}(1972)]{more1972}
Mor{\'e},~J.~J. {Nonlinear Generalizations of Matrix Diagonal Dominance with Applications To Gauss-Seidel Iteration}. \emph{SIAM J. Numer. Anal.} \textbf{1972}, \emph{9}, 357--378\relax
\mciteBstWouldAddEndPuncttrue
\mciteSetBstMidEndSepPunct{\mcitedefaultmidpunct}
{\mcitedefaultendpunct}{\mcitedefaultseppunct}\relax
\EndOfBibitem
\bibitem[Porshing(1971)]{porshing1971}
Porshing,~T.~A. {On the Rates of Convergence of Jacobi and Gauss-Seidel Methods for M-functions}. \emph{SIAM J. Numer. Anal.} \textbf{1971}, \emph{8}, 575--582\relax
\mciteBstWouldAddEndPuncttrue
\mciteSetBstMidEndSepPunct{\mcitedefaultmidpunct}
{\mcitedefaultendpunct}{\mcitedefaultseppunct}\relax
\EndOfBibitem
\bibitem[L\'opez-Villellas \latin{et~al.}(2023)L\'opez-Villellas, Mikkelsen, Galano, Marco-Sola, Alastruey, Ib\'a\~nez Mar\'in, Moret\'o, Sancho, and Garc\'ia-Risue\~no]{ilves_2023}
L\'opez-Villellas,~L.; Mikkelsen,~C.; Galano,~J.; Marco-Sola,~S.; Alastruey,~J.; Ib\'a\~nez Mar\'in,~P.; Moret\'o,~M.; Sancho,~J.; Garc\'ia-Risue\~no,~P. Accurate and efficient constrained molecular dynamics of polymers using Newton's method and special purpose code. \emph{Computer Physics Communications} \textbf{2023}, \emph{288}, 108742\relax
\mciteBstWouldAddEndPuncttrue
\mciteSetBstMidEndSepPunct{\mcitedefaultmidpunct}
{\mcitedefaultendpunct}{\mcitedefaultseppunct}\relax
\EndOfBibitem
\bibitem[Phillips \latin{et~al.}(2005)Phillips, Braun, Wang, Gumbart, Tajkhorshid, Villa, Chipot, Skeel, Kal\'e, and Schulten]{namd_paper}
Phillips,~J.~C.; Braun,~R.; Wang,~W.; Gumbart,~J.; Tajkhorshid,~E.; Villa,~E.; Chipot,~C.; Skeel,~R.~D.; Kal\'e,~L.; Schulten,~K. Scalable {M}olecular {D}ynamics with {N}{A}{M}{D}. \emph{J. Comput. Chem.} \textbf{2005}, \emph{26}, 1781--1802\relax
\mciteBstWouldAddEndPuncttrue
\mciteSetBstMidEndSepPunct{\mcitedefaultmidpunct}
{\mcitedefaultendpunct}{\mcitedefaultseppunct}\relax
\EndOfBibitem
\bibitem[Phillips \latin{et~al.}(2020)Phillips, Hardy, Maia, Stone, Ribeiro, Bernardi, Buch, Fiorin, H{\'e}nin, Jiang, \latin{et~al.} others]{namd2}
Phillips,~J.~C.; Hardy,~D.~J.; Maia,~J.~D.; Stone,~J.~E.; Ribeiro,~J.~V.; Bernardi,~R.~C.; Buch,~R.; Fiorin,~G.; H{\'e}nin,~J.; Jiang,~W.; others Scalable molecular dynamics on {C}{P}{U} and {G}{P}{U} architectures with {N}{A}{M}{D}. \emph{J. Chem. Phys.} \textbf{2020}, \emph{153}\relax
\mciteBstWouldAddEndPuncttrue
\mciteSetBstMidEndSepPunct{\mcitedefaultmidpunct}
{\mcitedefaultendpunct}{\mcitedefaultseppunct}\relax
\EndOfBibitem
\bibitem[Pearlman \latin{et~al.}(1995)Pearlman, Case, Caldwell, Ross, Cheatham~III, DeBolt, Ferguson, Seibel, and Kollman]{Pea1995CoPC}
Pearlman,~D.~A.; Case,~D.~A.; Caldwell,~J.~W.; Ross,~W.~R.; Cheatham~III,~T.~E.; DeBolt,~S.; Ferguson,~D.; Seibel,~G.; Kollman,~P. {AMBER}, a computer program for applying molecular mechanics, normal mode analysis, {M}olecular {D}ynamics and free energy calculations to elucidate the structures and energies of molecules. \emph{Comp. Phys. Commun.} \textbf{1995}, \emph{91}, 1--41\relax
\mciteBstWouldAddEndPuncttrue
\mciteSetBstMidEndSepPunct{\mcitedefaultmidpunct}
{\mcitedefaultendpunct}{\mcitedefaultseppunct}\relax
\EndOfBibitem
\bibitem[Brooks \latin{et~al.}(2009)Brooks, Brooks~III, MacKerell, Nilsson, Petrella, Roux, Won, Archontis, Bartels, Boresch, Caflisch, Caves, Cui, Dinner, Feig, Fischer, Gao, Hodoscek, Im, Kuczera, Lazaridis, Ma, Ovchinnikov, Paci, Pastor, Post, Pu, Schaefer, Tidor, Venable, Woodcock, Wu, Yang, York, and Karplus]{Bro2009JCC}
Brooks,~B.~R. \latin{et~al.}  {CHARMM}: {T}he biomolecular simulation program. \emph{J. Comput. Chem.} \textbf{2009}, \emph{30}, 1545--1615\relax
\mciteBstWouldAddEndPuncttrue
\mciteSetBstMidEndSepPunct{\mcitedefaultmidpunct}
{\mcitedefaultendpunct}{\mcitedefaultseppunct}\relax
\EndOfBibitem
\bibitem[Ortega and Rheinboldt(1970)Ortega, and Rheinboldt]{ortega_iterative_1970}
Ortega,~J.~M.; Rheinboldt,~W.~C. \emph{Iterative solution of nonlinear equations in several variables}; Computer science and applied mathematics; Academic Press: New York, 1970\relax
\mciteBstWouldAddEndPuncttrue
\mciteSetBstMidEndSepPunct{\mcitedefaultmidpunct}
{\mcitedefaultendpunct}{\mcitedefaultseppunct}\relax
\EndOfBibitem
\bibitem[Kelley(1995)]{kelley_iterative_1995}
Kelley,~C.~T. \emph{Iterative methods for linear and nonlinear equations}; Frontiers in applied mathematics 16; Society for Industrial and Applied Mathematics: Philadelphia, 1995\relax
\mciteBstWouldAddEndPuncttrue
\mciteSetBstMidEndSepPunct{\mcitedefaultmidpunct}
{\mcitedefaultendpunct}{\mcitedefaultseppunct}\relax
\EndOfBibitem
\bibitem[Lou(1989)]{lou1989}
Lou,~G. Parallel methods for solving linear systems via overlapping decomposition. M.Sc.\ thesis, University of Illinois at Urbana-Champaign, 1989\relax
\mciteBstWouldAddEndPuncttrue
\mciteSetBstMidEndSepPunct{\mcitedefaultmidpunct}
{\mcitedefaultendpunct}{\mcitedefaultseppunct}\relax
\EndOfBibitem
\bibitem[Golub and Van~Loan(1993)Golub, and Van~Loan]{Gol1989BOOK}
Golub,~G.~H., Van~Loan,~C.~F., Eds. \emph{Matrix Computations}, 2nd ed.; The Johns Hopkins University Press: Baltimore and London, 1993\relax
\mciteBstWouldAddEndPuncttrue
\mciteSetBstMidEndSepPunct{\mcitedefaultmidpunct}
{\mcitedefaultendpunct}{\mcitedefaultseppunct}\relax
\EndOfBibitem
\bibitem[Karypis and Kumar(1998)Karypis, and Kumar]{Karypis1998}
Karypis,~G.; Kumar,~V. Multilevelk-way partitioning scheme for irregular graphs. \emph{Journal of Parallel and Distributed Computing} \textbf{1998}, \emph{48}, 96--129\relax
\mciteBstWouldAddEndPuncttrue
\mciteSetBstMidEndSepPunct{\mcitedefaultmidpunct}
{\mcitedefaultendpunct}{\mcitedefaultseppunct}\relax
\EndOfBibitem
\bibitem[Amestoy \latin{et~al.}(1996)Amestoy, Davis, and Duff]{amestoy1996approximate}
Amestoy,~P.~R.; Davis,~T.~A.; Duff,~I.~S. An approximate minimum degree ordering algorithm. \emph{SIAM Journal on Matrix Analysis and Applications} \textbf{1996}, \emph{17}, 886--905\relax
\mciteBstWouldAddEndPuncttrue
\mciteSetBstMidEndSepPunct{\mcitedefaultmidpunct}
{\mcitedefaultendpunct}{\mcitedefaultseppunct}\relax
\EndOfBibitem
\bibitem[Zlatev \latin{et~al.}(2018)Zlatev, Dimov, Faragó, and Ágnes Havasi]{zahari2018richardson}
Zlatev,~Z.; Dimov,~I.; Faragó,~I.; Ágnes Havasi \emph{Richardson Extrapolation: Practical Aspects and Applications}; De Gruyter, 2018\relax
\mciteBstWouldAddEndPuncttrue
\mciteSetBstMidEndSepPunct{\mcitedefaultmidpunct}
{\mcitedefaultendpunct}{\mcitedefaultseppunct}\relax
\EndOfBibitem
\bibitem[Kjelgaard~Mikkelsen and L{\'o}pez-Villellas(2025)Kjelgaard~Mikkelsen, and L{\'o}pez-Villellas]{ppam3_paper}
Kjelgaard~Mikkelsen,~C.~C.; L{\'o}pez-Villellas,~L. The need for accuracy and smoothness in numerical simulations. Parallel Processing and Applied Mathematics, 15th International Conference PPAM 2024. 2025; pp 3--16\relax
\mciteBstWouldAddEndPuncttrue
\mciteSetBstMidEndSepPunct{\mcitedefaultmidpunct}
{\mcitedefaultendpunct}{\mcitedefaultseppunct}\relax
\EndOfBibitem
\bibitem[Garc\'ia-Risue\~no and Ib\'a\~nez(2012)Garc\'ia-Risue\~no, and Ib\'a\~nez]{reviewibanez}
Garc\'ia-Risue\~no,~P.; Ib\'a\~nez,~P.~E. A review of high performance computing foundations for scientists. \emph{International Journal of Modern Physics C} \textbf{2012}, \emph{23}, 1230001\relax
\mciteBstWouldAddEndPuncttrue
\mciteSetBstMidEndSepPunct{\mcitedefaultmidpunct}
{\mcitedefaultendpunct}{\mcitedefaultseppunct}\relax
\EndOfBibitem
\bibitem[Martin \latin{et~al.}(1999)Martin, Richard, Salem, Hartley, and Mauguen]{structurebarnase}
Martin,~C.; Richard,~V.; Salem,~M.; Hartley,~R.; Mauguen,~Y. Refinement and structural analysis of Barnase at 1.5 Å resolution. \emph{Acta Crystallographica Section D: Biological Crystallography} \textbf{1999}, \emph{55}, 386--398\relax
\mciteBstWouldAddEndPuncttrue
\mciteSetBstMidEndSepPunct{\mcitedefaultmidpunct}
{\mcitedefaultendpunct}{\mcitedefaultseppunct}\relax
\EndOfBibitem
\bibitem[Mackerell~Jr. \latin{et~al.}(2004)Mackerell~Jr., Feig, and Brooks~III]{charmm27}
Mackerell~Jr.,~A.~D.; Feig,~M.; Brooks~III,~C.~L. Extending the treatment of backbone energetics in protein force fields: Limitations of gas-phase quantum mechanics in reproducing protein conformational distributions in molecular dynamics simulations. \emph{Journal of Computational Chemistry} \textbf{2004}, \emph{25}, 1400--1415\relax
\mciteBstWouldAddEndPuncttrue
\mciteSetBstMidEndSepPunct{\mcitedefaultmidpunct}
{\mcitedefaultendpunct}{\mcitedefaultseppunct}\relax
\EndOfBibitem
\bibitem[Jorgensen \latin{et~al.}(1983)Jorgensen, Chandrasekhar, Madura, Impey, and Klein]{tip3p}
Jorgensen,~W.~L.; Chandrasekhar,~J.; Madura,~J.~D.; Impey,~R.~W.; Klein,~M.~L. {Comparison of simple potential functions for simulating liquid water}. \emph{The Journal of Chemical Physics} \textbf{1983}, \emph{79}, 926--935\relax
\mciteBstWouldAddEndPuncttrue
\mciteSetBstMidEndSepPunct{\mcitedefaultmidpunct}
{\mcitedefaultendpunct}{\mcitedefaultseppunct}\relax
\EndOfBibitem
\bibitem[Cotton \latin{et~al.}(1977)Cotton, Hazen~Jr., and Legg]{PDB_2SNS}
Cotton,~F.; Hazen~Jr.,~E.; Legg,~M. Staphylococcal nuclease. Proposed mechanism of action based on structure of enzyme-thymidine 3',5'-biphosphate-calcium ion complex at 1.5-angstroms resolution. 1977; \url{https://www.rcsb.org/structure/2SNS}, RCSB Protein Data Bank entry, Accessed: 2023-09-27\relax
\mciteBstWouldAddEndPuncttrue
\mciteSetBstMidEndSepPunct{\mcitedefaultmidpunct}
{\mcitedefaultendpunct}{\mcitedefaultseppunct}\relax
\EndOfBibitem
\bibitem[Douangamath \latin{et~al.}(2020)Douangamath, Fearon, Gehrtz, Krojer, Lukacik, Owen, Resnick, Strain-Damerell, Aimon, \latin{et~al.} others]{articuloproteinacovid}
Douangamath,~A.; Fearon,~D.; Gehrtz,~P.; Krojer,~T.; Lukacik,~P.; Owen,~C.~D.; Resnick,~E.; Strain-Damerell,~C.; Aimon,~A.; others Crystallographic and electrophilic fragment screening of the {SARS}-{C}o{V}-2 main protease. \emph{Nat. Comm.} \textbf{2020}, \emph{11}, 1--11\relax
\mciteBstWouldAddEndPuncttrue
\mciteSetBstMidEndSepPunct{\mcitedefaultmidpunct}
{\mcitedefaultendpunct}{\mcitedefaultseppunct}\relax
\EndOfBibitem
\bibitem[N{\'u}{\~n}ez-Rojas \latin{et~al.}(2018)N{\'u}{\~n}ez-Rojas, Flores-Ruiz, and Alejandre]{mexicoorganicsolvents}
N{\'u}{\~n}ez-Rojas,~E.; Flores-Ruiz,~H.~M.; Alejandre,~J. Molecular dynamics simulations to separate benzene from hydrocarbons using polar and ionic liquid solvents. \emph{Journal of Molecular Liquids} \textbf{2018}, \emph{249}, 591--599\relax
\mciteBstWouldAddEndPuncttrue
\mciteSetBstMidEndSepPunct{\mcitedefaultmidpunct}
{\mcitedefaultendpunct}{\mcitedefaultseppunct}\relax
\EndOfBibitem
\bibitem[Schmid \latin{et~al.}(2011)Schmid, Eichenberger, Choutko, Riniker, Winger, Mark, and van Gunsteren]{gromos54a7}
Schmid,~N.; Eichenberger,~A.~P.; Choutko,~A.; Riniker,~S.; Winger,~M.; Mark,~A.~E.; van Gunsteren,~W.~F. Definition and testing of the GROMOS force-field versions 54A7 and 54B7. \emph{European Biophysics Journal} \textbf{2011}, \emph{40}, 843--856\relax
\mciteBstWouldAddEndPuncttrue
\mciteSetBstMidEndSepPunct{\mcitedefaultmidpunct}
{\mcitedefaultendpunct}{\mcitedefaultseppunct}\relax
\EndOfBibitem
\bibitem[Wu \latin{et~al.}(2010)Wu, Mohideen, Vasudevan, and Davey]{Wu2010}
Wu,~B.; Mohideen,~K.; Vasudevan,~D.; Davey,~C.~A. Structural Insight into the Sequence Dependence of Nucleosome Positioning. \emph{Structure} \textbf{2010}, \emph{18}, 528–536\relax
\mciteBstWouldAddEndPuncttrue
\mciteSetBstMidEndSepPunct{\mcitedefaultmidpunct}
{\mcitedefaultendpunct}{\mcitedefaultseppunct}\relax
\EndOfBibitem
\bibitem[Tian \latin{et~al.}(2020)Tian, Kasavajhala, Belfon, Raguette, Huang, Migues, Bickel, Wang, Pincay, Wu, and Simmerling]{ff19sb}
Tian,~C.; Kasavajhala,~K.; Belfon,~K. A.~A.; Raguette,~L.; Huang,~H.; Migues,~A.~N.; Bickel,~J.; Wang,~Y.; Pincay,~J.; Wu,~Q.; Simmerling,~C. ff19SB: Amino-Acid-Specific Protein Backbone Parameters Trained against Quantum Mechanics Energy Surfaces in Solution. \emph{Journal of Chemical Theory and Computation} \textbf{2020}, \emph{16}, 528--552\relax
\mciteBstWouldAddEndPuncttrue
\mciteSetBstMidEndSepPunct{\mcitedefaultmidpunct}
{\mcitedefaultendpunct}{\mcitedefaultseppunct}\relax
\EndOfBibitem
\bibitem[Zgarbová \latin{et~al.}(2021)Zgarbová, Šponer, and Jurečka]{ol21}
Zgarbová,~M.; Šponer,~J.; Jurečka,~P. Z-DNA as a Touchstone for Additive Empirical Force Fields and a Refinement of the Alpha/Gamma DNA Torsions for AMBER. \emph{Journal of Chemical Theory and Computation} \textbf{2021}, \emph{17}, 6292--6301\relax
\mciteBstWouldAddEndPuncttrue
\mciteSetBstMidEndSepPunct{\mcitedefaultmidpunct}
{\mcitedefaultendpunct}{\mcitedefaultseppunct}\relax
\EndOfBibitem
\bibitem[Izadi \latin{et~al.}(2014)Izadi, Anandakrishnan, and Onufriev]{opc4}
Izadi,~S.; Anandakrishnan,~R.; Onufriev,~A.~V. Building Water Models: A Different Approach. \emph{The Journal of Physical Chemistry Letters} \textbf{2014}, \emph{5}, 3863--3871\relax
\mciteBstWouldAddEndPuncttrue
\mciteSetBstMidEndSepPunct{\mcitedefaultmidpunct}
{\mcitedefaultendpunct}{\mcitedefaultseppunct}\relax
\EndOfBibitem
\bibitem[Li \latin{et~al.}(2015)Li, Song, and Merz]{Li2015}
Li,~P.; Song,~L.~F.; Merz,~K.~M. Systematic Parameterization of Monovalent Ions Employing the Nonbonded Model. \emph{Journal of Chemical Theory and Computation} \textbf{2015}, \emph{11}, 1645–1657\relax
\mciteBstWouldAddEndPuncttrue
\mciteSetBstMidEndSepPunct{\mcitedefaultmidpunct}
{\mcitedefaultendpunct}{\mcitedefaultseppunct}\relax
\EndOfBibitem
\bibitem[Sengupta \latin{et~al.}(2021)Sengupta, Li, Song, Li, and Merz]{Sengupta2021}
Sengupta,~A.; Li,~Z.; Song,~L.~F.; Li,~P.; Merz,~K.~M. Parameterization of Monovalent Ions for the OPC3, OPC, TIP3P-FB, and TIP4P-FB Water Models. \emph{Journal of Chemical Information and Modeling} \textbf{2021}, \emph{61}, 869–880\relax
\mciteBstWouldAddEndPuncttrue
\mciteSetBstMidEndSepPunct{\mcitedefaultmidpunct}
{\mcitedefaultendpunct}{\mcitedefaultseppunct}\relax
\EndOfBibitem
\bibitem[Sui \latin{et~al.}(2001)Sui, Han, Lee, Walian, and Jap]{Sui2001}
Sui,~H.; Han,~B.-G.; Lee,~J.~K.; Walian,~P.; Jap,~B.~K. Structural basis of water-specific transport through the AQP1 water channel. \emph{Nature} \textbf{2001}, \emph{414}, 872–878\relax
\mciteBstWouldAddEndPuncttrue
\mciteSetBstMidEndSepPunct{\mcitedefaultmidpunct}
{\mcitedefaultendpunct}{\mcitedefaultseppunct}\relax
\EndOfBibitem
\bibitem[Huang \latin{et~al.}(2017)Huang, Rauscher, Nawrocki, Ran, Feig, de~Groot, Grubmüller, and MacKerell]{charmm36m}
Huang,~J.; Rauscher,~S.; Nawrocki,~G.; Ran,~T.; Feig,~M.; de~Groot,~B.~L.; Grubmüller,~H.; MacKerell,~A. D.~J. {CHARMM36m: an improved force field for folded and intrinsically disordered proteins}. \emph{Nature Methods} \textbf{2017}, \emph{14}, 71--73\relax
\mciteBstWouldAddEndPuncttrue
\mciteSetBstMidEndSepPunct{\mcitedefaultmidpunct}
{\mcitedefaultendpunct}{\mcitedefaultseppunct}\relax
\EndOfBibitem
\bibitem[Huang and MacKerell~Jr(2013)Huang, and MacKerell~Jr]{charmm36}
Huang,~J.; MacKerell~Jr,~A.~D. CHARMM36 all-atom additive protein force field: Validation based on comparison to NMR data. \emph{Journal of Computational Chemistry} \textbf{2013}, \emph{34}, 2135--2145\relax
\mciteBstWouldAddEndPuncttrue
\mciteSetBstMidEndSepPunct{\mcitedefaultmidpunct}
{\mcitedefaultendpunct}{\mcitedefaultseppunct}\relax
\EndOfBibitem
\bibitem[Kim \latin{et~al.}(2021)Kim, Oh, and Kim]{benzeneapplications}
Kim,~J.; Oh,~J.~H.; Kim,~D. Recent advances in single-benzene-based fluorophores: Physicochemical properties and applications. \emph{Organic \& Biomolecular Chemistry} \textbf{2021}, \emph{19}, 933--946\relax
\mciteBstWouldAddEndPuncttrue
\mciteSetBstMidEndSepPunct{\mcitedefaultmidpunct}
{\mcitedefaultendpunct}{\mcitedefaultseppunct}\relax
\EndOfBibitem
\bibitem[Levental and Lyman(2023)Levental, and Lyman]{membranes2}
Levental,~I.; Lyman,~E. Regulation of membrane protein structure and function by their lipid nano-environment. \emph{Nature Reviews Molecular Cell Biology} \textbf{2023}, \emph{24}, 107--122\relax
\mciteBstWouldAddEndPuncttrue
\mciteSetBstMidEndSepPunct{\mcitedefaultmidpunct}
{\mcitedefaultendpunct}{\mcitedefaultseppunct}\relax
\EndOfBibitem
\bibitem[Jo \latin{et~al.}(2008)Jo, Kim, Iyer, and Im]{Jo2008}
Jo,~S.; Kim,~T.; Iyer,~V.~G.; Im,~W. CHARMM‐GUI: A web‐based graphical user interface for CHARMM. \emph{J. Comput. Chem.} \textbf{2008}, \emph{29}, 1859–1865\relax
\mciteBstWouldAddEndPuncttrue
\mciteSetBstMidEndSepPunct{\mcitedefaultmidpunct}
{\mcitedefaultendpunct}{\mcitedefaultseppunct}\relax
\EndOfBibitem
\bibitem[Lee \latin{et~al.}(2015)Lee, Cheng, Swails, Yeom, Eastman, Lemkul, Wei, Buckner, Jeong, Qi, Jo, Pande, Case, Brooks, MacKerell, Klauda, and Im]{Lee2015}
Lee,~J. \latin{et~al.}  CHARMM-GUI Input Generator for NAMD, GROMACS, AMBER, OpenMM, and CHARMM/OpenMM Simulations Using the CHARMM36 Additive Force Field. \emph{Biophysical Journal} \textbf{2015}, \emph{12}, 405–413\relax
\mciteBstWouldAddEndPuncttrue
\mciteSetBstMidEndSepPunct{\mcitedefaultmidpunct}
{\mcitedefaultendpunct}{\mcitedefaultseppunct}\relax
\EndOfBibitem
\bibitem[Wu \latin{et~al.}(2014)Wu, Cheng, Jo, Rui, Song, Dávila-Contreras, Qi, Lee, Monje-Galvan, Venable, Klauda, and Im]{Wu2014}
Wu,~E.~L.; Cheng,~X.; Jo,~S.; Rui,~H.; Song,~K.~C.; Dávila-Contreras,~E.~M.; Qi,~Y.; Lee,~J.; Monje-Galvan,~V.; Venable,~R.~M.; Klauda,~J.~B.; Im,~W. CHARMM-GUI {\it Membrane Builder} toward realistic biological membrane simulations. \emph{J. Comput. Chem.} \textbf{2014}, \emph{35}, 1997–2004\relax
\mciteBstWouldAddEndPuncttrue
\mciteSetBstMidEndSepPunct{\mcitedefaultmidpunct}
{\mcitedefaultendpunct}{\mcitedefaultseppunct}\relax
\EndOfBibitem
\bibitem[Jo \latin{et~al.}(2007)Jo, Kim, and Im]{Jo2007}
Jo,~S.; Kim,~T.; Im,~W. Automated builder and database of protein/membrane complexes for molecular dynamics simulations. \emph{PLoS ONE} \textbf{2007}, \emph{2}, e880\relax
\mciteBstWouldAddEndPuncttrue
\mciteSetBstMidEndSepPunct{\mcitedefaultmidpunct}
{\mcitedefaultendpunct}{\mcitedefaultseppunct}\relax
\EndOfBibitem
\bibitem[Jorgensen \latin{et~al.}(1983)Jorgensen, Chandrasekhar, Madura, Impey, and Klein]{Jorgensen1983}
Jorgensen,~W.~L.; Chandrasekhar,~J.; Madura,~J.~D.; Impey,~R.~W.; Klein,~M.~L. Comparison of simple potential functions for simulating liquid water. \emph{J. Chem. Phys.} \textbf{1983}, \emph{79}, 926–935\relax
\mciteBstWouldAddEndPuncttrue
\mciteSetBstMidEndSepPunct{\mcitedefaultmidpunct}
{\mcitedefaultendpunct}{\mcitedefaultseppunct}\relax
\EndOfBibitem
\bibitem[Izadi \latin{et~al.}(2014)Izadi, Anandakrishnan, and Onufriev]{izadi2014building}
Izadi,~S.; Anandakrishnan,~R.; Onufriev,~A.~V. Building water models: a different approach. \emph{J. Phys. Chem. Lett.} \textbf{2014}, \emph{5}, 3863--3871\relax
\mciteBstWouldAddEndPuncttrue
\mciteSetBstMidEndSepPunct{\mcitedefaultmidpunct}
{\mcitedefaultendpunct}{\mcitedefaultseppunct}\relax
\EndOfBibitem
\bibitem[P\'all and Hess(2013)P\'all, and Hess]{Pll2013}
P\'all,~S.; Hess,~B. A flexible algorithm for calculating pair interactions on {S}{I}{M}{D} architectures. \emph{Comput. Phys. Comm.} \textbf{2013}, \emph{184}, 2641–2650\relax
\mciteBstWouldAddEndPuncttrue
\mciteSetBstMidEndSepPunct{\mcitedefaultmidpunct}
{\mcitedefaultendpunct}{\mcitedefaultseppunct}\relax
\EndOfBibitem
\bibitem[Essmann \latin{et~al.}(1995)Essmann, Perera, Berkowitz, Darden, Lee, and Pedersen]{essmann1995smooth}
Essmann,~U.; Perera,~L.; Berkowitz,~M.~L.; Darden,~T.; Lee,~H.; Pedersen,~L.~G. A smooth particle mesh Ewald method. \emph{J. Chem. Phys.} \textbf{1995}, \emph{103}, 8577--8593\relax
\mciteBstWouldAddEndPuncttrue
\mciteSetBstMidEndSepPunct{\mcitedefaultmidpunct}
{\mcitedefaultendpunct}{\mcitedefaultseppunct}\relax
\EndOfBibitem
\bibitem[Haug \latin{et~al.}(1976)Haug, Arora, and Matsui]{Haug1976}
Haug,~E.~J.; Arora,~J.~S.; Matsui,~K. A steepest-descent method for optimization of mechanical systems. \emph{Journal of Optimization Theory and Applications} \textbf{1976}, \emph{19}, 401–424\relax
\mciteBstWouldAddEndPuncttrue
\mciteSetBstMidEndSepPunct{\mcitedefaultmidpunct}
{\mcitedefaultendpunct}{\mcitedefaultseppunct}\relax
\EndOfBibitem
\bibitem[Bussi \latin{et~al.}(2009)Bussi, Zykova-Timan, and Parrinello]{papervrescale}
Bussi,~G.; Zykova-Timan,~T.; Parrinello,~M. Isothermal-isobaric {M}olecular {D}ynamics using stochastic velocity rescaling. \emph{J. Chem. Phys.} \textbf{2009}, \emph{130}, 074101\relax
\mciteBstWouldAddEndPuncttrue
\mciteSetBstMidEndSepPunct{\mcitedefaultmidpunct}
{\mcitedefaultendpunct}{\mcitedefaultseppunct}\relax
\EndOfBibitem
\bibitem[Berendsen \latin{et~al.}(1984)Berendsen, Postma, van Gunsteren, DiNola, and Haak]{Berendsen1984}
Berendsen,~H. J.~C.; Postma,~J. P.~M.; van Gunsteren,~W.~F.; DiNola,~A.; Haak,~J.~R. Molecular dynamics with coupling to an external bath. \emph{J. Chem. Phys.} \textbf{1984}, \emph{81}, 3684–3690\relax
\mciteBstWouldAddEndPuncttrue
\mciteSetBstMidEndSepPunct{\mcitedefaultmidpunct}
{\mcitedefaultendpunct}{\mcitedefaultseppunct}\relax
\EndOfBibitem
\bibitem[Parrinello and Rahman(1981)Parrinello, and Rahman]{Parrinello1981}
Parrinello,~M.; Rahman,~A. Polymorphic transitions in single crystals: A new molecular dynamics method. \emph{J. Appl. Phys.} \textbf{1981}, \emph{52}, 7182–7190\relax
\mciteBstWouldAddEndPuncttrue
\mciteSetBstMidEndSepPunct{\mcitedefaultmidpunct}
{\mcitedefaultendpunct}{\mcitedefaultseppunct}\relax
\EndOfBibitem
\bibitem[Galano-Frutos and Sancho(2019)Galano-Frutos, and Sancho]{GalanoFrutos2019}
Galano-Frutos,~J.~J.; Sancho,~J. Accurate Calculation of Barnase and SNase Folding Energetics Using Short Molecular Dynamics Simulations and an Atomistic Model of the Unfolded Ensemble: Evaluation of Force Fields and Water Models. \emph{Journal of Chemical Information and Modeling} \textbf{2019}, \emph{59}, 4350–4360\relax
\mciteBstWouldAddEndPuncttrue
\mciteSetBstMidEndSepPunct{\mcitedefaultmidpunct}
{\mcitedefaultendpunct}{\mcitedefaultseppunct}\relax
\EndOfBibitem
\bibitem[Galano-Frutos \latin{et~al.}(2023)Galano-Frutos, Nerín-Fonz, and Sancho]{GalanoFrutos2023}
Galano-Frutos,~J.~J.; Nerín-Fonz,~F.; Sancho,~J. Calculation of Protein Folding Thermodynamics Using Molecular Dynamics Simulations. \emph{Journal of Chemical Information and Modeling} \textbf{2023}, \emph{63}, 7791–7806\relax
\mciteBstWouldAddEndPuncttrue
\mciteSetBstMidEndSepPunct{\mcitedefaultmidpunct}
{\mcitedefaultendpunct}{\mcitedefaultseppunct}\relax
\EndOfBibitem
\bibitem[Estrada \latin{et~al.}(2009)Estrada, Bernadó, Blackledge, and Sancho]{protsa}
Estrada,~J.; Bernadó,~P.; Blackledge,~M.; Sancho,~J. ProtSA: a web application for calculating sequence specific protein solvent accessibilities in the unfolded ensemble. \emph{BMC Bioinformatics} \textbf{2009}, \emph{10}, 104\relax
\mciteBstWouldAddEndPuncttrue
\mciteSetBstMidEndSepPunct{\mcitedefaultmidpunct}
{\mcitedefaultendpunct}{\mcitedefaultseppunct}\relax
\EndOfBibitem
\bibitem[Becktel and Schellman(1987)Becktel, and Schellman]{gibbshelmholtz}
Becktel,~W.~J.; Schellman,~J.~A. Protein stability curves. \emph{Biopolymers} \textbf{1987}, \emph{26}, 1859--1877\relax
\mciteBstWouldAddEndPuncttrue
\mciteSetBstMidEndSepPunct{\mcitedefaultmidpunct}
{\mcitedefaultendpunct}{\mcitedefaultseppunct}\relax
\EndOfBibitem
\bibitem[Van~Gunsteren and Karplus(1982)Van~Gunsteren, and Karplus]{vGu1982MM}
Van~Gunsteren,~W.~F.; Karplus,~M. Effects of constraints on the dynamics of macromolecules. \emph{Macromolecules} \textbf{1982}, \emph{15}, 1528--1544\relax
\mciteBstWouldAddEndPuncttrue
\mciteSetBstMidEndSepPunct{\mcitedefaultmidpunct}
{\mcitedefaultendpunct}{\mcitedefaultseppunct}\relax
\EndOfBibitem
\end{mcitethebibliography}

\end{document}

\newpage 

\begin{abstract}
This document presents relevant information about the ILVES algorithms, which do not appear in the main article due to space limitations. It is divided into three parts: theory, methods, and results. In the former (Theory), we present an overview of the constraint solvers for molecular dynamics, the mathematical background of constrained molecular dynamics, a complete derivation of the equations behind ILVES, as well as the concepts behind its implementation. Additionally, we provide a more in-depth discussion on why constraints should be solved as accurately as possible. In the second part (Methods), we present detailed information about the infrastructure on which our simulations were performed and the input data. Finally, in the third part (Results), we present calculations of observable quantities that prove that ILVES does not distort the dynamics more than SHAKE or LINCS, as well as extended outcomes about ILVES' performance.
\end{abstract}

\section{Table of Contents}

\noindent{\bf Part A: Theory}

\noindent{1. Overview of Constraint Solvers for Molecular Dynamics\hfill\pageref{sec:overview}}

\noindent{2. Mathematical Foundations of Constrained Molecular Dynamics\hfill\pageref{sec:math-foundation}}

\noindent{3. Extended Description of ILVES-M and ILVES-F\hfill\pageref{sec:ilves-impl}}

\noindent{4. Further Discussion on the Need to Solve the Constraint Equations Accurately\hfill\pageref{sec:need-to-solve-acc}}

\noindent{\\\bf Part B: Methods}

\noindent{5. Experimental Setup\hfill\pageref{sec:experimental-setup}}

\noindent{6. Simulations\hfill\pageref{sec:simulations}}

\noindent{\\\bf Part C: Results}

\noindent{7. Reliability Study\hfill\pageref{sec:reliability}}

\noindent{8. Extended Performance Results\hfill\pageref{sec:ext-performance}}

\newpage

\section{1. Overview of Constraint Solvers for Molecular Dynamics}\label{sec:overview}

The most popular algorithms for imposing holonomic constraints in the context of molecular dynamics are decades old: SHAKE was presented in 1977, while LINCS was presented in 1997; its parallel version (P-LINCS) appeared in 2008. 
Apart from SHAKE, LINCS and P-LINCS, several algorithms to impose constraints on molecules have been proposed~\cite{And1983JCOP,settle,lambrakosshake,wiggle,Gon2006JCP,ciccotti1986molecular,ryckaert2001molecular,Eas2010JCTC1,Kra2000JCC,flexshake}. 
Among them, we highlight the profusely cited RATTLE~\cite{And1983JCOP} and SETTLE~\cite{settle}. The former enforces constraints for both positions and velocities, while the latter imposes constraints on water molecules, treating them as rigid bodies. 
Other ones are typically slower than SHAKE~\cite{wiggle} or have quadratic ($\mathcal{O}(n^2)$) or cubic ($\mathcal{O}(n^3)$) complexity~\cite{Eas2010JCTC1,Kra2000JCC,ciccotti1986molecular,Gon2006JCP}. 

Some proposed constraint algorithms~\cite{Kra2000JCC,Bai2008JCoP,Bai2009JCC} rely on direct ({\it analytical}) solvers~\cite{Maz2007JPAMT} for the involved linear systems of equations:
M-SHAKE~\cite{Kra2000JCC} performs Gaussian elimination without applying sparse techniques, resulting in a cubic ($\mathcal{O}(n^3)$) numerical complexity in the number of constraints;
MILC-SHAKE and MILCH-SHAKE ~\cite{Bai2008JCoP,Bai2009JCC} are only applicable to linear chains and n-alkanes, respectively.
Application of sparse libraries for solving constraints, 
as well as parallel versions of SHAKE~\cite{shakeparallel1, shakeparallel2} have been proposed in the past~\cite{Bar1995JCoP}, yet not widely applied.
Further algorithms have been designed to impose constraints on bond angles~\cite{Gon2009JCP,pechlaner2021method,Eas2010JCTC1} and dihedral angles~\cite{pechlaner2020algorithms,dihedral_solver_2}.
%The simplest manner to impose constraints on bond angles is through distance constraints; this is, if atoms A and B are bonded, atoms B and C are bonded, and bond length constraints are imposed, then one defines a constraint on the angle between atoms A,B,C by merely setting a distance constraint between atoms A and C.
Refs.~\cite{pechlaner2021method,Gon2009JCP} use more sophisticated expressions than distance constraints for angular constraints but solve the equations one after the other (i.e. {\it \`a la SHAKE}, solving the first equation first, then the second equation, etc.). 
This approach converges slowly and is challenging to parallelize. The execution time of
$\theta$-SHAKE~\cite{Gon2009JCP} %which has linear ($\mathcal{O}(n)$) scaling
has the same order of magnitude as SHAKE's, which is significant for constraints on bond angles. Moreover, its efficiency relies on the low connectivity of the bonds of the tackled molecules.
Though there are exceptions~\cite{usageBAconstraints4,usageBAconstraints1,usageBAconstraints2,usageBAconstraints3,martinicoarsegrain,hammonds}, bond angles are not frequently constrained in MD simulations. This is likely due to the slowness of SHAKE in performing the corresponding calculations~\cite{note2024a}. 

\section{2. Mathematical Foundations of Constrained Molecular Dynamics}\label{sec:math-foundation} %aaa

In this section, we present the mathematical fundamentals of constrained MD.
We explicitly state the SHAKE, (P-)LINCS and ILVES algorithms. 

\subsection{2.1 Notation}

We begin by defining our notation.
Vectors are written using bold lowercase letters, e.g., $\bm{x}$, while matrices are written using bold uppercase letters, e.g., $\bm{A}$. 
If $\bm{A} = [a_{ij}]$ is any matrix, then the transpose matrix $\bm{A}^T$ is the matrix $\bm{B} = [b_{ij}]$ given by $b_{ij} = a_{ji}$.
By default, all vectors are column vectors and when a row vector is needed a column vector is explicitly transposed.
The Euclidean inner product $s = \bm{x} \cdot \bm{y}$ between two vectors $\bm{x} = (x_1, x_2, \dots, x_n)^T \in \mathbb{R}^n$ and $\bm{y} = (y_1, y_2, \dots, y_n)^T \in \mathbb{R}^n$ is given by $s = \bm{x}^T \bm{y} = \sum_{j=1}^n x_i y_i$.
The corresponding Euclidean norm $\|\bm{x}\|_2$ is the non-negative real number given by $\|\bm{x}\|_2 = \sqrt{\bm{x}^T \bm{x}} = \sqrt{\sum_{j=1}^n x_j^2}$.
If the function $\bm{f} = (f_1, f_2, \dotsc, f_l)^T : \mathbb{R}^n \rightarrow \mathbb{R}^l$ is differentiable, then the Jacobian $\bm{F}(\bm{x})$ of $\bm{f}$ at the point $\bm{x} \in \mathbb{R}^n$ is the matrix $\bm{A} = [a_{ij}] \in \mathbb{R}^{l \times n}$ of first order partial derivatives, i.e.,
\begin{equation}
 a_{ij} := \frac{\partial f_i}{\partial x_j}(\bm{x}) \, .
\end{equation}
In a system of $s$ atoms, let $m_i > 0$ denote the mass of the $i$th atom and let $\bm{m}_i \in \mathbb{R}^3$ and the diagonal mass matrix $\bm{M} \in \mathbb{R}^{3s \times 3s}$ be given by
\begin{equation}
  \bm{m}_i := (m_i,m_i,m_i)^T, \quad \bm{M} := \text{diag} ( \bm{m}_1^T, \bm{m}_2^T, \dotsc, \bm{m}_n^T ) \, .
\end{equation}
In addition, let $\bm{q}_i, \bm{v}_i \in \mathbb{R}^3$ denote the position and velocity of the $i$th atom, and let
$\bm{f}_i \in \mathbb{R}^3$ 
represent the force acting on it (due to any atom or field).
The vectors $\bm{q}, \bm{v}, \bm{f} \in \mathbb{R}^{3s}$, which gather all positions, velocities, and forces are defined as follows 

\begin{equation}
      \bm{q} := (\bm{q}_1^T, \bm{q}_2^T, \ldots, \bm{q}_s^T)^T \, ; \quad 
      \bm{v} := (\bm{v}_1^T, \bm{v}_2^T, \ldots, \bm{v}_s^T)^T \, ; \quad 
      \bm{f} := (\bm{f}_1^T, \bm{f}_2^T, \ldots, \bm{f}_s^T)^T \, .
\end{equation}
Throughout this work, $s$ is the number of atoms (hence, the total number of Cartesian coordinates for positions is $3s$) and $n$ is the number of imposed constraints.

\subsection{2.2 Equations of motion}

The phenomena of interest for MD span a wide range of time scales. 
The subset that can be simulated is limited by the time step ($\Delta t$), which
is defined as the gap between consecutive simulated times; 
for force fields-based MD $\Delta t$ this is typically in the range of femtoseconds (fs). Therefore, setting it to values that are as high as possible (without distorting the dynamics of the simulated system) is a major task. To accomplish this, a widely used method involves imposing constraints on the molecular internal degrees of freedom. Numerous algorithms to perform the corresponding calculations have been proposed, with SHAKE~\cite{shake_paper}, LINCS~\cite{lincs_paper}, and P-LINCS~\cite{plincs_paper} being the most frequent ones to be applied to molecules other than water. 

Molecular dynamics is based on Newton's second law, which can be written as follows for a system consisting of $s$ atoms: 
\begin{subequations}\label{eqsNewton1}
\begin{align}
  \dot{\bm{q}}(t) &= \bm{v}(t)  \, , \\
  \bm{M} \dot{\bm{v}}(t) &= \bm{f}(t) \, ,
\end{align}
\end{subequations}
where $\bm{q}$, $\bm{v}$, $\bm{f}$ $\in \mathbb{R}^{3s}$ are the vectors of the atomic positions, atomic velocities and total forces acting on the atoms; $\bm{M}$ is the matrix of (constant) atomic masses and 
the dot on top indicates time ($t$) derivative. 

If we impose a set of $n$ bond constraints on the system (\ref{eqsNewton1}) then the equations of motion change to a system of differential-algebraic equations:
\begin{subequations} \label{equ:multi-body:dae}
  \begin{align}
   \dot{\bm{q}}(t) &=  \bm{v}(t) \, , \\
   \bm{M}  \dot{\bm{v}}(t) &=  \bm{f}_{\text{ext}}(t) - \bm{G}( \bm{q}(t))^T \bm{\lambda}(t) \, , \\
   \bm{g}( \bm{q}(t)) &=  \bm{0} \, .
  \end{align}
\end{subequations}
Here $\bm{f}_{\text{ext}}$ is the total external force acting on the atoms (which can be calculated using a force field); $\bm{g} : \mathbb{R}^{3s} \rightarrow \mathbb{R}^n$ is the constraint function
\begin{equation}
\bm{q} \rightarrow (g_1(\bm{q}), g_2(\bm{q}), \ldots, g_n(\bm{q}))
\end{equation}
and $\bm{G}(\bm{q}) \in \mathbb{R}^{n \times 3s}$ is the Jacobian of $\bm{g}$ with respect to the vector $\bm{q}$ of atomic coordinates.
If the $i$th constraint is a bond length constraint between atoms $a_i$ and $b_i$, then we may choose
\begin{equation}\label{ecconstr}
  g_i(\bm{q}) = \frac{1}{2} \left (  d^2_i - \| \bm{q}_{a_i} - \bm{q}_{b_i} \|^2 \right).
\end{equation}
where $d_i$ (commonly a constant) is the desired distance between the atoms $a_i$ and $b_i$. This particular choice ensures that the constraint function is infinitely differentiable at every point $\bm{q}$. 

The vast majority of computer simulations rely on discrete time steps. Among the standard integrators used to update positions (whether strictly following Newton's second law or a modification of it) are the Verlet\cite{Ver1967PR} algorithms (e.g., Leapfrog and Velocity Verlet), as well as the Stochastic Dynamics (SD) and Langevin Dynamics algorithms~\cite{hairer2006geometric}. 

Below (Secs. 2.3 and 2.4), we present a summary of the state-of-the-art algorithms for constrained MD (of molecules other than water), i.e., SHAKE, ILVES, and (P-)LINCS.

\subsection{2.3 The SHAKE/ILVES algorithms}

The SHAKE and ILVES algorithms use Verlet integration to solve the same system of differential-algebraic equations. 
They solve the constraint equations using different techniques. 
However, if the constraint equations are solved exactly and if all calculations are done without rounding errors, then they will return the same results.

The two algorithms use a pair of staggered uniform grids with fixed time step $h$ (being $h:= \Delta t$). The central equations can be stated as follows
\begin{subequations} \label{equ:nonlinear:1}
  \begin{align}
  \bm{v}_{k+1/2} &=  \bm{v}_{k-1/2} + h  \bm{M}^{-1} \big(  \bm{f}_{ext}( t_k) -  \bm{G} ( \bm{q}_k)^T \bm{\lambda}_k \big) \, , \label{equ:nonlinear:1a}\\
   \bm{q}_{k+1} &=  \bm{q}_k + h  \bm{v}_{k + 1/2} \, , \label{equ:nonlinear:1b}\\
%     \bm{v}_{k} &=  \left( \bm{v}_{k-1/2} + \bm{v}_{k+1/2} \right)/2  \, , \label{equ:nonlinear:1b2}\\
% CCKM: I eliminated the line defining v_k as it is not necessary. It is true that the velocity can be approximated in this manner and that the error is O(h^2), but the point is not important here and including the equation requires an explanation.
   \bm{g}( \bm{q}_{k+1}) &= \bm{0} \, ,   \label{equ:nonlinear:1c}
\end{align}
\end{subequations}
where $k \in \mathbb{N}$ is the time index, i.e., 
$$ t_k = kh, \quad t_{k+\frac{1}{2}} = \left(t + \frac{1}{2}\right) h $$
and 
\begin{equation}
\bm{q}_k \approx \bm{q}(t_k), \quad \bm{v}_{k+\frac{1}{2}} \approx \bm{v}\left(t_{k+\frac{1}{2}}\right).
\end{equation}
Equation \eqref{equ:nonlinear:1c} is usually a non-linear equation for the unknown vector of Lagrange multipliers $\bm{\lambda}_k$, namely $\bm{g}(\bm{\phi}_k(\bm{\lambda})) = \bm{0}$ where $\bm{\phi}_k$ is a function obtained by combining (\ref{equ:nonlinear:1a}), (\ref{equ:nonlinear:1b}):
\begin{equation}\label{eq6}
  \bm{\phi}_k(\bm{\lambda}) := \bm{q}_k + h \left(\bm{v}_{k-\frac{1}{2}} + h \bm{M}^{-1} \big( \bm{f}_{ext}(t_k) - \bm{G}(\bm{q}_k )^T \bm{\lambda} \big) \right) \, .
\end{equation}

The SHAKE algorithm solves the nonlinear constraint equations using the nonlinear Gauss-Seidel iteration. 
This method converges slowly and locally, subject to very modest conditions; see \cite{more1972, porshing1971} and the references therein. 
In contrast, the ILVES algorithms use either Newton's method or a quasi-Newton method to solve the constraint equations rapidly. 
We now begin our exposition of the ILVES algorithms by stating Newton's method in the general case.
Let $\bm{f} : \mathbb{R}^{n} \rightarrow \mathbb{R}^n$ be a differentiable function and consider the problem of solving the following system of equations:
\begin{equation}
    \bm{f}(\bm{x}) = \bm{0},
\end{equation}
with respect to $\bm{x} \in \mathbb{R}^{n}$. If the Jacobian $\bm{F}(\bm{x}) $ of $\bm{f}$ at any point $\bm{x}$ is nonsingular, then Newton's method is given by
\begin{subequations}\label{defeqnewnton}
\begin{align}
  \bm{F}(\bm{x}_l) \bm{z}_l &= \bm{f}(\bm{x}_l)  \, , \\
  \bm{x}_{l+1} &= \bm{x}_{l} - \bm{z}_l  \, .
\end{align}
\end{subequations}
where the initial approximation $\bm{x}_0$ must be chosen by the user. In general, Newton's method will converge locally and quadratically to a zero $\bm{z}$ of $\bm{f}$ if the Jacobian $\bm{F}(\bm{z})$ at $\bm{z}$ is nonsingular, i.e., there exists a constant $C$ such that
\begin{equation}
    \frac{\| \bm{z} - \bm{x}_{l+1}\|_2}{\|\bm{z} - \bm{x}_l\|^2_2} \rightarrow C, \quad l \rightarrow \infty, \quad l \in \mathbb{N}.
\end{equation}
for all choices of $\bm{x}_0$, provided that $\|\bm{z} - \bm{x}_0\|_2$ is sufficiently small.
We shall now state Newton's method for the non-linear constraint equation \eqref{equ:nonlinear:1c}. 
We first use the chain rule of differentiation to compute the Jacobian of the function 
$
\bm{\lambda}\rightarrow \bm{g}(\bm{\phi}_k(\bm{\lambda}))
$
and find that
\begin{equation}
    \frac{\partial}{\partial \bm{\lambda}} (\bm{g} \circ \bm{\phi}_k)(\bm{\lambda}) = \bm{G}(\bm{\phi}_k(\bm{\lambda})) \frac{\partial \bm{\phi}_k}{\partial \bm{\lambda}}(\bm{\lambda}) = - h^2 \bm{G}(\bm{\phi}_k(\bm{\lambda}))\bm{M}^{-1} \bm{G}(\bm{q}_k)^T
\end{equation}
which is why we define the matrix function $\bm{A} : \mathbb{R}^{3s} \times \mathbb{R}^{3s} \rightarrow \mathbb{R}^{n \times n}$ given by
\begin{equation}\label{defA}
    \bm{A}(\bm{x},\bm{y}) := - h^2 \bm{G}(\bm{x}) \bm{M}^{-1} \bm{G}(\bm{y})^T \, .
\end{equation}
In general, each matrix $\bm{A}(\bm{x},\bm{y}) = [a_{ij}]$ is both sparse and \textit{structurally} symmetric because $a_{ij} = a_{ji} = 0$ if the $i$th and $j$th bond do not share an atom~\cite{ilves_2023}. However, we cannot expect that $\bm{A}$ is symmetric in the sense that $\bm{A} = \bm{A}^T$, unless, of course, $\bm{x} = \bm{y}$.

Newton's method for solving equation \eqref{equ:nonlinear:1c} with respect to the Lagrange multipliers $\bm{\lambda}_k$ can be written as
\begin{subequations} \label{equ:newton-constraint-equation}
\begin{align} 
  \bm{A} \big(\bm{\phi}_k(\bm{\lambda}_{k,l}), \bm{q}_k \big) \bm{z}_{k,l} &= \bm{g}  \big(\bm{\phi}_k(\bm{\lambda}_{k,l}) \big), \label{equ:newton-constraint-equation1}\\
  \bm{\lambda}_{k,l+1} &= \bm{\lambda}_{k,l} -  \bm{z}_{k,l}, \label{equ:newton-constraint-equation2}
\end{align}
\end{subequations}
It is common to initialize the search for $\bm{\lambda}_k$ using $\bm{\lambda}_{k,0} = \bm{0}$ and we have used this value to good effect.
We emphasize that equation \eqref{equ:newton-constraint-equation1} is a \emph{linear} equation with respect to the unknown vector $\bm{z}_{k,l}$ and the solution of several such systems is usually required to advance the simulation from time $t_k$ to time $t_{k+1}$.

The ILVES algorithms solve these linear systems by approximating the local action of the inverse of $\bm{A}$ using a different approximation for every processing unit. 
The technical details are provided in Section 3.

\subsection{2.4 The (P-)LINCS algorithms}

The P-LINCS algorithm is a parallel version of LINCS. In the absence of any rounding errors, they produce identical output.
The (P-)LINCS algorithms are also based on Verlet integration and use the same pair of staggered grids with a uniform time step $h$ as SHAKE and ILVES. 
The (P-)LINCS algorithms apply to bond constraints functions $\bm{g} : \mathbb{R}^{3N} \rightarrow \mathbb{R}^n$ where the individual components $g_i$ of $\bm{g}$ must have the form
\begin{equation}
    g_i(\bm{q}) = \|\bm{q}_{a_i} - \bm{q}_{b_i} \|_2 - d_i, \quad i = 1,2,\dotsc,n.
\end{equation}
Here $d_i$ is the length of the bond between atoms $a_i$ and $b_i$ whose coordinates are $\bm{q}_{a_i}$ and $\bm{q}_{b_i}$ respectively. 
The Jacobian $\bm{G}$ of this specific choice of constraints satisfy the elementary identity
\begin{equation}\label{eqforall}
    \forall \, \bm{q}, \bm{s} \in \mathbb{R}^{3s} \: : \: (G(\bm{q}) \bm{s})_i = \left(\frac{\bm{q}_{a_i} - \bm{q}_{b_i}}{\| \bm{q}_{a_i} - \bm{q}_{b_i}\|} \right)^T (\bm{s}_{a_i} - \bm{s}_{b_i}), \quad i=1,2,\dotsc, n,
\end{equation}
and this result is used repeatedly to derive the algorithm \cite{lincs_paper}. 
We shall now summarize the LINCS algorithm. 
Given the current positions $\bm{q}_k$ and velocities $\bm{v}_{k-\frac{1}{2}}$ we compute the next vectors $\bm{q}_{k+1}$ and $\bm{v}_{k+\frac{1}{2}}$ in the following manner. An initial approximation $\bm{q}_{k+1}^*$ is computed using
\begin{equation}
    \bm{q}_{k+1}^* = \bm{q}_k + h \bm{v}_{k-\frac{1}{2}} + h^2 \bm{M}^{-1} \bm{f}_k.
\end{equation}
where $\bm{f}_k = \bm{f}_{\text{ext}}(\bm{q}_k)$ is the total external force acting on the atoms at the current positions $\bm{q}_k$.
This definition corresponds to a single step of the Verlet method for the differential equation
\begin{align}
\dot{\bm{q}}(t) &= \bm{v}(t), \\
\bm{M} \dot{\bm{v}}(t) &= \bm{f}_{\text{ext}}(t).
\end{align}
In general, one cannot expect that $\bm{q}_{k+1}^*$ solves the constraint equation $\bm{g}(\bm{q}) = \bm{0}$, which is why LINCS performs at least one additional step. Specifically, the algorithm computes 
\begin{equation}
\bm{q}_{k+1}^0 = \bm{q}_{k+1}^* - \bm{T}_k   \left(\bm{G}_k \bm{q}_{k+1}^* - \bm{d} \right)  
\end{equation}
where the central matrix $\bm{T}_k$ and the vector $\bm{d}$ are given by
\begin{align}
    \bm{T}_k &= \bm{M}^{-1} \bm{G}_k^T (\bm{G}_k \bm{M}^{-1} \bm{G}_k^T)^{-1}, \\
    \bm{d} &= (d_1, d_2, \dotsc, d_n)^T,
\end{align}
and then (optionally) continues with the iteration
\begin{equation}
    \bm{q}_{k+1}^{l+1} = \bm{q}_{k+1}^l - \bm{T}_k \left(\bm{G}_k \bm{q}_{k+1}^l - \bm{p}( \bm{q}_{k+1}^l) \right), \quad l = 0, 1,2, \dotsc,
\end{equation}
until the norm of the residual $\bm{g}(\bm{q}_{k+1}^l)$ is sufficiently small. 
Here the central function $\bm{p} : \mathbb{R}^{3s} \rightarrow \mathbb{R}^n$ is given by
\begin{equation}
  p_i(\bm{q}) = \sqrt{2d_i^2 - \|\bm{q}_{a_i}- \bm{q}_{b_i}\|_2^2}
\end{equation}
and the next state vectors are computed as follows 
\begin{align}
   \bm{q}_{k+1} &= \bm{q}_{k+1}^{l_{\max}} \\
   \bm{v}_{k+\frac{1}{2}} &= \frac{\bm{q}_{k+1} - \bm{q}_k}{h}.
\end{align}
where $l_{\max}$ is the final value of the index $l$.
We are still investigating the deep mathematical reasons behind the fact that the LINCS sequence $\{\bm{q}_{k+1}^l \})_{l=0}^\infty$ converges to a solution of the constraint equation $\bm{g}(\bm{q}) = \bm{0}$, and how this property was formally established in the papers defining LINCS and P-LINCS \cite{lincs_paper,plincs_paper}. 
However, in our numerical experiments, we have observed that the 2-norm of the residual $\bm{g}(\bm{q}_{k+1}^l)$ decays linearly until we reach the limits of floating point arithmetic and additional iterations are pointless.
The LINCS algorithm hinges on our ability to apply the operator $\bm{T}_n$. Specifically, it is necessary to solve linear systems of the form
\begin{equation}
    \left( \bm{G}_k \bm{M}^{-1} \bm{G}_k^T \right) \bm{x} = \bm{y}
\end{equation}
where the right-hand side vector $\bm{y}$ and the unknown vector $\bm{x}$ are both elements of $\mathbb{R}^s$. Since the inception of the algorithm, the solution $\bm{x}$ has been approximated using a truncated Neumann series. 
Specifically, the coefficient matrix $\bm{G}_n \bm{M}^{-1} \bm{G}_n^T$ is subjected to a diagonal scaling, i.e.,
\begin{equation}
    \bm{G}_k \bm{M}^{-1} \bm{G}_k^T = \bm{S}^{-1} \left(\bm{S} \bm{G}_k \bm{M}^{-1} \bm{G}_k^T \bm{S} \right) \bm{S}^{-1}
\end{equation}
where $\bm{S} \in \mathbb{R}^{s \times s}$ is a diagonal matrix such the diagonal entries of the matrix $\bm{S} \bm{G}_n \bm{M} \bm{G}_k^T \bm{S}$ are all equal to one. This matrix is then viewed as a perturbation of the identity matrix $I_s$, i.e.,
\begin{equation}
   (\bm{S} \bm{G}_k \bm{M}^{-1} \bm{G}_k^T \bm{S}) = \bm{I}_s - \bm{A}_k
\end{equation}
and the action of the inverse of $\bm{S} \bm{G}_k \bm{M}^{-1} \bm{G}_k^T \bm{S}$ is approximated using a truncated Neumann series, i.e.,
\begin{multline} \label{equ:neumann}
   \bm{x} = (\bm{G}_k \bm{M}^{-1} \bm{G}_k^T)^{-1} \bm{y} = \bm{S}(\bm{S} \bm{G}_k \bm{M}^{-1} \bm{G}_k^T \bm{S})^{-1} \bm{S} \bm{y} \\
   = \bm{S} (\bm{I}_s - \bm{A}_n)^{-1} \bm{S} \bm{y} \approx \bm{S} \left( \sum_{j=0}^{I-1} \bm{A}_k^j\right) \bm{S} \bm{y}.
\end{multline} 
for a modest value of the positive integer $I$ with is typically chosen as $4$ or $8$. The parameter $I$ is known as \texttt{lincs-order} in \textsc{Gromacs}.
It is important to appreciate 
the fact that while there are known molecules for which this series does not converge ---see the original paper \cite{lincs_paper}---, there is ample experience to suggest that the approximation works well in practice. 

The P-LINCS algorithm is a straightforward parallelization of the LINCS algorithm and hinges on the fact that it is relatively easy to parallelize the matrix-vector multiplications needed to evaluate the right-hand side of equation \eqref{equ:neumann}.

\section{3. Extended Description of ILVES-M and ILVES-F}\label{sec:ilves-impl}

In this section, we provide a detailed description of the ILVES-M and ILVES-F algorithms as well as their parallel implementations which we have integrated into \textsc{Gromacs}. This software was used to analyze the performance and accuracy of the algorithms presented in this study. While the implementation is fine-tuned for \textsc{Gromacs}, it can be straightforwardly migrated to any MD packages, such as \textsc{Namd}~\cite{namd_paper,namd2}, {\textsc{Amber}}~\cite{Pea1995CoPC}, \textsc{Charmm}~\cite{Bro2009JCC}, \textsc{Lammps} or \textsc{Desmond}.

ILVES-M solves the same system of equations \eqref{equ:nonlinear:1} as SHAKE, but uses a quasi-Newton method to solve the constraint equations.
If ILVES-M and SHAKE were all executed using exact arithmetic and if the constraint equations were solved exactly, then these algorithms would return the same results.
Why is this? If we assume that the Jacobian of the constraint function is of full rank, then the Inverse Function Theorem implies that the constraint equations have a unique solution when the time step is sufficiently small. It follows inductively that the ILVES-M and SHAKE algorithms must produce identical output.

ILVES-F also solves the same system of equations \eqref{equ:nonlinear:1} as SHAKE, but it uses a quasi-Newton method to solve the constraint equations. The central matrix $\bm{A} \big(\bm{\phi}_k(\bm{\lambda}_{k,l}), \bm{q}^0_k \big)$ is a function of $l$, but it can be approximated by a constant matrix, i.e.,
\begin{equation}
\bm{A} \big(\bm{\phi}_k(\bm{\lambda}_{k,l}), \bm{q}^0_k \big) \approx \bm{A} \big(\bm{q}^0_k, \bm{q}^0_k \big)
\end{equation}
simply because $\bm{\phi}_k(\bm{\lambda}) = \bm{q}_k + \mathcal{O}(h)$.
It is therefore natural to replace Newton's method with the quasi-Newton method
\begin{align} 
  \bm{A} \big( \bm{q}^0_k , \bm{q}^0_k \big) \bm{z}_{k,l} &= \bm{g}  \big(\bm{\phi}_k(\bm{\lambda}_{k,l}) \big) \, , \label{equ:quasi-newton-constraint-equation1}\\
  \bm{\lambda}_{k,l+1} &= \bm{\lambda}_{k,l} -  \bm{z}_{k,l}  \,  
\end{align}
This approach is standard within the field of numerical analysis and different names such as the ``simplified Newton's method'' \cite{ortega_iterative_1970} or the ``chord method'' \cite{kelley_iterative_1995} are used. 
It is the foundation of the Constant Constraint Matrix approach used by Ref. \cite{Eas2010JCTC1} in the context of molecular dynamics. As described in this section, this approach enables ILVES-F to outperform ILVES-M.

\begin{figure}[htbp]
    \centering
    \begin{minipage}[t]{0.25\textwidth}
        \centering
        \includegraphics[width=\textwidth]{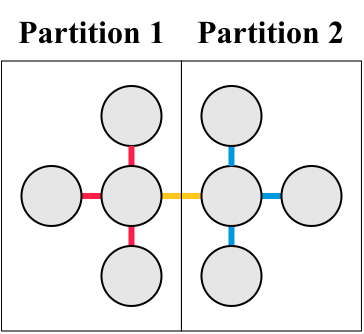}
        \renewcommand{\thefigure}{a}
        \\ (a) Partitioning of graph
        \label{fig:example_partitioning}
    \end{minipage}

    \begin{minipage}[t]{0.18\textwidth}
        \centering
        \includegraphics[width=\textwidth]{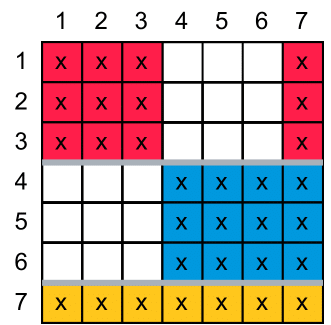}
        \renewcommand{\thefigure}{b}
        \\ (b) Initial $A$
    \end{minipage}
    \hfill
    \begin{minipage}[t]{0.18\textwidth}
        \centering
        \includegraphics[width=\textwidth]{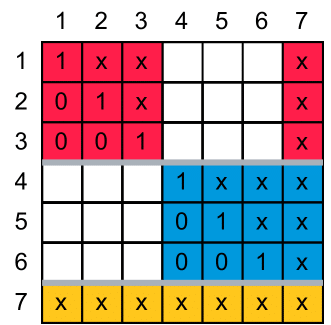}
        \renewcommand{\thefigure}{c}
        \\ (c) Step 1
    \end{minipage}
    \hfill
    \begin{minipage}[t]{0.18\textwidth}
        \centering
        \includegraphics[width=\textwidth]{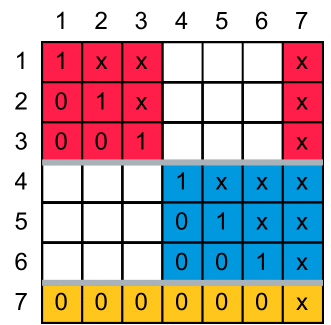}
        \renewcommand{\thefigure}{d}
        \\ (d) Step 2
    \end{minipage}
    \hfill
    \begin{minipage}[t]{0.18\textwidth}
        \centering
        \includegraphics[width=\textwidth]{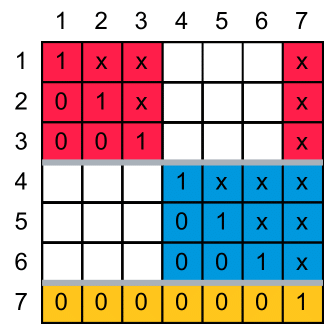}
        \renewcommand{\thefigure}{e}
        \\ (e) Step 3
    \end{minipage}
    \hfill
    \begin{minipage}[t]{0.18\textwidth}
        \centering
        \includegraphics[width=\textwidth]{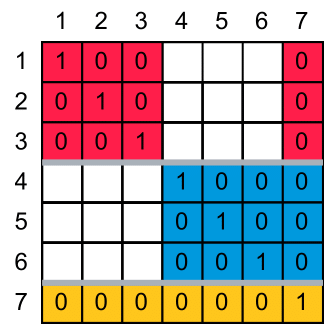}
        \renewcommand{\thefigure}{f}
        \\ (f) Step 4
    \end{minipage}
    \caption{A possible partitioning of a simple graph (a), 
    the corresponding coordinate matrix ($\bm{A}$) of the linear system of equations 
    %(Eq.~(\ref{equ:newton-constraint-equation1})) 
    ($\bm{A} \big(\bm{\phi}_k(\bm{\lambda}_{k,l}), \bm{q}^0_k \big) \bm{z}_{k,l} = \bm{g}  \big(\bm{\phi}_k(\bm{\lambda}_{k,l}) \big) $)
    (b), and the steps to solve the linear system in parallel through Gauss-Jordan elimination and the Schur complement method (c-f). An analogous approach applies to solving the system with LU/LDLT decompositions and backward/forward substitution. Colors denote partitions, and each '$\mathsf{x}$'  character represents a non-zero entry in the linear system. The red rows (1–3) and blue rows (4–6) represent private partitions for threads 1 and 2, respectively, while the yellow row (7) is the shared partition. To solve the system in parallel, each thread begins by zeroing the subdiagonal part of its private partition and setting the diagonal entries to ones (c). Next, the threads add their contributions to the shared partition in mutual exclusion (d). The master thread then performs Gauss-Jordan elimination on the shared partition (e), modifying only the last diagonal entry in this example. Once the master thread completes this step, all threads proceed in parallel to zero out the upper diagonal elements in their private partitions (f).}
    \label{fig:example_linear_system_and_partitioning}
\end{figure}

ILVES-M and ILVES-F leverage three levels of parallelism: distributed memory parallelism, shared memory parallelism, and vectorization.
To exploit distributed memory parallelism, our algorithms use a natural extension of the Overlapping Partitioning Method (OPM)~\cite{lou1989} from banded to general sparse systems. 
% This is a technique that is distinctly different from the one that is used by (P-)LINCS, but it is compatible with the domain decomposition used by (P-)LINCS.
In distributed-memory simulations with \textsc{Gromacs}, each domain also stores information about neighboring atoms, i.e., those within a distance of up to $I$ edges from any atom local to the domain, where $I$ is a user-defined parameter (\texttt{lincs-order} in the case of P-LINCS). 
Using both the local and neighboring atoms in each domain, a local approximation of the matrix-vector product $\mathbf{z}={\mathbf{A}}^{-1}\mathbf{g}$ is computed. This approximation is computed in each domain using a multi-thread implementation of Gaussian elimination based on the Schur method~\cite{Gol1989BOOK}. As detailed in the paragraphs that follow, this approach partitions the linear system, distributing segments across the domain's threads for parallel processing. An example of how a domain partitions a simple graph and the steps followed to solve its corresponding linear system is presented in Fig.~\ref{fig:example_linear_system_and_partitioning}.

It is important to note that, in distributed memory simulations, \textsc{Gromacs} frequently rebuilds the list of atoms assigned to each domain, a data structure known as the neighbor list. This list determines which atoms are considered local to each domain and which neighboring atoms must be included for calculations such as constraints. The frequency of these updates is controlled by the \texttt{nstlist} parameter, whose default value is 10 steps. Each time the neighbor list is recomputed, the constraint solver must be reinitialized to reflect possible displacements of atoms between domains. Since the neighbor list is frequently recomputed, the performance of the initialization has a significant impact on the overall algorithm performance. Hence, we designed and implemented an initialization of ILVES-M and ILVES-F, which minimizes the time spent in this stage.

The atoms and bonds handled by the constraint solver in each domain can be represented as an undirected graph $G = (V,E)$. The vertices $V$ represent the atoms connected with at least one constrained bond and the edges $E \subseteq V \times V$ represent the constrained bonds. Since the simulation can contain more than one molecule and it is common to constrain only a subset of the bonds; for example, bonds that connect a hydrogen bond (H-bonds), $G$ is usually a disconnected graph. From $G$ we can compute its line-graph $L(G)$ in which each vertex represents a constrained bond, and two constrained bonds are connected if they share one vertex in $G$. 
The adjacency matrix of the line-graph has the same sparsity pattern as the coordinate matrices of the linear systems solved by ILVES-M and ILVES-F.

The graph $L(G)$ can be partitioned such that all the edges of $L(G)$ that do not cross partitions comprise each thread's private partition, while edges that connect vertices of $L(G)$ which lie in different private partitions comprise the shared partition. The data in the private partitions can be processed simultaneously, as they are independent of one another. In contrast, the data in the shared partition must be processed separately, in mutual exclusion from the private partitions, to ensure correctness and avoid conflicts.

While both P-LINCS and ILVES-M/F employ a partitioning of $L(G)$ for parallel execution, their methodologies differ significantly. 
P-LINCS partitions $L(G)$ directly dividing its vertices into roughly equal sets, followed by computation to identify edges crossing partitions. 
On the other hand, ILVES-M/F tries to minimize the edges crossing partitions using a greedy algorithm to partition $G$ rather than $L(G)$. Graphs in MD exhibit very low connectivity and, additionally, {\textsc{Gromacs}} assigns close indexes to connected atoms. 
Our algorithm sequentially assigns vertices of $G$ to partition $p$ while the partition size remains below a target value plus an offset. 
The final vertex in partition $p$ is chosen based on the number of edges cut and the similarity of the partition size to the target size. 
This greedy algorithm runs in $\mathcal{O}(|E|)$ time, where $|E|$ is the number of edges in $G$, and produces partitions comparable to those generated by general but slower graph-partitioning libraries like \textsc{Metis}~\cite{Karypis1998}, often cutting a similar or smaller number of edges relative to the number of partitions. 
Our partitioning approach produces significantly fewer cuts than P-LINCS', which results in notably less serial work and less synchronization between threads in cases where zero-cut partitions are possible, such as simulations of several small molecules. 

\begin{figure}[htbp]
    \caption{Greedy algorithm for partitioning a graph into $k$ pieces}
    \label{alg:kway_partition}
    \includegraphics[width=\textwidth]{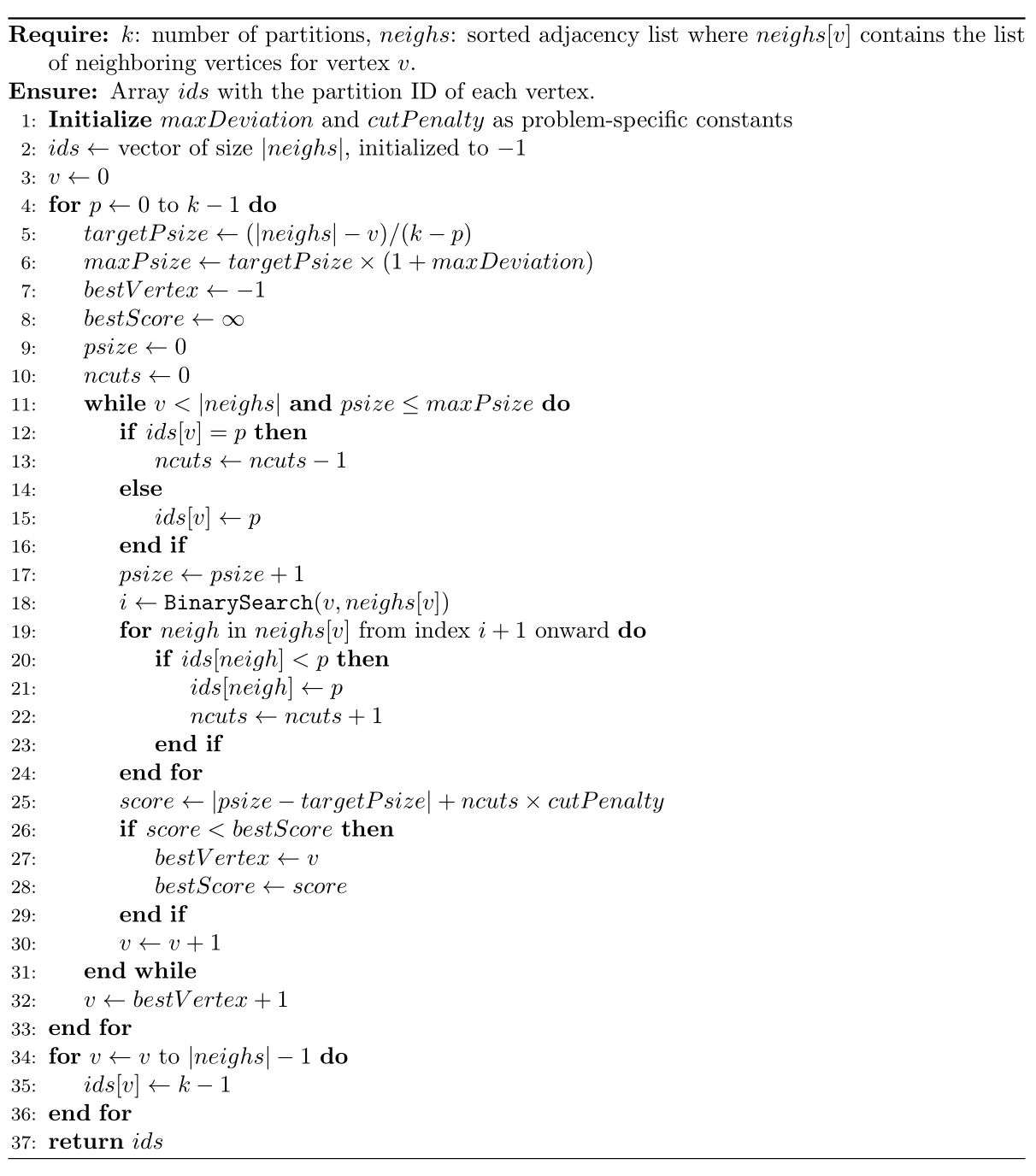}
\end{figure}

The pseudocode of our greedy approach for partitioning a graph into $k$ partitions is presented in Fig.~\ref{alg:kway_partition}.
For the results discussed in this study, the parameter {\texttt{maxDeviation}} was set to 0.05, and {\texttt{cutPenalty}} was assigned a value of 10.

After computing the partitioning of $L(G)$ we use a modified parallel implementation of the Approximate Minimum Degree ordering algorithm (AMD)~\cite{amestoy1996approximate} to minimize the fillins in the part of the linear system corresponding to each partition (a {\it fillin} is a zero entry of the coordinate matrix which becomes non-zero when computing a factorization such as an LU or an LDLT factorization which is needed for Gaussian elimination). This step completes the initialization of the ILVES algorithm, which is then ready to apply the constraints in each step until reinitialization is required after \textsc{Gromacs} recomputes its neighbor list.

ILVES-M and ILVES-F enforce constraints at each step through three primary phases: i) constructing the relevant linear systems and calculating the maximum relative error, ii) solving the linear systems, and iii) updating atomic positions. These phases are iteratively executed until the desired tolerance for the maximum relative error is reached.

During the first phase, each thread constructs the corresponding portion of the linear system for its private partition, while the master thread also handles the shared partition. In this phase, the private and shared partitions can be computed concurrently. The computation of the maximum relative error is overlapped with the construction of the linear system. In distributed-memory simulations, domains exchange their maximum relative errors after this step is complete. Note that for ILVES-F, the coordinate matrix of the linear system of equations to solve ($\bm{A}$) %defined in Eq. (\ref{defA})) 
remains constant for all iterations and is thus computed just once per simulated time. This step is vectorized using the \textsc{Gromacs} SIMD intrinsics interface.

Next, the Lagrange multipliers $\bm{\lambda}$ are calculated by solving the linear system in parallel %(\ref{equ:newton-constraint-equation1}) 
through LU/LDLT factorization for ILVES-M/F. The factorization is followed by forward and backward substitutions. ILVES-F requires a single LDLT factorization per step because $\bm{A}$ remains unchanged for all the iterations in a given simulated time. This process is detailed in Fig.~\ref{fig:example_linear_system_and_partitioning}. 

Finally, the positions of the atoms are updated.
%as indicated by Eqs. (\ref{equ:nonlinear:1a}), (\ref{equ:nonlinear:1b}). 
Updates to the positions of atoms in private partitions are processed in mutual exclusion to updates of atoms in the shared partition. In distributed-memory simulations, the updated atomic positions are exchanged between domains, ensuring that each domain has the latest positions of its neighboring atoms.

\section{4. Further Discussion on the Need to Solve the Constraint Equations Accurately}\label{sec:need-to-solve-acc}

We have presented arguments in favor of an accurate solving of the constraint equations in the main paper and in a recent publication \cite{ilves_2023}. In this section we present further arguments, supported by numerical evidence, to reinforce this thesis.

First, let us note that if the constraint equations are solved exactly, then the SHAKE algorithm is second-order accurate in the time step.
In practice, it is impossible to solve the constraint equations exactly and we accept an approximation if the relative error is less than the tolerance $\textrm{Tol}$ passed to the constraint solver. In this section, we shall explain why the only safe value of $\textrm{Tol}$ is tiny and why we should run the constraint solver until stagnation.
We shall present arguments that should appeal to both the numerical analyst and the experimental scientist who seeks to validate a model of physical reality.

It is important to distinguish between the exact numbers that are defined by the SHAKE algorithm and the numbers that are computed when the algorithm is executed using finite precision arithmetic and tolerance $\textrm{Tol} > 0$.

\begin{table}[ht]
  \centering
    \caption{Parameters defining a simple pendulum.} \label{tab:pendulum}
  \begin{tabular}{@{}lcc@{}}
  \hline
    Parameter & Symbol &  Value \\ \hline
    Mass   & $m$ & $1~kg$ \\
    Length & $l$ & $1~ m$ \\
    Gravitational const. & $g$ & $9.82~ m/s^2$ \\
    Initial velocity & $v_0$ & $(1,0)~m/s$ \\
    Initial position & $r_0$ & $(0,0)$\\ \hline
  \end{tabular}
\end{table}

We begin by demonstrating that if the tolerance parameter $\textrm{Tol}$ is small enough, then the computed numbers behave in a manner that is indistinguishable from the numbers that are defined by the SHAKE algorithm. 
To this end, we shall study a simple pendulum. This is a standard test problem.
The necessary files to execute the calculations and regenerate our figures from scratch can be found in the public GitHub repository 
\begin{center}
\url{https://github.com/spockcc/_PAPER_ILVES_pendulum}
\end{center}
We shall demonstrate that if the tolerance in solving is small enough, then the total energy is very nearly preserved and the error is $\mathcal{O}(h^2)$. 
Our pendulum is defined by Tab.~\ref{tab:pendulum}. 

\begin{figure}
\includegraphics[width=\textwidth]{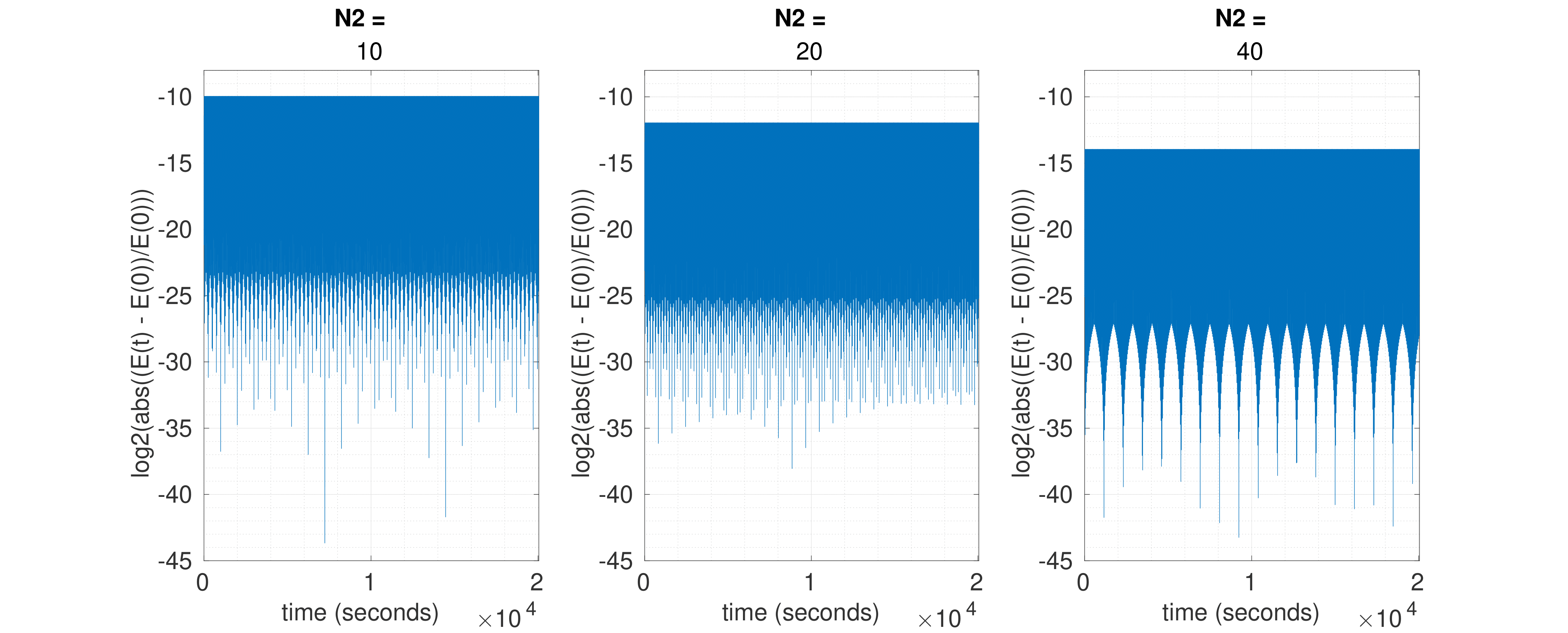}
\caption{Total energy of a simple pendulum as a function of time for three different values of timestep.} \label{fig:good_pendulum}
\end{figure}

\begin{table}[htbp]
  \centering
  \caption{The parameters for the first set of simulations of the pendulum given by Tab.~\ref{tab:pendulum}. The evolution of the total energy is shown in Fig.~\ref{fig:good_pendulum}.} \label{tab:good_pendulum}
  \begin{tabular}{@{}lcc@{}}
  \hline
    Description & Name & Value \\ \hline
    Number of periods & \texttt{np} & 10,000 \\
    Number of recorded state vectors per period & \texttt{nsp} & 10 \\
    Number of recorded state vectors & \texttt{N1} & $\texttt{N1} = \texttt{np}\cdot\texttt{nsp}$ \\
    Number of steps between recorded state vectors & \texttt{N2} & $\{10, 20, 40\}$ \\
    Tolerance passed to constraint solver & $\textrm{Tol}$ & $10^{-12}$
    \\ \hline
  \end{tabular} 
\end{table}

The \textsc{MATLAB} function \texttt{shake\_pendulum} is included in the GitHub mentioned above. It does a sequence of 3 simulations of the same pendulum using different values of the time step and generates Fig.~\ref{fig:good_pendulum}. 
The exact parameters of these simulations are given by Tab.~\ref{tab:good_pendulum}. 

We note that the simulations do $\texttt{nsp} \cdot \texttt{N2} \in \{100, 200, 400\}$ steps per period, which is far above the rule of thumb that suggests at least 5 steps per period. In Fig.~\ref{fig:good_pendulum} we see that the relative error of the total energy, i.e.,
\begin{equation}
R(t) = \frac{E(t) - E(0)}{E(0)}
\end{equation}
is bounded as a function of the time $t$ and that the maximum value of the absolute value of relative error is approximately $\{2^{-10}, 2^{-12}, 2^{-14} \}$ for $\texttt{N2} = \{10, 20, 40\}$. 
This is experimental evidence in support of the hypothesis that the maximum relative error is $\mathcal{O}(h^2)$. 
This is the behavior that we would observe if the algorithm was executed using exact arithmetic and the constraint equations were solved exactly.

\begin{figure}
    \centering
    \includegraphics[width=\linewidth]{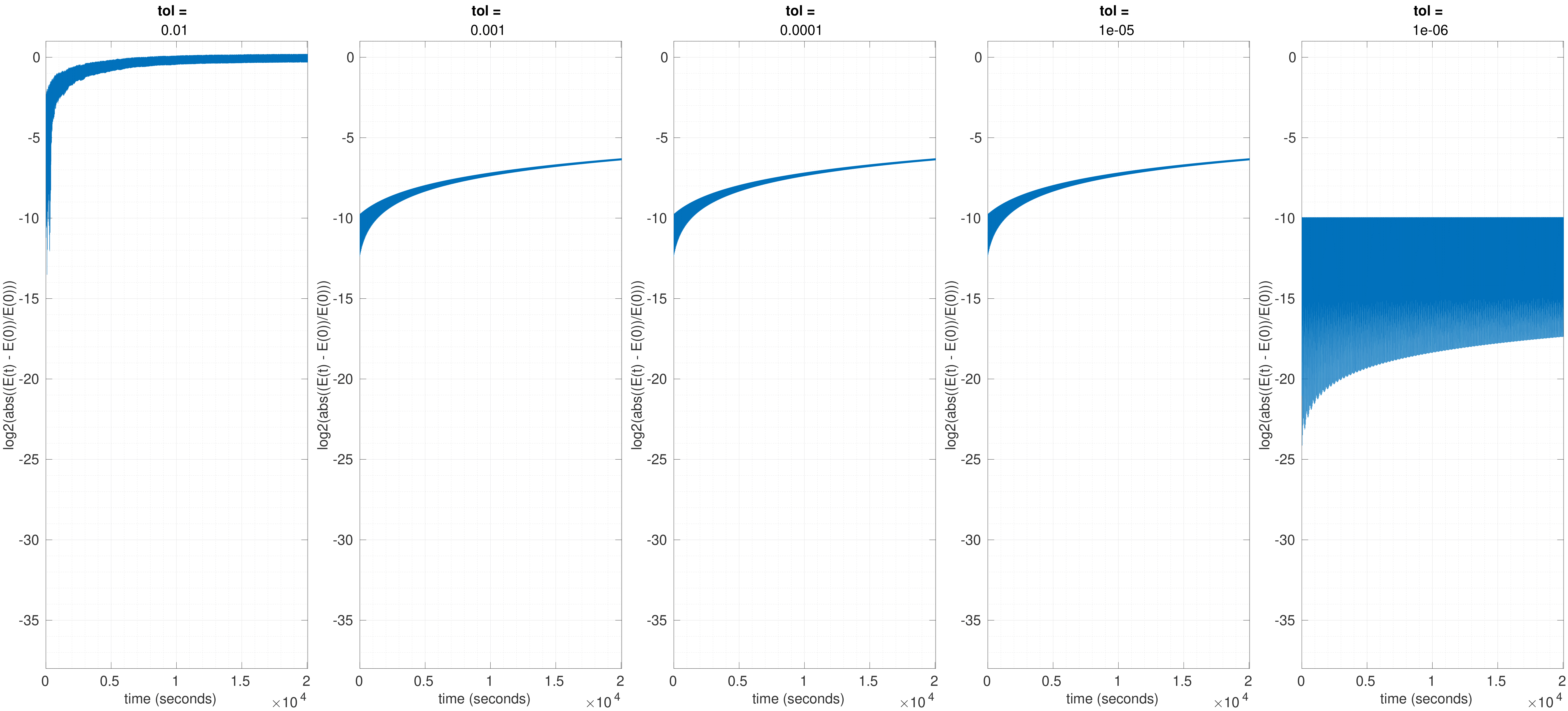}
    \caption{The evolution of the total energy of a pendulum over time for different values of the tolerance $\textrm{Tol}$ passed to the constraint solver.}
    \label{fig:bad_pendulum}
\end{figure}

 \begin{table}[htbp]
  \centering
  \caption{The parameters for the second set of simulations of a pendulum given by Tab.~\ref{tab:pendulum}. The evolution of the total energy is displayed in Fig.~\ref{fig:bad_pendulum}.} \label{tab:bad_pendulum}
  \begin{tabular}{@{}lcc@{}}
  \hline
    Description & Name & Value \\ \hline
    Number of periods & \texttt{np} & 10,000 \\
    Number of recorded state vectors per period & \texttt{nsp} & 10 \\
    Number of recorded state vectors & \texttt{N1} & $\texttt{N1} = \texttt{np}\cdot\texttt{nsp}$ \\
    Number of steps between recorded state vectors & \texttt{N2} & 10 \\
    Tolerance passed to constraint solver & $\textrm{Tol}$ & $\{10^{-2}, 10^{-3}, \dotsc, 10^{-6}\}$ \\
    Constraint solver & \texttt{method} & \texttt{quasi} \\ \hline
  \end{tabular} 
  \end{table}

The \textsc{MATLAB} function \texttt{shake\_pendulum} does a second sequence of 10 simulations of the same pendulum using different values of the tolerance \text{Tol}. 
The exact parameters are given by Tab.~\ref{tab:bad_pendulum}.
The constraint solver is a quasi-Newton's method, specifically the simplified Newton's method that uses a fixed value of the Jacobian for each step. 
The linear convergence of this method ensures that the constraints are solved with a relative error that is essentially equal to the tolerance passed to the solver. 
Fig.~\ref{fig:bad_pendulum} shows how the relative error $R(t)$ of the energy evolves over time for different values of the tolerance $\textrm{Tol}$. 
In particular, we see that the relative error is a strictly increasing function of $t$ for $ $\textrm{Tol}$ \in \{10^{-2}, 10^{-3}, 10^{-4}, 10^{-5} \}$. 
For these values of the tolerance $\textrm{Tol}$, we cannot claim that the simulation corresponds to an isolated system because the fundamental principle of the conservation of energy is clearly violated. 
However, for $\tau = 10^{-6}$, the maximum relative error is bounded as a function of $t$. Again, this is exactly the behavior we would observe if the algorithm were executed in exact arithmetic and the constraint equations were solved exactly. 

These two numerical experiments show that if the tolerance parameter $\textrm{Tol}$ is sufficiently small, then the total energy of the pendulum is very nearly preserved, and the computed numbers behave in a manner that is consistent with a method that is second-order accurate in the time step. 
If the tolerance is not sufficiently small, then the absolute value of the relative error of the energy is an increasing function of time.

In the absence of any deep understanding of the relationship between any system, the time step, the tolerance, and the evolution of the relative error of the energy, the only safe policy is to run the constraint solver until it stagnates due to the limitation of finite precision arithmetic. 
Naturally, this process is not practical unless the constraint solver converges rapidly.

\begin{table}[htbp]
  \centering
  \caption{The parameters of a third set of simulations of the pendulum given by Tab.~\ref{tab:pendulum}. The results are shown in Fig.~\ref{fig:flip}.} \label{tab:flip}
  \begin{tabular}{@{}lcc@{}}
  \hline
    Parameter & Name & Value (s) \\ \hline
    Number of periods & \texttt{np} & 10,000 \\
    Number of recorded state vectors per period & \texttt{nsp} & 10 \\
    Number of recorded state vectors & \texttt{N1} & $\texttt{N1} = \texttt{np}\cdot\texttt{nsp}$ \\
    Number of steps between recorded state vectors & \texttt{N2} & 10 \\
    Tolerance passed to constraint solver & $\textrm{Tol}$ & $10^{-4}$ \\
    Constraint solver & \texttt{method} & $\{$\texttt{newton},\texttt{quasi}$\}$ \\ \hline
  \end{tabular} 
  \end{table}

\begin{figure}
    \centering
    \includegraphics[width=\linewidth]{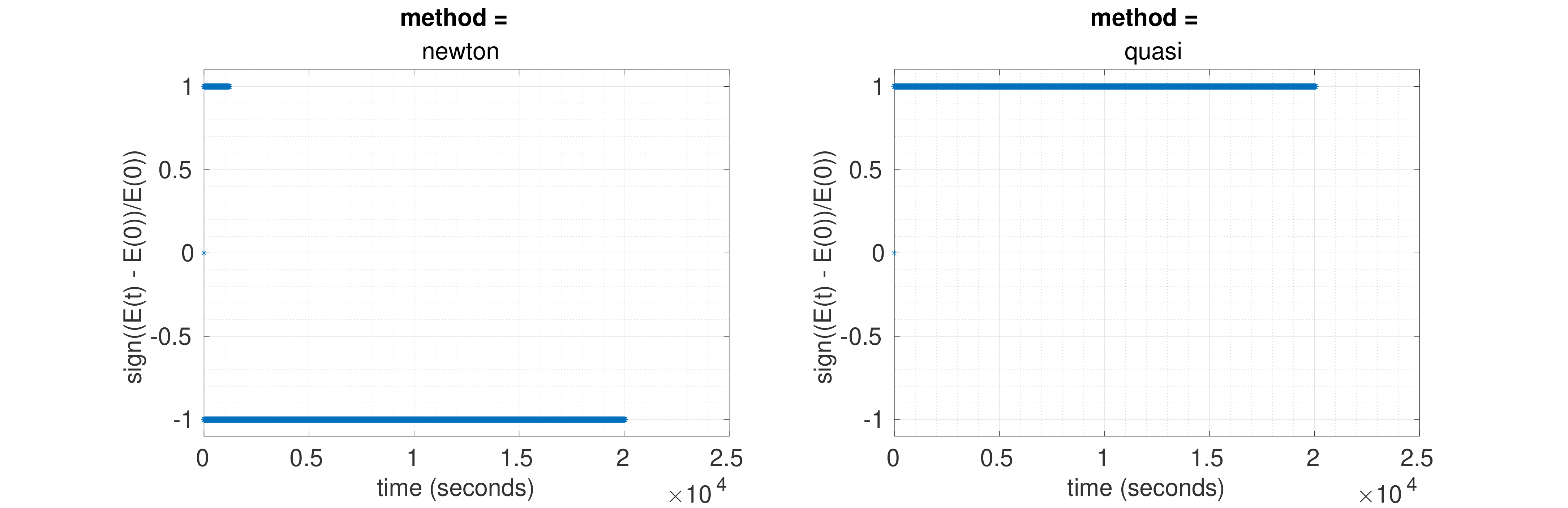}
    \caption{Sign of the relative error of the total energy for pendulum for two different constraint solvers as a function of the time: Newton's method (left) and the simplified Newton's method (right).}
    \label{fig:flip}
\end{figure}

The \textsc{MATLAB} function \texttt{shake\_pendulum} also conducts a third sequence of numerical experiments. 
Here, the focus is on the sign of the relative error rather than on its absolute value. 
The two experiments reveal that the sign of the relative error can depend on the choice of the constraint solver.
The exact parameters of these simulations are given by Tab.~\ref{tab:flip}. 

Fig.~\ref{fig:flip} shows the evolution of the sign of relative error of the energy. 
The scientist who uses Newton's method to probe the inner mysteries of the pendulum will (almost certainly) conclude that the system loses energy over time, whereas the scientist who uses the simplified Newton's method will conclude that the system gains energy over time.
We find it intolerably dangerous that the physical conclusions drawn from a numerical experiment might depend on the algorithm used to solve the central equations.

We shall now argue further in favor of solving the constraint equations as accurately as the hardware allows. 
To this end, we ask the reader to consider a fundamental question: What is the purpose of an MD simulation? 
In general, we seek to obtain numbers that cannot be measured in the laboratory either because we lack the equipment or because the procedure is too expensive. However, before these experiments can be conducted it is necessary to validate our model and determine if it accurately captures the critical aspects of physical reality.
In general, we do not care for the solution of a specific initial value problem for the system of differential algebraic equations (DAEs); rather we seek the value of a physical quantity that can be computed in terms of several such trajectories. 
Let $T$ denote the value that would be obtained if we could solve the relevant systems of DAEs exactly. 
Let $A_h$ be the value that would be obtained if we could solve the corresponding discrete-time equations exactly and let $\hat{A}_h$ denote the value of $A_h$ that is returned by the computer. 
It is important to recognize that the three numbers $T$, $A_h$, and $\hat{A}_h$ are almost certainly different. 
The discretization error $T-A_h$ is not zero because we have replaced the system of DAEs with a set of discrete equations.
Moreover, the computational error $A_h - \hat{A}_h$ is not zero because we cannot solve the constraint equations exactly (truncation error) and because we are forced to use finite precision arithmetic (rounding error).
Why should we care about these errors? 
We validate our model by comparing $T$ against the physical reality $P$.
We cannot compute $T$ directly, and a small value of the observed error $P-\hat{A}_h$ does not necessarily imply that the modeling error $P - T$ is small. Why is that?
This issue can be understood using the elementary identity
\begin{equation}
\underset{\text{modeling error}}{ \underbrace{(P - T)} } = \underset{\text{observed error}}{\underbrace{(P - \hat{A}_h)}} - \underset{\text{computational error}}{\underbrace{(A_h - \hat{A}_h)}} - \underset{\text{discretization error}}{\underbrace{(T - A_h)}}.
\end{equation}
From this identity, we conclude that $T$ is a good approximation of $P$ if the computational error and the discretization error are both negligible compared with a small value of the observed error $P - \hat{A}_h$.
It is, therefore, critical to control the size of the computational error and the discretization error.
Frequently, but not universally, it is possible to use Richardson extrapolation~\cite{zahari2018richardson, ppam3_paper} to estimate the discretization error $T-A_h$ by repeating the same simulation using different values of the time step $h$.
In particular, Richardson extrapolation cannot be applied unless the force fields are sufficiently smooth~\cite{ppam3_paper}.
Moreover, Richardson's error estimate is not reliable unless the computational error $A_h - \hat{A}_h$ is sufficiently small~\cite{ppam3_paper}. 
The size of the computational error $A_h - \hat{A}_h$ is controlled by the size of the unit roundoff $u$ and the tolerance $\textrm{Tol}$ that is passed to the constraint solver. 
In the absence of any information about the size of the discretization error $T-A_h$, the safest course of action is, therefore, to use the smallest practical value of the unit roundoff $u$ and the tolerance $\textrm{Tol}$. 
This choice minimizes the computational error and maximizes the chance that the computational error is so small that we can reliably estimate the discretization error $T-A_h$.
Therefore, it is prudent to favor double precision over single precision and tiny values of $\textrm{Tol}$ over larger values of $\textrm{Tol}$.
In ILVES we measure the relative constraint violation and iterate until 
$$\frac{1}{2} \frac{\|\bm{q}_{a_i} - \bm{q}_{b_i}\|_2^2 - d_i^2}{d_i^2} \leq \textrm{Tol} $$
for all bonds or the maximum number of iterations is reached.
If the relative constraint violation is small, then 
\begin{equation}
    \frac{1}{2} \frac{\|\bm{q}_{a_i} - \bm{q}_{b_i}\|_2^2 - d_i^2}{d_i^2} = \frac{1}{2} \frac{(\|\bm{q}_{a_i}-\bm{q}_{b_i}\|_2 - d_i)(\|\bm{q}_{a_i}-\bm{q}_{b_i}\|_2 + d_i) }{d_i^2} \approx \frac{\|\bm{q}_{a_i}-\bm{q}_{b_i}\|_2 - d_i}{d_i}.
\end{equation}
On a machine with unit roundoff $u$ we cannot hope to reliably reduce the right-hand side below $u$. Therefore, the smallest value of $\textrm{Tol}$ that should be considered is literally the unit roundoff $\textrm{Tol} = u$. 

In conclusion, we can say the following: If we do not solve the constraint equations accurately, then we cannot claim that our calculations are $m$-th order accurate in the time step, we cannot assume that the fundamental principle of conversation of energy is respected, and we cannot reliably estimate the discretization error and perform the validation needed to develop faith in our model of the physical reality. This strong reason supports solving the constraints as accurately as possible.

\section{5. Experimental Setup}\label{sec:experimental-setup}

\begin{table}[htbp]
\centering
%\begin{adjustbox}{width=\textwidth}
\caption{Main features of the computing nodes used to execute the simulations.}
\label{tab:experimental.setup}
\begin{tabular}{ll}
%\toprule
\hline
Processor & 2 $\times$ Intel Xeon Platinum 8480+\\
Cores     & 2 $\times$ 56 \\
Frequency & 2~GHz \\
%\midrule
\hline
L1i cache (I + D) & 32~KB + 48~KB (per core) \\
L2 cache  & 2 MB (per core)  \\
LLC       & 105 MB (shared)  \\
%\midrule
\hline
Main Memory & 256 GB DDR5 (16 $\times$ 16~GB 4800~MHz DIMMs) \\
%\midrule
\hline
Interconnection & ConnectX-7 NDR200 InfiniBand (100~Gb/s per node) \\
%\bottomrule 
\hline
\end{tabular}
\end{table}

The simulations presented in this text were performed on a multi-node cluster. Each node is equipped with two Intel Xeon Platinum 8480+ chips, each with 56 cores. The main features of the nodes are summarized in Tab.~\ref{tab:experimental.setup} (see \cite{reviewibanez} for a presentation of high-performance computing). We used \textsc{Gromacs} 2021.0 compiled with GCC-10.1.0 for the numerical experiments.

\section{6. Simulations}\label{sec:simulations}

\begin{table}[htbp]
\centering
\caption{Molecular systems simulated in this article. In the Force Field column, C22+CMAP is Charmm22+CMAP, C36m is Charmm36m and C36 is Charmm36. Although the ff19SB force field defines residue-specific CMAP corrections, the protein topology of the DNA-protein system contains a generic CMAP matrix (taken from the \texttt{frcmod.ff19SB\_XXX} internal file of AmberTools24) for all residues, due to the use of the \texttt{amb2gro\_top\_gro.py} script for conversion from Amber to \textsc{Gromacs} format, which ensures compatibility with versions prior to 2025.\\}
\label{tab:simulations}
\begin{tabular}{@{}lccccc@{}}
\hline
  \textbf{System} &
  \textbf{\begin{tabular}[c]{@{}c@{}}\# Molec.\\entities \\(PDB ID)\end{tabular}} &
  \textbf{\# Mol.} &
  \textbf{\begin{tabular}[c]{@{}c@{}}\# Resid./\\Atoms\end{tabular}} &
  \textbf{\begin{tabular}[c]{@{}c@{}}\# Constr.\\all-bonds/\\ H-bonds\end{tabular}} &
  \textbf{\begin{tabular}[c]{@{}c@{}}Force Field/\\Water Model\end{tabular}} \\ 
\hline
  \begin{tabular}[c]{@{}l@{}}Solvated\\barnase~\cite{structurebarnase}\end{tabular} &
  \begin{tabular}[c]{@{}c@{}}barnase (1A2P) \\ Water \\ NA$^+$, CI$^-$\end{tabular} &
  \begin{tabular}[c]{@{}c@{}} 1 \\ 62,848 \\ 4/11\end{tabular} &
  \begin{tabular}[c]{@{}c@{}} 108/1,705 \\ 188,544 \\ 15\end{tabular} &
  \begin{tabular}[c]{@{}c@{}} 1,721/836 \\ --- \\ ---\end{tabular} &
  \begin{tabular}[c]{@{}c@{}} C22+CMAP~\cite{charmm27} \\ Tip3p~\cite{tip3p} \\ C22+CMAP~\cite{charmm27}\end{tabular} \\
\hline
  \begin{tabular}[c]{@{}l@{}}Solvated\\ nuclease\cite{PDB_2SNS}\end{tabular} &
  \begin{tabular}[c]{@{}c@{}}nuclease (2SNS) \\ Water \\ CI$^-$\end{tabular} &
  \begin{tabular}[c]{@{}c@{}} 1 \\ 115,602 \\ 10\end{tabular} &
  \begin{tabular}[c]{@{}c@{}} 149/2,398 \\ 346,806 \\ 10\end{tabular} &
  \begin{tabular}[c]{@{}c@{}} 2,459/1,224 \\ --- \\ ---\end{tabular} &
  \begin{tabular}[c]{@{}c@{}} C22+CMAP~\cite{charmm27} \\ Tip3p~\cite{tip3p} \\ C22+CMAP~\cite{charmm27}\end{tabular} \\
\hline
  \begin{tabular}[c]{@{}l@{}}Solvated\\ COVID\\ protease~\cite{articuloproteinacovid}\end{tabular} &
  \begin{tabular}[c]{@{}c@{}}protease (5R7Y) \\ Water \\ NA$^+$, CI$^-$\end{tabular} &
  \begin{tabular}[c]{@{}c@{}} 1 \\ 61,958 \\ 8/4\end{tabular} &
  \begin{tabular}[c]{@{}c@{}} 304/4,645 \\ 185,874 \\ 12\end{tabular} &
  \begin{tabular}[c]{@{}c@{}} 4,697/2,297 \\ --- \\ ---\end{tabular} &
  \begin{tabular}[c]{@{}c@{}} C22+CMAP~\cite{charmm27} \\ Tip3p~\cite{tip3p} \\ C22+CMAP~\cite{charmm27}\end{tabular} \\
\hline
\begin{tabular}[c]{@{}l@{}}Pure\\ benzene\cite{mexicoorganicsolvents}\end{tabular} &
  benzene &
  2000 &
  24,000 &
  24,000/12,000 &
  Gromos54a7~\cite{gromos54a7} \\
\hline
  \begin{tabular}[c]{@{}l@{}}Solvated\\ DNA-\\ proteins\cite{Wu2010}\end{tabular} &
  \begin{tabular}[c]{@{}c@{}}histones (3LEL) \\ DNA \\ Water \\ NA$^+$\end{tabular} &
  \begin{tabular}[c]{@{}c@{}} 16 \\ 4 \\ 83,872 \\ 428\end{tabular} &
  \begin{tabular}[c]{@{}c@{}} 1,540/25,257 \\ 588/18,712 \\ 251,616 \\ 428\end{tabular} &
  \begin{tabular}[c]{@{}c@{}} 25,933/13,078 \\ 20,178/6,670 \\ --- \\ ---\end{tabular} &
  \begin{tabular}[c]{@{}c@{}} ff19SB~\cite{ff19sb} \\ OL21~\cite{ol21} \\ OPC4~\cite{opc4} \\ Ions~\cite{Li2015, Sengupta2021}\end{tabular} \\
\hline
  \begin{tabular}[c]{@{}l@{}}Aquaporin 1\\ embedded in \\ lipid bilayer\cite{Sui2001}\end{tabular} &
  \begin{tabular}[c]{@{}c@{}}Aqp1 (1J4N) \\ MMPC \\ Water \\ NA$^+$, CI$^-$\end{tabular} &
  \begin{tabular}[c]{@{}c@{}} 4 \\ 1,129 \\ 80,598 \\ 223/235\end{tabular} &
  \begin{tabular}[c]{@{}c@{}} 996/15,072 \\ 133,222 \\ 241,794 \\ 458\end{tabular} &
  \begin{tabular}[c]{@{}c@{}} 15,480/7,755 \\ 131,837/81,193 \\ --- \\ ---\end{tabular} &
  \begin{tabular}[c]{@{}c@{}} C36m~\cite{charmm36m} \\ C36~\cite{charmm36} \\ Tip3p~\cite{tip3p} \\ C36~\cite{charmm36}\end{tabular} \\
\hline
\end{tabular}
\end{table}

\begin{figure}[htbp]
    \centering
    \includegraphics[width=0.96\textwidth]{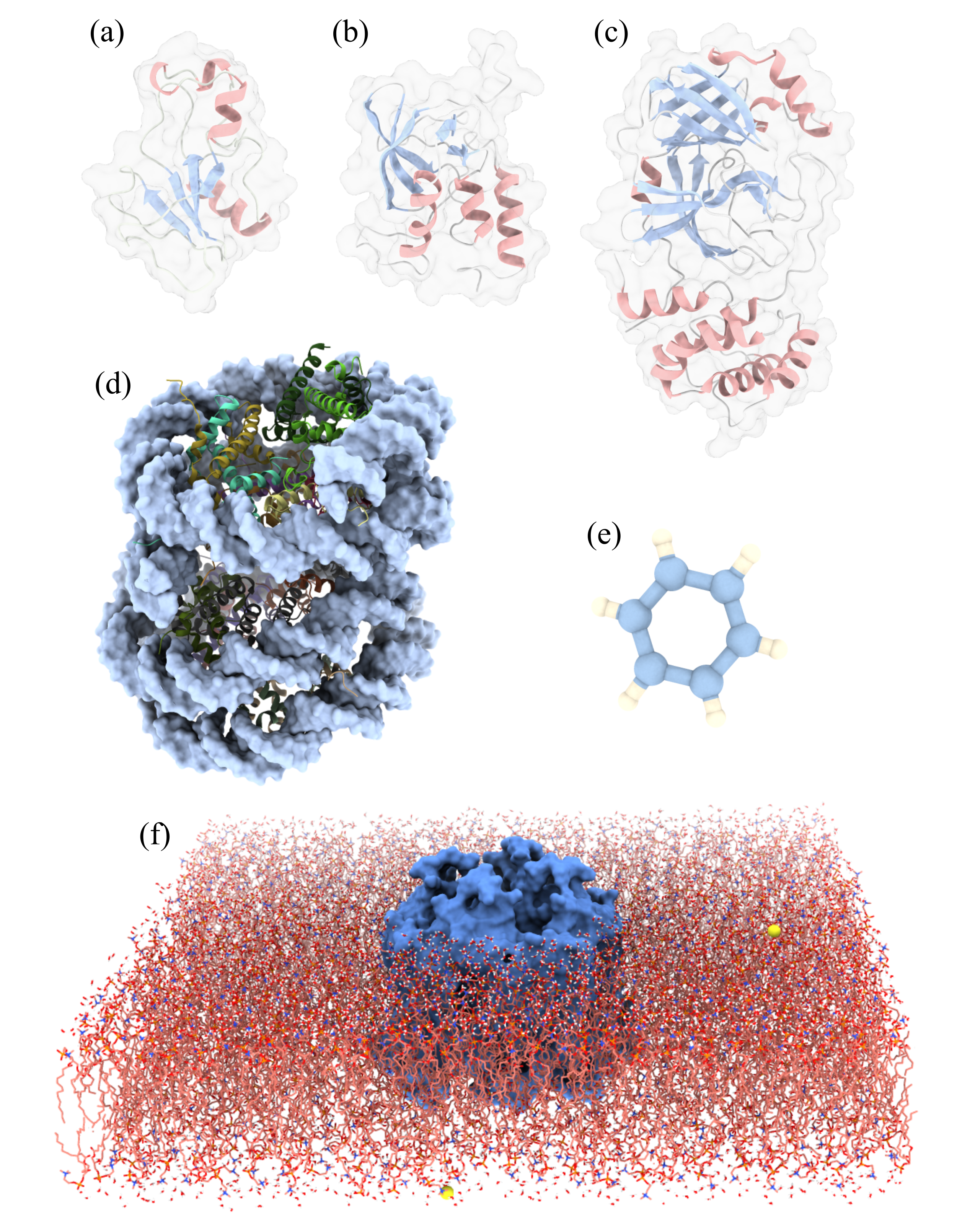}
    \caption{Representations of the simulated molecules, covering solvated proteins: (a) barnase (PDB 1A2P), (b) nuclease (PDB 2SNS) and (c) Covid-19 main protease (PDB 5R7Y); (d) a solvated protein-DNA complex (PDB 3LEL); (e) an organic solvent (benzene); and (f) a tetrameric protein embedded in a lipid bilayer. }
    \label{fig:systems-images}
\end{figure}

We have conducted simulations on a diverse set of molecular systems, including three solvated proteins, a solvated protein-DNA complex, an organic solvent, and a membrane protein embedded in a lipid bilayer surrounded by water.
Tab.~\ref{tab:simulations} summarizes the key characteristics of the systems, while Fig.~\ref{fig:systems-images} provides a visual representation.
Biological macromolecules, such as DNA and proteins, have long been of interest for medical and biotechnological purposes.
Benzene, as a representation of small molecules, has important chemical and industrial applications~\cite{benzeneapplications}.
Proteins embedded in a lipid bilayer include channels, transporters, and signal receptors, among others, which play a vital role in cell biology and medical research\cite{membranes2}. This diversity of molecules allowed us to evaluate the performance of our ILVES algorithms across a wide range of molecular structures.

\begin{table}[htbp]
    \centering
    \caption{Relevant input parameters employed in the stages of the simulations.}
    \vspace{30pt}
    \label{tab:sim-parameters}
    \rotatebox{90}{
        \makebox[0.90\textheight][c]{
    \renewcommand{\arraystretch}{0.8}
    \begin{tabular}{@{}llllll@{}}
    \hline
    \textbf{Step} &
      \textbf{General Settings} &
      \textbf{\begin{tabular}[c]{@{}l@{}}Thermodynamic\\ Ensemble (baths)\end{tabular}} &
      \textbf{\begin{tabular}[c]{@{}l@{}}Physical\\ Conditions\end{tabular}} &
      \textbf{\begin{tabular}[c]{@{}l@{}}Time step /\\ Simulated\\ Time\end{tabular}} \\ \hline
    Min. &
      \begin{tabular}[c]{@{}l@{}}Integrator: steepest descent\\ Neighboring searching: grid\\ rcoul (PME): 1.0 (0.9)$^b$ nm\\ rvdw (cut-off): 1.0 (0.9)$^b$ nm\\PBC: xyz\\ Constr: none\end{tabular} &
      --- &
      \begin{tabular}[c]{@{}l@{}}Emtol: 1 kJ/mol\\ Temp: 0 K\\ Press: 0 atm\end{tabular} &
      max 20K steps \\ \hline
    Heating &
      \begin{tabular}[c]{@{}l@{}}Integrator: md\\ Neighboring searching: grid\\ rcoul (PME): 1.0 (0.9)$^b$ nm\\ rvdw (cut-off): 1.0 (0.9)$^b$ nm\\ vdw-mod: potential-shift-verlet\\PBC: xyz\\ Constr: all-bonds\end{tabular} &
      \begin{tabular}[c]{@{}l@{}}Termost: Berendsen\\ $\tau_T$ = 0.1 ps\end{tabular} &
      \begin{tabular}[c]{@{}l@{}}Temp: Ini-T + \\ ramp (n $\times$ 50 K)\\ Press: 0 atm\end{tabular} &
      1 fs / n $\times$ 50 ps \\ \hline
    Equil. 1 &
      Same as Heating &
      \begin{tabular}[c]{@{}l@{}}NVT\\ Termost: v-rescale\\ $\tau_T$ = 0.1 ps\end{tabular} &
      \begin{tabular}[c]{@{}l@{}}Temp: Final-T\\ Press: 0 atm\end{tabular} &
      1 fs / 100 ps \\ \hline
    Equil. 2 &
      Same as Heating &
      \begin{tabular}[c]{@{}l@{}}NPT\\ Termost: v-rescale\\ $\tau_T$ = 0.1 ps\\ Barost: Berendsen\\ $\tau_P$ = 2.0 ps\end{tabular} &
      \begin{tabular}[c]{@{}l@{}}Temp: Final-T\\ Press: 1 atm\end{tabular} &
      2 fs / 100 ps \\ \hline
    Equil. 3 &
      Same as Heating &
      \begin{tabular}[c]{@{}l@{}}NPT\\ Termost: v-rescale\\ $\tau_T$ = 0.1 ps\\ Barost: Parrinello-Rahman\\ $\tau_P$ = 2.0 ps\end{tabular} &
      \begin{tabular}[c]{@{}l@{}}Same as \\ Equil. 2\end{tabular} &
      2 fs / 200 ps \\ \hline
    Prod. &
      \begin{tabular}[c]{@{}l@{}}Same as Heating + \\ Constr: h-bonds, \\ all-bonds or h-angles\end{tabular} &
      Same as Equil. 3 &
      \begin{tabular}[c]{@{}l@{}}Same as \\ Equil. 2\end{tabular} &
      \begin{tabular}[c]{@{}l@{}}2 fs or 3.5 fs / \\ 0.1 ns or 2 ns / \\ or 5.25 ns\end{tabular} \\ \hline
    \end{tabular}
    }}
\end{table}

Below we outline the procedures on which the simulations are based. All the systems, except the protein (human aquaporin-1) embedded in a lipid bilayer, were prepared through a standard protocol involving solvation, minimization, heating, and equilibration. 
The most important parameters used in the preparation and also in the production phase of the simulations are listed in Tab.~\ref{tab:sim-parameters}. The membrane protein system was prepared using the membrane-builder wizard from the \texttt{CHARMM-GUI} online tool~\cite{Jo2008, Lee2015, Wu2014, Jo2007}. We used different force fields for the simulations (Tab.~\ref{tab:simulations}), and explicit Tip3p water molecules~\cite{Jorgensen1983} were used to solvate the macromolecular systems, except for the protein-DNA complex, which was solvated with the OPC4 water model~\cite{izadi2014building}. The ionizable residues of proteins were protonated at pH 7 (\textsc{Gromacs} default), except for the case of barnase, which was set to pH 4.1 because the experimental thermodynamic observables used as a control were obtained under this condition. Chloride and/or sodium ions were added to neutralize the systems or to achieve the desired ionic strength. 
A cubic box was used for the membrane system, while a truncated dodecahedral box was used for the rest of the systems. The Verlet cutoff scheme~\cite{Ver1967PR,Pll2013} was applied for van der Waals interactions, and the Particle Mesh Ewald (PME) method~\cite{essmann1995smooth} was used to account for electrostatics. Both methods were used with a radius cutoff of 1.0 nm (0.9 nm for the membrane protein), while all the preparation steps imposed periodic boundary conditions (PBC). 

The minimization step was performed using the steepest descent algorithm~\cite{Haug1976}, followed by an NVT gradual heating protocol, increasing temperature by 50 K every 50 ps with a 1 fs time step, until the target temperature was reached. The equilibration process began with a 100 ps NVT simulation with a 2 fs time step using the V-rescale thermostat~\cite{papervrescale} ($\tau_T$ = 0.1 ps). During this step, position restraints were applied to all the heavy atoms (protein, DNA, lipids, and water). This was followed by a 100 ps simulation in the NPT ensemble using a 2 fs time step and the heavy atoms now unrestrained. V-rescale thermostat ($\tau_T$ = 0.1 ps) and the Berendsen barostat~\cite{Berendsen1984} (p = 1 atm, $\tau_P$ = 2.0 ps) were settled. Finally, the system underwent 200-ps NPT simulation with a 2 fs time step, employing the V-rescale thermostat ($\tau_T$ = 0.1 ps) and the Parrinello-Rahman barostat~\cite{Parrinello1981} (p = 1 atm, $\tau_P$ = 2.0 ps). Temperature corrections due to the thermostat were applied every 10 steps ({\textsc{Gromacs'}} \texttt{nsttcouple} default value). 
Input files from our simulations can be retrieved from the provided {\texttt{GitHub}} repository ({\texttt{https://github.com/LorienLV/$\_$PAPER$\_$ILVES}}).

The final equilibrated configuration served as the starting point for the production phase of the simulations performed in our reliability and performance studies.

\section{7. Reliability Study}\label{sec:reliability}

In a previous work~\cite{GalanoFrutos2019, GalanoFrutos2023}, one of the authors presented an MD-based method that has proven accurate in reproducing protein folding thermodynamics using both SHAKE and P-LINCS. In this paper, we applied that approach to evaluate the reliability of our constraint solvers by calculating three thermodynamic observable quantities: namely, the change in enthalpy upon unfolding $\Delta$H$_{unf}$ (or upon folding, with opposite sign); the change in heat capacity upon the same event, $\Delta$Cp$_{unf}$; and the corresponding change in Gibbs free energy (i.e., the protein conformational stability), $\Delta$G$_{unf}$. The analysis was conducted by simulating two well-studied proteins: barnase and nuclease (see Tab.~\ref{tab:simulations}), which have been similarly modeled in Refs.~\cite{GalanoFrutos2019, GalanoFrutos2023}. Tab.~\ref{tab:sim-parameters} includes the main parameters used in the preparation and production simulations carried out with these systems.

\begin{figure}[htbp]
    \centering
    \includegraphics[width=0.7\textwidth]{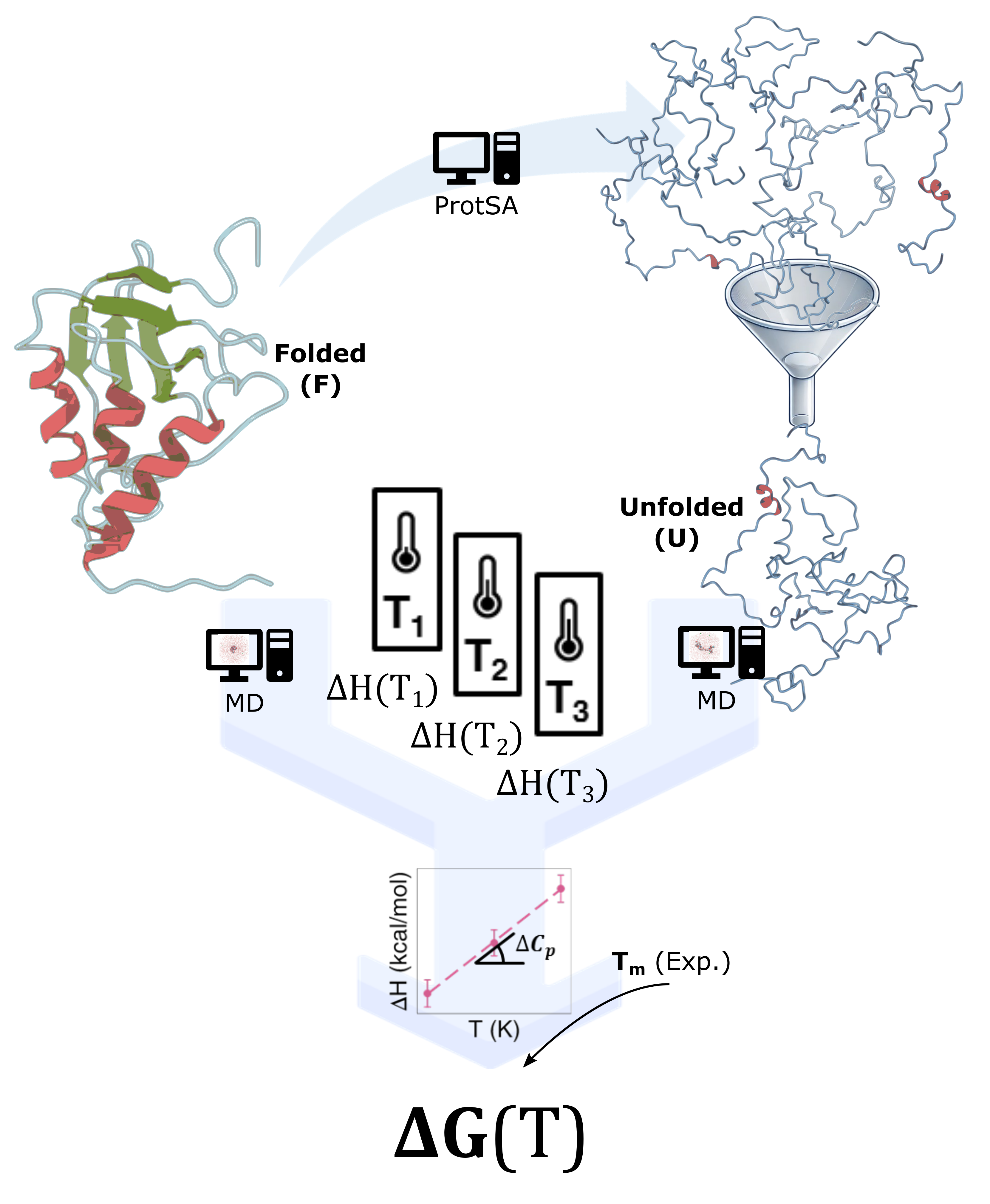}
    \caption{Workflow of the calculation of observable quantities.} 
    \label{fig:scheme_si}
\end{figure}

\begin{table}[htbp]
  \centering
  \caption{Experimentally measured and calculated (MD) thermodynamic quantities of unfolding ($\Delta H_{unf}$ and $\Delta Cp_{unf}$) for barnase ($T_{m} = 323.0 \pm 0.7$ at pH $\sim$4.1~\cite{GalanoFrutos2023}) and nuclease ($T_{m} = 326.9 \pm 0.4$ at pH $\sim$7.0~\cite{GalanoFrutos2023}). The listed Gibbs free energies ($\Delta G_{unf}$) are calculated through the Gibbs-Helmholtz equation %(eq.~\ref{gibbs-helmholtz}) 
  using either the measured or MD thermodynamics. P-LINCS is configured with \textsc{Gromacs} default parameters, while MP-LINCS uses \texttt{lincs-order=4}. All of the simulations are performed with \texttt{constraints=h-bonds}.}
  \label{tab:thermodynamics}
  \begin{tabular}{@{}llcccc@{}}
  %\toprule
  \hline
  \begin{tabular}[c]{@{}c@{}}\textbf{Protein} \\ \textbf{(Method)}\end{tabular} &
  \textbf{Solver} &
  \textbf{Tol.} &
  \begin{tabular}[c]{@{}c@{}}\boldmath$\Delta H_{unf}$ \textbf{at $T_{m}$} \\ \textbf{(kcal/mol)} \end{tabular} &
  \begin{tabular}[c]{@{}c@{}}\boldmath$\Delta Cp_{unf}$\\ \textbf{(kcal/mol} \boldmath$\cdot$ \textbf{K)}\end{tabular} &
  \begin{tabular}[c]{@{}c@{}}\boldmath$\Delta G_{unf}$ \textbf{at 298 K} \\ \textbf{(kcal/mol)}\end{tabular} \\  
  %\midrule
  \hline
  Barnase (Exp.)            & --- & ---             & $118.7 \pm 4.9$ & $1.4 \pm 0.1$ & $7.8 \pm 1.6$ \\
  %\midrule
  \hline
  Barnase (MD) &
  \begin{tabular}[c]{@{}l@{}}SHAKE \\ SHAKE \\ MP-LINCS \\ MP-LINCS \\ P-LINCS \\ ILVES-M \\ ILVES-M \\ ILVES-F \\ ILVES-F\end{tabular} &
  \begin{tabular}[c]{@{}c@{}}$10^{-4}$ \\ $10^{-10}$ \\ $10^{-4}$ \\ $10^{-10}$ \\ --- \\ $10^{-4}$ \\ $10^{-10}$ \\ $10^{-4}$ \\ $10^{-10}$\end{tabular} &
  \begin{tabular}[c]{@{}c@{}}$113.9 \pm 3.1$ \\ $115.2 \pm 3.1$ \\ $116.9 \pm 3.1$ \\ $118.6 \pm 3.1$ \\ $116.3 \pm 3.1$ \\ $113.6 \pm 3.1$ \\ $117.4 \pm 3.1$ \\ $116.4 \pm 3.1$ \\ $115.6 \pm 3.1$\end{tabular} &
  \begin{tabular}[c]{@{}c@{}}$1.1 \pm 0.1$ \\ $1.3 \pm 0.1$ \\ $1.2 \pm 0.0$ \\ $1.3 \pm 0.1$ \\ $1.2 \pm 0.0$ \\ $1.0 \pm 0.1$ \\ $1.1 \pm 0.2$ \\ $1.2 \pm 0.0$ \\ $1.0 \pm 0.1$ \end{tabular} &
  \begin{tabular}[c]{@{}c@{}}$7.8 \pm 1.1$ \\ $7.6 \pm 1.4$ \\ $7.9 \pm 1.2$ \\ $7.9 \pm 1.3$ \\ $7.9 \pm 1.2$ \\ $7.8 \pm 1.2$ \\ $8.0 \pm 1.3$ \\ $7.8 \pm 1.2$ \\ $7.9 \pm 1.2$\end{tabular} \\
  \hline
  Nuclease (Exp.)            & --- & ---             & $82.1 \pm 4.7$ & $2.3 \pm 0.1$ & $4.2 \pm 1.6$ \\
  \hline
  Nuclease (MD) &
  \begin{tabular}[c]{@{}l@{}}SHAKE \\ SHAKE \\ MP-LINCS \\ MP-LINCS \\ P-LINCS \\ ILVES-M \\ ILVES-M \\ ILVES-F \\ ILVES-F\end{tabular} &
  \begin{tabular}[c]{@{}c@{}}$10^{-4}$ \\ $10^{-10}$ \\ $10^{-4}$ \\ $10^{-10}$ \\ --- \\ $10^{-4}$ \\ $10^{-10}$ \\ $10^{-4}$ \\ $10^{-10}$\end{tabular} &
  \begin{tabular}[c]{@{}c@{}}$78.1 \pm 4.8$ \\ $73.5 \pm 4.8$ \\ $74.6 \pm 4.8$ \\ $71.6 \pm 4.8$ \\ $70.8 \pm 4.8$ \\ $75.8 \pm 4.8$ \\ $74.6 \pm 4.8$ \\ $74.4 \pm 4.8$ \\ $73.1 \pm 4.8$\end{tabular} &
  \begin{tabular}[c]{@{}c@{}}$2.0 \pm 0.1$ \\ $1.9 \pm 0.1$ \\ $1.8 \pm 0.1$ \\ $1.7 \pm 0.2$ \\ $1.6 \pm 0.0$ \\ $2.0 \pm 0.1$ \\ $2.0 \pm 0.1$ \\ $2.0 \pm 0.0$ \\ $1.8 \pm 0.1$\end{tabular} &
  \begin{tabular}[c]{@{}c@{}}$4.2 \pm 1.4$ \\ $4.1 \pm 1.3$ \\ $4.2 \pm 1.3$ \\ $4.1 \pm 1.4$ \\ $4.2 \pm 1.1$ \\ $4.0 \pm 1.4$ \\ $4.0 \pm 1.4$ \\ $4.0 \pm 1.3$ \\ $4.1 \pm 1.3$\end{tabular} \\
  \hline
\end{tabular}
\end{table}

Tab.~\ref{tab:thermodynamics} displays the results of our reliability study. The methodology of the study involves separately simulating---in the NPT ensemble---the folded and unfolded states of a solvated protein (see Fig.~\ref{fig:scheme_si}) under identical conditions (force field, type and number of water molecules, ions, pH, temperature, pressure, time step, simulation length, etc.). To achieve precise averaging of thermodynamic quantities, the method requires representative sampling of the conformational space of both states. We used 40 simulation replicas for the folded state and 100 replicas for the unfolded state, with a time step of 2 fs. The 100 protein structures representative of the unfolded state is a filtered sample of a larger set of completely unfolded conformations generated by the ProtSA server~\cite{protsa}. The average enthalpies obtained for the two states are subtracted to yield the enthalpy change upon unfolding, $\Delta$H$_{unf}$. From this calculation, the error % (standard error of the mean) %Tab.~\ref{tab:thermodynamics}) 
is obtained as the sum of both errors: that corresponding to the folded ensemble and that of the unfolded ensemble. The calculation of $\Delta$H$_{unf}$ is performed at three temperatures (295 K, 315 K, and 335 K), allowing for the determination of the enthalpy change with temperature, which is typically linear. This linear relationship, whose slope corresponds to the change in heat capacity, $\Delta$Cp$_{unf}$, allows interpolating the enthalpy change at the mid-denaturation temperature (T$_{m}$, taken from the literature), which is the enthalpy change that is commonly reported in scientific articles. %(see Tab.~\ref{tab:thermodynamics}).
The reported error in $\Delta$Cp$_{unf}$ calculation is that obtained from the linear fitting. Once these two thermodynamic quantities are calculated ($\Delta$H$_{unf}$ at the T$_{m}$ and $\Delta$Cp$_{unf}$), the Gibbs free energy change can be obtained through the Gibbs-Helmholtz equation\cite{gibbshelmholtz}:
\begin{equation}\label{gibbs-helmholtz}
    \Delta G(T) = \Delta H_m \left( 1 - \frac{T}{T_m} \right) - \Delta C_p \left[ T_m - T + T \ln\left(\frac{T}{T_m}\right) \right] .
\end{equation}

The reported error in $\Delta$G$(T)$ calculations (Tab.~\ref{tab:thermodynamics}) 
is obtained by error propagation. In this assessment, we used
the combination of force field and water model that has been shown to best reproduce experimental protein thermodynamics~\cite{GalanoFrutos2023, GalanoFrutos2019}: Charmm22+CMAP\cite{charmm27} and Tip3p\cite{tip3p}. 

The results displayed in Tab.~\ref{tab:thermodynamics} indicate that for ILVES and state-of-the-art solvers, the evaluated thermodynamics $\Delta$H$_{unf}(T_{m})$ and $\Delta$G$_{unf}$(298K) fall within the error ranges defined by experimental measurements, for both proteins. In the case of $\Delta$Cp$_{unf}$, the MD-calculated values fall outside the experimental ranges in some of the setups assessed, but the differences are similar for all constraint solvers. This outcome supports the thesis that ILVES algorithms preserve system dynamics and do not introduce distortions larger than those due to traditionally used constraint solvers.

\section{8. Extended Performance Results}\label{sec:ext-performance}

In this section, we complement the performance results presented in the main paper.

\begin{figure}[htbp]
    \centering
    \subfloat[all-bonds]{
        \includegraphics[width=0.98\textwidth]{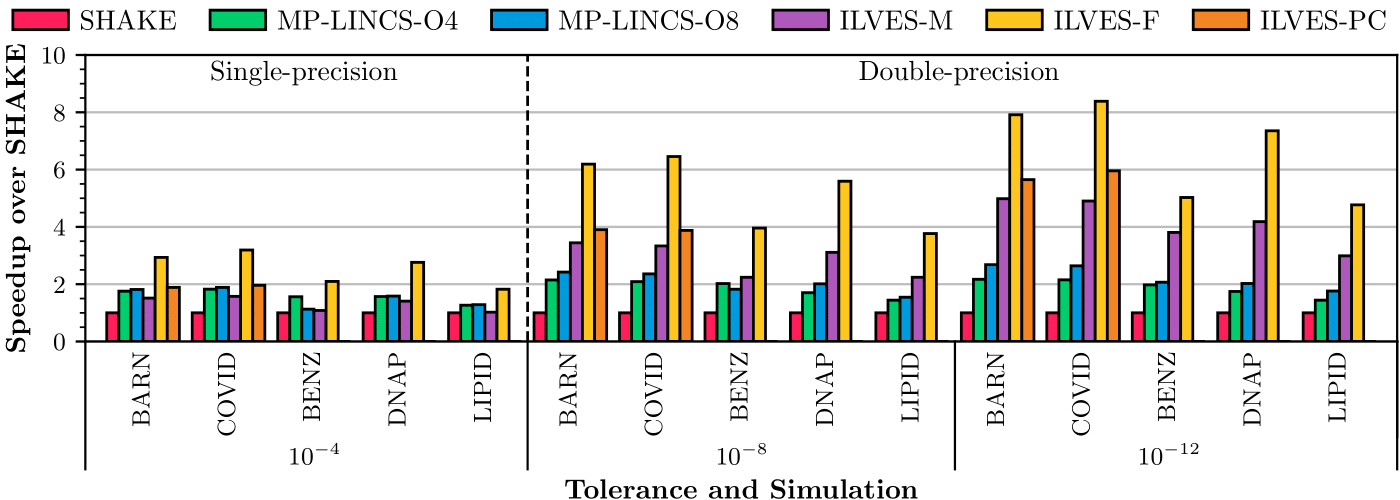}
        \label{fig:speedup_over_shake_1t_all_bonds}
    }
    
    \subfloat[H-bonds]{
        \includegraphics[width=0.98\textwidth]{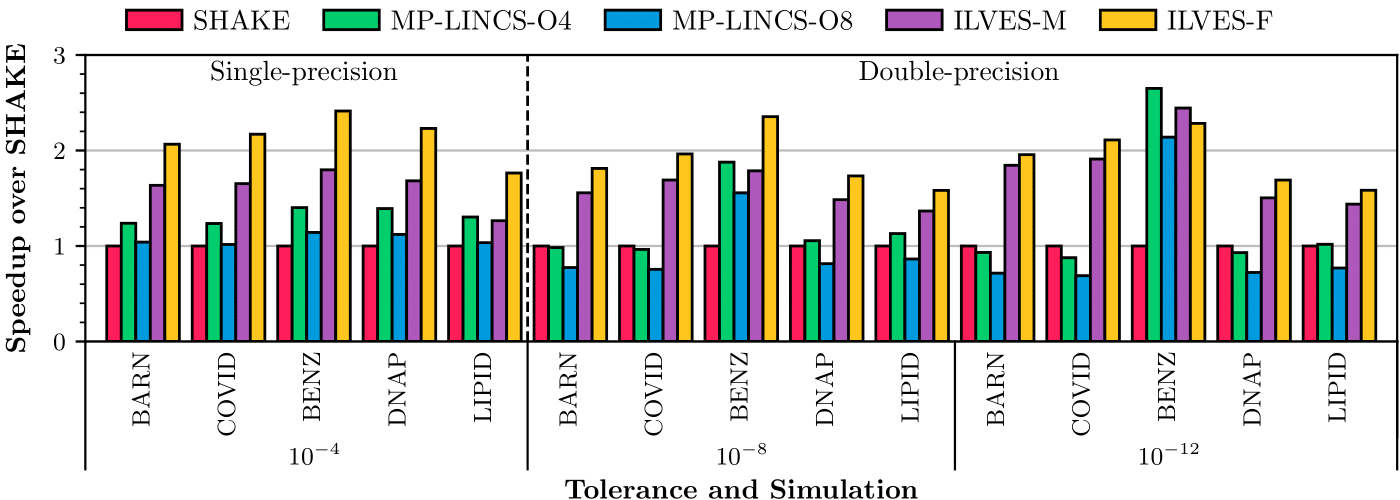}
        \label{fig:speedup_over_shake_1t_h_bonds}
    }
    \caption{Single-thread speedup over SHAKE of MP-LINCS (\texttt{lincs-order=4} and \texttt{lincs-order=8}), ILVES, ILVES-F and ILVES-PC.}
    \label{fig:speedup_over_shake_1t}
\end{figure}

Fig.~\ref{fig:speedup_over_shake_1t} displays the {\it single-thread} speedup over SHAKE. This metric is defined as the ratio of the single-thread execution time of SHAKE to the single-thread execution time of the solver. While ILVES-PC is tailored to protein simulations, it delivers only modest speedups over ILVES-M, which incorporates optimizations absent in ILVES-PC, the most significant being the vectorization of linear-system construction. When constraining all-bonds and using $\textrm{Tol} = 10^{-4}$, ILVES-M and ILVES-PC achieve speedups between $1\times$ and $1.8\times$ compared to SHAKE, similar to MP-LINCS. For the same tolerance, ILVES-F outperforms both, delivering speedups between $1.9\times$ and $3.2\times$ over SHAKE and between $1.5\times$ and $1.8\times$ over MP-LINCS. As the tolerance becomes more demanding, all ILVES versions show greater speedups, with a maximum of $8.4\times$ over SHAKE and $4.2\times$ over MP-LINCS. For \texttt{constraints=h-bonds}, MP-LINCS performs similarly to SHAKE in most simulations, while ILVES-M and ILVES-F show consistent speedups over both SHAKE and MP-LINCS across all tolerances, achieving maximum speedups of $2.4\times$ and $3\times$ over SHAKE and MP-LINCS, respectively. 
The only case in Fig.~\ref{fig:speedup_over_shake_1t} where MP-LINCS outperforms ILVES-F is the BENZ simulation at the tightest tolerance. In it, each benzene molecule has only six constraints, and MP-LINCS achieves low errors in its projection step, sometimes avoiding any iteration of its correction step.
Note that the benzene molecule with constraints on just H-bonds is a very exceptional case where every constraint is fully decoupled from the rest. This makes the matrix that must be inverted (in the involved linear systems of equations) diagonal because there exists no atom that belongs to two different constraints, which makes the linear system suitable for a faster analytical solution.
Relevant research works recommend constraining both bond lengths and bond angles in benzene \cite{martinicoarsegrain}.

\begin{figure}[htbp]
    \centering
    \begin{minipage}[b]{0.98\textwidth}
        \centering
        \includegraphics[width=\textwidth]{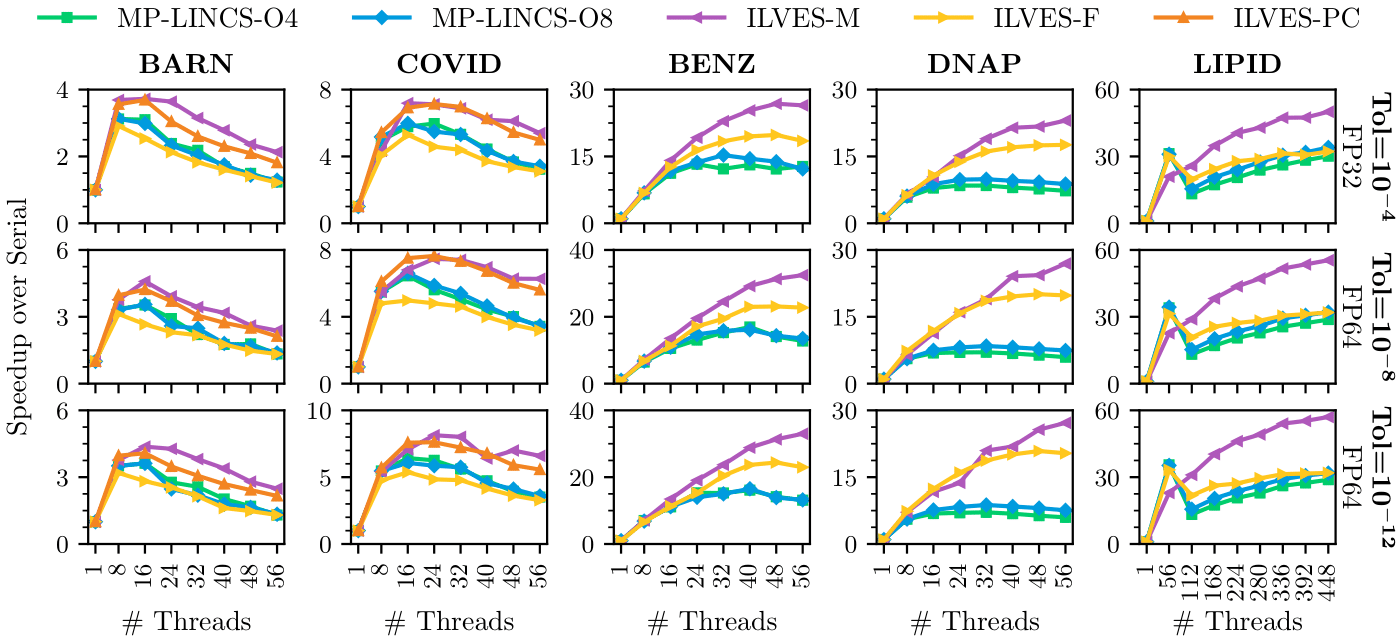}
        \renewcommand{\thefigure}{a}
        \\ (a) All bonds
        \label{fig:scalability_all_bonds}
    \end{minipage}

    \medskip

    \begin{minipage}[b]{0.98\textwidth}
        \centering
        \includegraphics[width=\textwidth]{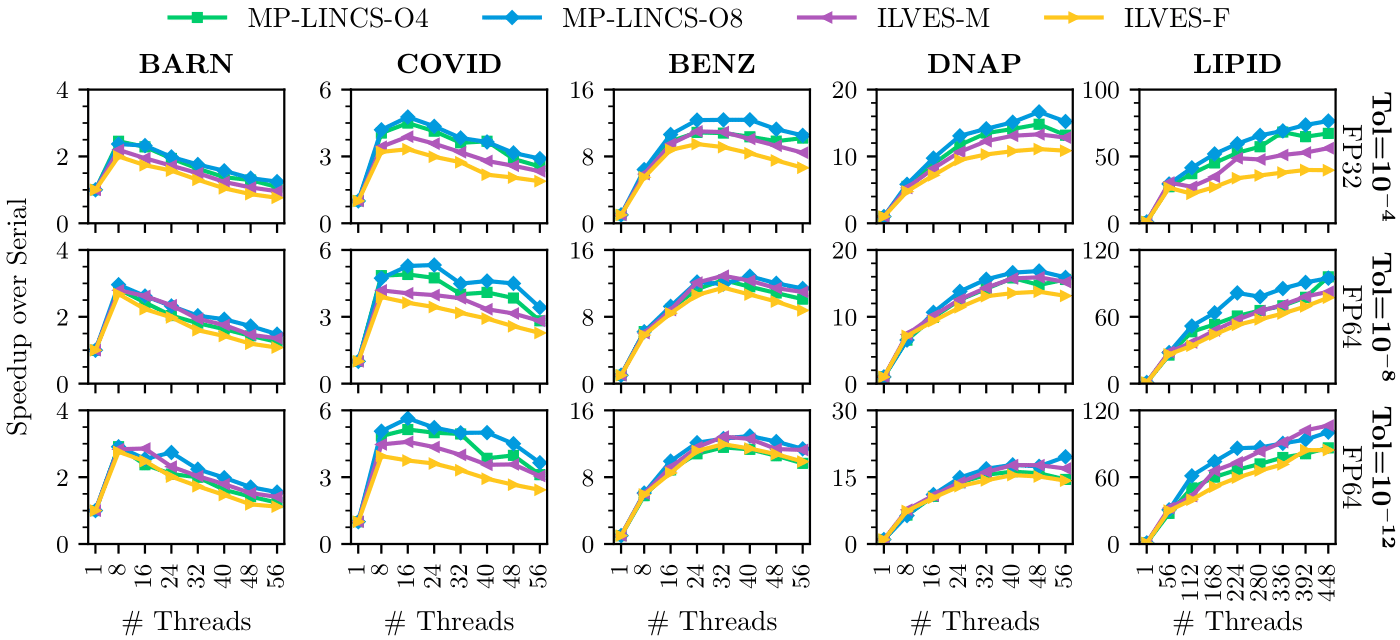}
        \renewcommand{\thefigure}{b}
        \\ (b) H-bonds
        \label{fig:scalability_h_bonds}
    \end{minipage}
    \caption{Multi-thread speedup over serial execution of MP-LINCS (\texttt{lincs-order=4} and \texttt{lincs-order=8}), ILVES-M, ILVES-F and ILVES-PC. The BARN, COVID, BENZ, and DNAP simulations are executed using a single task in a single chip of a computing node. The LIPID simulation is executed using up to 8 tasks (up to 4 nodes and 8 chips). (a) Constraints imposed on all bonds; (b) On H-bonds.}
    \label{fig:scalability}
\end{figure}

Fig.~\ref{fig:scalability} illustrates the parallel scalability of the constraint algorithms. The scalability of a solver is defined as the ratio of the execution time of the solver using $N$ threads to the execution time of the solver using 1 thread. The parallel scalability of the ILVES algorithms is significantly better in the BENZ and DNAP simulations with \texttt{constraints=all-bonds}. 
In these simulations, the graph partitioning produced by ILVES-M results in substantially less serial work compared to MP-LINCS. 
In the case of BENZ, the partitioning eliminates serial work entirely. The strong scalability of the BENZ system may seem counter-intuitive given that benzene molecules are small. However, the system consists of 2,000 benzene molecules, leading to a substantial parallel workload. While in other simulations the scalability of the ILVES algorithms and MP-LINCS is more similar, it is worth noting that ILVES delivers better performance than MP-LINCS in most cases. 
Consequently, for shorter execution times, the overhead of thread synchronization in ILVES becomes more pronounced, ultimately affecting its parallel scalability.

In most parallel simulations peak performance is not achieved using the maximum number of threads available. As more threads are added, synchronization overhead increases, and if zero-cut partitions are not possible, the amount of serial code executed also rises. When the parallel work becomes insufficient to offset the synchronization cost, the performance begins to decline. This behavior is particularly evident in the lipid bilayer simulation with \texttt{constraints=all-bonds}, where a performance drop occurs beyond 56 threads. This decline corresponds to the transition from a single task to multiple tasks, which increases synchronization and serial execution. In contrast, this drop is absent with \texttt{constraints=h-bonds} due to the lack of bonds between atoms in different tasks, thereby eliminating the need for inter-task communication.

\begin{figure}[htbp]
    \centering
    \begin{minipage}[b]{0.98\textwidth}
        \centering
        \includegraphics[width=\textwidth]{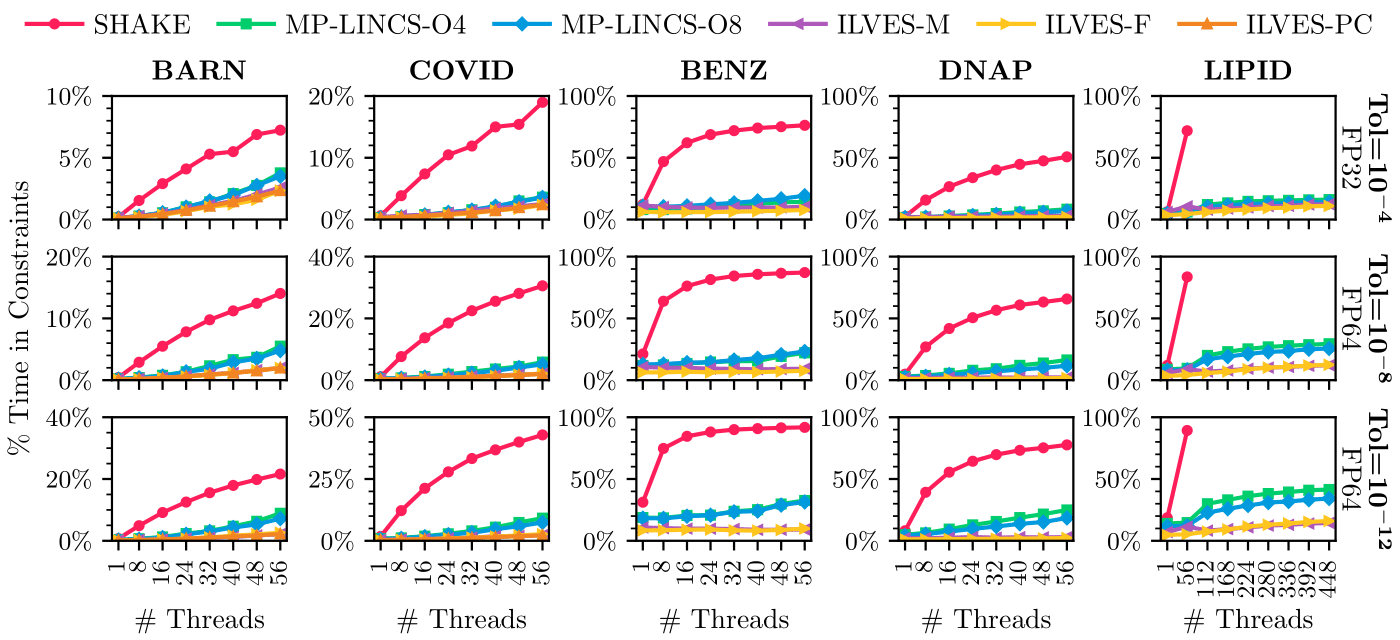}
        \renewcommand{\thefigure}{a}
        \\ (a) All bonds
        \label{fig:ntime_constr_all_bonds}
    \end{minipage}

    \medskip

    \begin{minipage}[b]{0.98\textwidth}
        \centering
        \includegraphics[width=\textwidth]{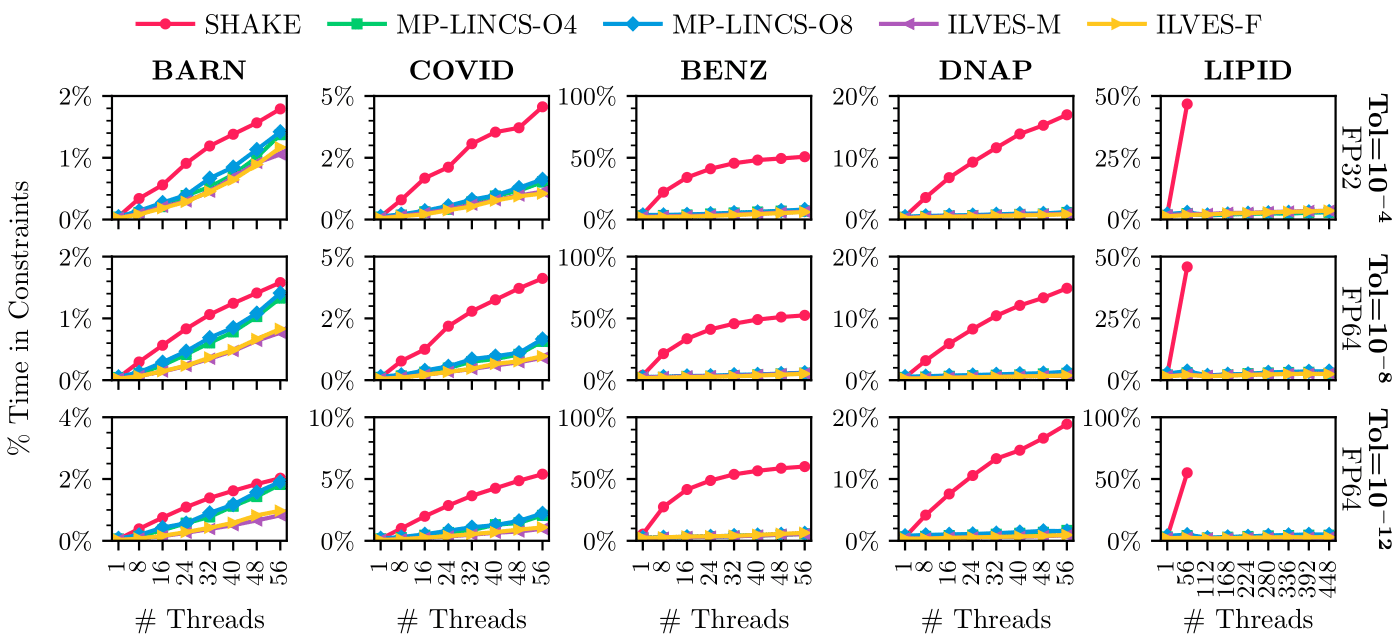}
        \renewcommand{\thefigure}{b}
        \\ (b) H-bonds
        \label{fig:ntime_constr_h_bonds}
    \end{minipage}
    \caption{Percentage of the total execution time of \texttt{Gromacs} spent in the constraint solver (ex water). (a) constraints imposed on all bonds; (b) On H-bonds.}
    \label{fig:ntime_constr}
\end{figure}

Fig.~\ref{fig:ntime_constr} displays the percentage of the total simulation execution time spent in the constraint solver (excluding water). In our multi-threaded simulations, with \texttt{constraints=all-bonds}, SHAKE accounts for up to 92\% of the total execution time, MP-LINCS up to 42\%, and ILVES up to 16\%. With \texttt{constraints=h-bonds}, SHAKE accounts for up to 60\%, MP-LINCS up to 6\%, and ILVES up to 5\%.

\clearpage
\bibliography{ilves}